\documentclass[aps,pre,onecolumn,10pt]{revtex4-1}

\expandafter\let\csname equation*\endcsname\relax
\expandafter\let\csname endequation*\endcsname\relax 

\usepackage[english]{babel}
\usepackage{amsmath}
\usepackage{amssymb}
\usepackage{graphicx}
\usepackage{color}
\usepackage{subfigure}

\usepackage{etoolbox}

\usepackage[utf8]{inputenc}
\usepackage{amsfonts}
\usepackage{amsthm}
\usepackage{placeins}
\usepackage{verbatim}
\usepackage{color,xcolor}
\usepackage{bm}
\usepackage{bbold}
\usepackage{nicefrac}
\definecolor{myorange}{HTML}{F9532B}
\definecolor{myblue}{HTML}{2c63bd}

\usepackage[pdfstartview=FitH,colorlinks=true,linkcolor=myblue,citecolor=myorange,urlcolor=orange]{hyperref}
\usepackage{comment}
\usepackage[all]{xy}
\def\Tr{\mathrm{Tr}}

\newcommand{\ra}{\right\rangle}
\newcommand{\la}{\left\langle}
\newcommand{\dd}{\mathrm{d}}
\newcommand{\I}{\mathcal{I}}

\DeclareMathOperator*{\intinf}{\int_{-\infty}^{+\infty}\!\!\!}
\DeclareMathOperator*{\Var}{Var}
\newcommand{\e}{\mathrm{e}}
\renewcommand{\Im}{\text{Im}}

\begin{document}

\title{Number statistics for $\beta$-ensembles of random matrices: applications to trapped fermions at zero temperature}

\author{Ricardo Marino}

\address{Department of Physics of Complex Systems, Weizmann Institute of Science, 76100 Rehovot, Israel.}

\author{Satya N. Majumdar, Gr\'egory Schehr}

\address{LPTMS, CNRS, Univ. Paris-Sud, Université Paris-Saclay, 91405 Orsay, France.}

\author{Pierpaolo Vivo}
\address{King's College London, Department of Mathematics, Strand, London WC2R 2LS, United Kingdom.}



\date{\today}

\begin{abstract}
Let $\mathcal{P}_{\beta}^{(V)} (N_\I)$ be the probability that a $N\times N$ $\beta$-ensemble of random matrices with confining potential $V(x)$ has $N_\I$ eigenvalues inside an interval $\I=[a,b]$ of the real line. We introduce a general formalism, based on the Coulomb gas technique and the resolvent method, to compute analytically $\mathcal{P}_{\beta}^{(V)} (N_\I)$ for large $N$. We show that this probability scales for large $N$ as $\mathcal{P}_{\beta}^{(V)} (N_\I)\approx \exp\left(-\beta N^2 \psi^{(V)}(N_\I /N)\right)$, where $\beta$ is the Dyson index of the ensemble. The rate function $\psi^{(V)}(k_\I)$, independent of $\beta$, is computed in terms of single integrals that can be easily evaluated numerically. The general formalism is then applied to the classical $\beta$-Gaussian ($\I=[-L,L]$), $\beta$-Wishart ($\I=[1,L]$) and $\beta$-Cauchy ($\I=[-L,L]$) ensembles. Expanding the rate function around its minimum, we find that generically the number variance $\Var(N_\I)$ exhibits a non-monotonic behavior as a function of the size of the interval, with a maximum that can be precisely characterized. These analytical results, corroborated by numerical simulations, provide the full counting statistics of many systems where random matrix models apply. In particular, we present results for the full counting statistics of zero temperature one-dimensional spinless fermions in a harmonic trap.

\end{abstract}

\maketitle

\tableofcontents

\section{Introduction}

There has been an intense activity in the field of cold atoms in the last two decades \cite{BloDalZwe08,PitStr03}. Since the first realizations of Bose-Einstein condensation \cite{AndEnsMat95,BraSacTol95,DavMewAnd95}, new experimental developments involving the cooling and confinement of particles in optical traps led to a new chapter in many-body quantum systems, where the particle statistics is the main interest rather than individual atoms.

After the achievement of cooling trapped fermions to the point where Fermi statistics becomes dominant \cite{DemJin99}, experiments were able to explore the remarkable properties of Fermi gases \cite{GioPitStr08}. Non-interacting fermions exhibit non-trivial quantum effects, arising from Pauli exclusion principle. While confined bosons may collapse to the lowest level of the trap, fermions are forced to dilute and behave as a strongly correlated system, regardless of their original interaction. This induces a Fermi motion of the particles, which is a purely quantum phenomenon: it downplays to some extent the importance of individual interactions among the fermions, which can be, in many cases, neglected or treated as a small perturbation \cite{PitStr03}. This turns the \emph{ideal} Fermi gas into the natural playground to explore the properties of more general Fermi gases.

Consider a one-dimensional Fermi gas of $N$ particles confined by a harmonic potential $V_Q(x)=\frac{1}{2}m\omega^2 x^2$, where the subscript $Q$ stands for the quantum potential. For simplicity, we set $m=\omega=\hbar=1$. The many-body Hamiltonian is then given by $H = \sum_{i=1}^N \left[-\frac{1}{2}\partial_{x_i}^2 + V_Q(x_i)\right]$. We want to calculate the many-body ground state wave function $\Psi_0(\vec{x})$, where $\vec x \equiv \{x_1,x_2,\cdots,x_N\}$ are the positions of the particles on a line. This wave function can be constructed from the single-particle eigenfunctions of a harmonic oscillator $\phi_n(x)\propto \e^{-x^2/2}H_n(x)$, (with energy eigenvalues $\epsilon_n =(n+1/2)$). Here $H_n(x)$ are Hermite polynomials of degree $n$. For example, $H_0(x) = 1$, $H_1(x) = 2x$, $H_2(x) = 2x^2-4$ etc. One then writes $\Psi_0(\vec x)$ as a Slater determinant $\Psi_0(\vec{x})=\det[\phi_i(x_j)]/\sqrt{N!}$, with $0\leq i \leq N-1$ and $1\leq j \leq N$. By construction, this wave function vanishes whenever $x_i = x_j$ for $i \neq j$, thus satisfying the Pauli exclusion principle. This corresponds to filling each single-particle energy level from $n=0$ to $n=N-1$ with a single fermion. The highest occupied energy level, $N-1$, is thus the Fermi energy. The ground state energy associated to this many-body wave function is $E_0 = \sum_{n=0}^{N-1}(n+1/2) = N^2/2$. Amazingly, this Slater determinant $\Psi_0(\vec{x})=\det[\phi_i(x_j)]/\sqrt{N!} = \frac{1}{\sqrt{N!}}\mathrm{e}^{-\sum_{i=1}^N x_i^2/2}\det[H_i(x_j)]$ can be evaluated explicitly, by noting that $\det[H_i(x_j)] \propto \prod_{i<j}(x_i-x_j)$, which is just a Vandermonde determinant. Thus the squared many-body ground state wave function can be written as~\cite{MarMajSch14} 	  
\begin{equation}
    |\Psi_0(\vec{x})|^2=\frac{1}{Z_N}\e^{-\sum_{i=1}^N x_i^2}\prod_{j<k}(x_k-x_j)^2,\label{eq:fermion wave eq}
\end{equation} 
where $Z_N$ is the normalization constant. This quantity $ |\Psi_0(\vec{x})|^2$ quantifies the quantum fluctuations in this system at zero temperature. It can indeed be interpreted as the probability density function of this fermionic system at zero temperature. Note that the Vandermonde term $\prod_{i<j}(x_i-x_j)^2$ in Eq. (\ref{eq:fermion wave eq}) makes the variables $x_i$'s strongly correlated. Here the physical origin of this strong correlation is due to the Pauli exclusion principle satisfied by the fermions. 

Precisely the same joint distribution also appears in the Gaussian Unitary Ensemble (GUE) of Random Matrix Theory (RMT). Consider an $N \times N$ random Hermitian matrix $X$ with complex entries drawn from the distribution $\Pr(X) \propto \mathrm{e}^{-\Tr(X^2)}$ which is invariant under a unitary transformation. For each realization of $X$, one has $N$ real eigenvalues $x_1,x_2,\cdots,x_N$ . The joint distribution of the eigenvalues $x_1,x_2,\cdots,x_N$ can be computed explicitly by making a change of variables from the entries to the eigenvalues and eigenvectors~\cite{Meh91}. The eigenvectors decouple and the joint distribution of eigenvalues has exactly the same form as in Eq. (\ref{eq:fermion wave eq}), where the Vandermonde term comes from the Jacobian of this change of variables. {\it Thus $N$ fermions in a harmonic trap at zero temperature provide a physical realization of the GUE eigenvalues} \cite{MarMajSch14}, the origin of the Vandermonde term in the two problems being however very different. This remarkable connection has allowed recently to explore various observables in the ground state of the trapped fermions system such as counting statistics and entanglement entropy \cite{CamVic10,CalMinVic11,Vic12,Eis13,EisPes14,Vic14,MarMajSch14,CalLedMaj14}.  

In RMT various ensembles have been studied extensively \cite{Meh91,For10}. In particular, for rotationally invariant ensembles, the joint probability distribution function (jpdf) of eigenvalues can be written explicitly as
\begin{equation}
    P(\vec{x})=\frac{1}{Z_{N,\beta}}\e^{-\beta N\sum_{i=1}^N V(x_i)}\prod_{j<k}|x_k-x_j|^\beta,\label{eq:eigen_jpdf_general_3}
\end{equation}
where $\beta > 0$ is the Dyson index of the ensemble and $V(x)$ is a potential growing suitably fast at infinity to ensure that the jpdf is normalizable. Usually, for rotationally invariant ensembles, the Dyson index is quantized: $\beta = 1, 2, 4$ respectively for real symmetric, complex Hermitian and quaternionic self dual matrices. However there are other matrix ensembles where $\beta > 0$ can be arbitrary \cite{DumEde02}. In the case $V(x) = x^2/2$ one recovers the $\beta$-Gaussian ensembles. The other well studied examples are $\beta$-Wishart ($V(x)={x}/{2}-\alpha\ln x$) and $\beta$-Cauchy ($V(x)=\left((N-1)/2+1/\beta\right)\ln(1+x^2)/N$), described in detail in section \ref{summary}. The fermions in a harmonic well at $T=0$ thus correspond to the special case of the RMT with $\beta = 2$ and $V(x)={x^2}/{2}$ (GUE). Note that in Eq. (\ref{eq:eigen_jpdf_general_3}) we have rescaled the potential $V(x)$ by a factor $N$. This convention ensures that the eigenvalues are of order ${\cal O}(1)$ (see the discussion later). With this convention, which we will use in the rest of the paper, the ground state squared wave function for fermions in Eq. (\ref{eq:fermion wave eq}) reads 
\begin{equation}
    |\Psi_0(\vec{x})|^2=\frac{1}{\tilde Z_N}\e^{-N\,\sum_{i=1}^N x_i^2}\prod_{j<k}(x_k-x_j)^2,\label{eq:fermion wave eq_scaled}
\end{equation} 
where $\tilde Z_N$ is the new normalization constant (in this particular case, this expression (\ref{eq:fermion wave eq_scaled}) is obtained from Eq. (\ref{eq:fermion wave eq}) just by a rescaling of the fermions' positions $x_i \to \sqrt{N} x_i$). In the context of step fluctuations in vicinal surfaces of a crystal, such a fermionic representation of GUE was already noticed (see \cite{Ein03} for a review). For the steps in presence of a hard wall, a similar fermionic representation was found that corresponds to the Wishart case of RMT with $\beta=2$ \cite{NaMa09}.

\begin{figure}[!htbp]
    \includegraphics[width=.55\textwidth]{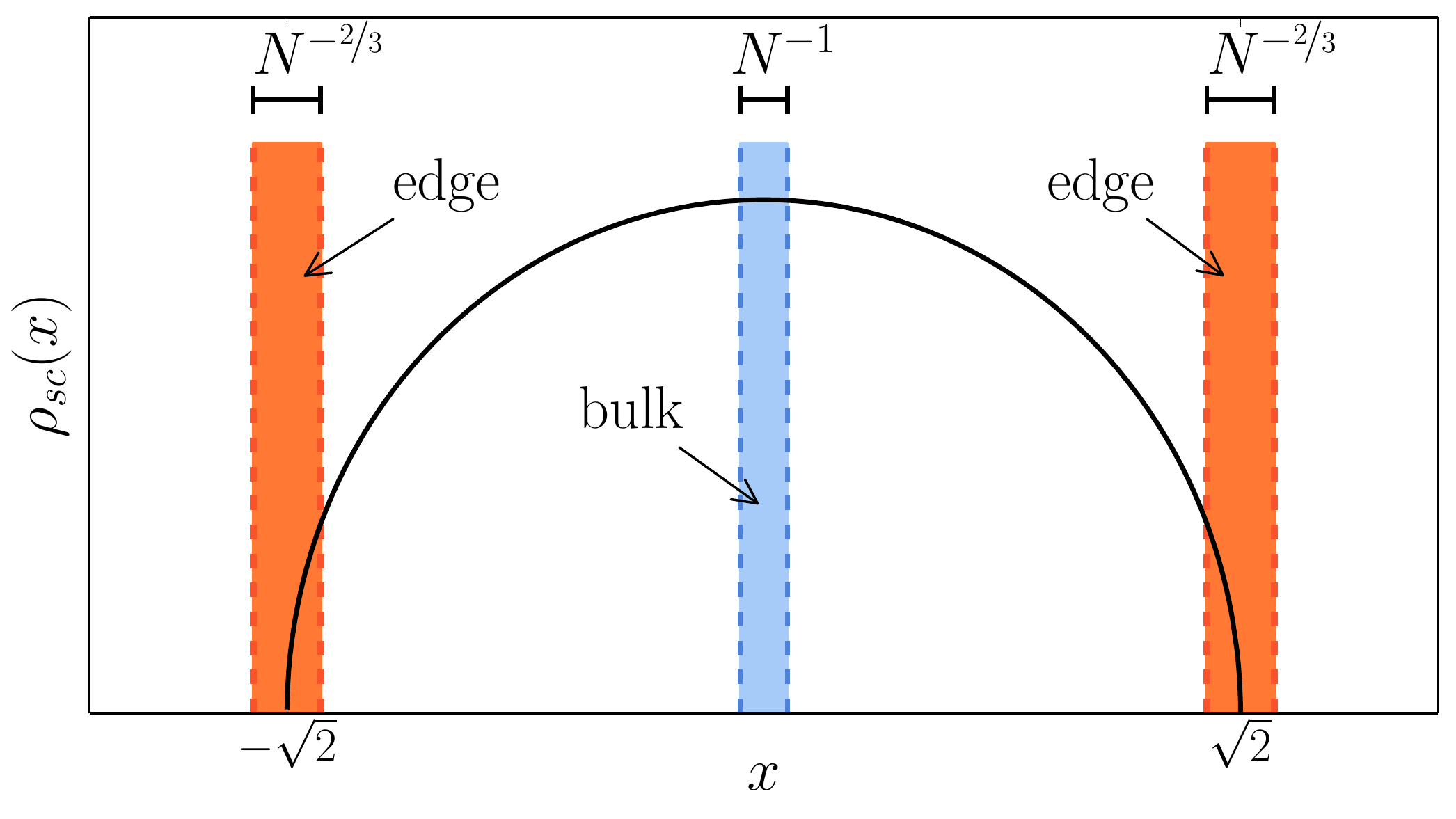}
    \caption{Average density of scaled Gaussian eigenvalues with a representation of bulk and edge regimes. The bulk is the scale of typical fluctuations of eigenvalues, whose width is of order $1/N$. The edge is the scale of typical fluctuations of the largest eigenvalue.}
    \label{fig:gaussian bulk and edge}
\end{figure}

One of the most natural observables in both RMT as well as in many body quantum physics is the global density 
\begin{equation}
    \rho_N(x)=\frac{1}{N}\sum_{i=1}^N\delta(x-x_i),\label{densrandom}
\end{equation}
where $\{ x_i\}$'s are the positions of eigenvalues or fermions on the line. Note that $\rho_N(x)$ is normalized to unity: $\int \rho_N(x)\,\dd x = 1$ and thus measures the fraction of eigenvalues in the interval $[x,x+dx]$ in any given sample. The average density of eigenvalues $\langle \rho_N(x)\rangle$, where $\langle \cdots \rangle$ stands for an average over the jpdf in Eq. (\ref{eq:eigen_jpdf_general_3}), converges to the celebrated Wigner semicircle law \cite{Wig58,Meh91} in the large $N$ limit
\begin{equation}
    \la \rho_N(x) \ra \xrightarrow{N\gg 1} \rho_{sc}\left(x\right),\quad \quad \rho_{sc}(x)=\frac{1}{\pi}\sqrt{2-x^2} \;.\label{eq:wigner semicircle intro}
\end{equation}
Therefore the eigenvalues are spread, on an average, over a finite interval $[-\sqrt{2},+\sqrt{2}]$. The typical spacing between eigenvalues, near the origin, is thus of order ${\cal O}(1/N)$. This regime near the center is known as the {\it bulk}, as explained in Fig. \ref{fig:gaussian bulk and edge}. In contrast, as one approaches the edges $\pm \sqrt{2}$ the density becomes smaller, indicating that the eigenvalues are sparser near the edges. Indeed, it is well known that the typical spacing between eigenvalues near the edges scales as ${\cal O}(N^{-2/3})$ for large $N$. This regime is known as the {\it edge} regime (see Fig. \ref{fig:gaussian bulk and edge}).

An important observable to characterize quantum fluctuations of cold fermions is their {\it number statistics}, or full counting statistics. This is the number $N_\I$ of fermions inside an interval $\I$ at zero temperature. For the ideal bosonic case at $T=0$, all particles occupy the ground state, centered at the minimum of the trap. They are uncorrelated and their number statistics can be easily determined. In the fermionic case, where Pauli principle applies, the statistics of the random variable $N_\I$ is instead particularly interesting, as it reflects the non trivial quantum correlations between the fermion positions. As a simple example, consider a symmetric interval $\I = [-L,L]$ for the fermions in a harmonic trap. In the context of RMT, this is just the number of eigenvalues in $[-L,L]$ in the GUE. This number $N_{[-L,L]}$ is a random variable whose mean can be easily computed, for large $N$, from the average density given in Eq. (\ref{eq:wigner semicircle intro})
\begin{equation}\label{eq:averageNI}
\langle N_{[-L,L]}\rangle = N\int_{-L}^{+L} \langle \rho_N(x) \rangle \, \dd x  \approx \frac{N}{\pi} \left(L\,\sqrt{2-L^2} +2 \sin ^{-1}\left(\frac{L}{\sqrt{2}}\right)\right) \;, \; 0 \leq L \leq \sqrt{2} \;.
\end{equation}
For $L \geq \sqrt{2}$, $\langle N_{[-L,L]}\rangle$ saturates to its maximum value, {\it i.e.}, $\langle N_{[-L,L]}\rangle \approx N$. 

\begin{figure}[!htbp]
    \includegraphics[width=.55\textwidth]{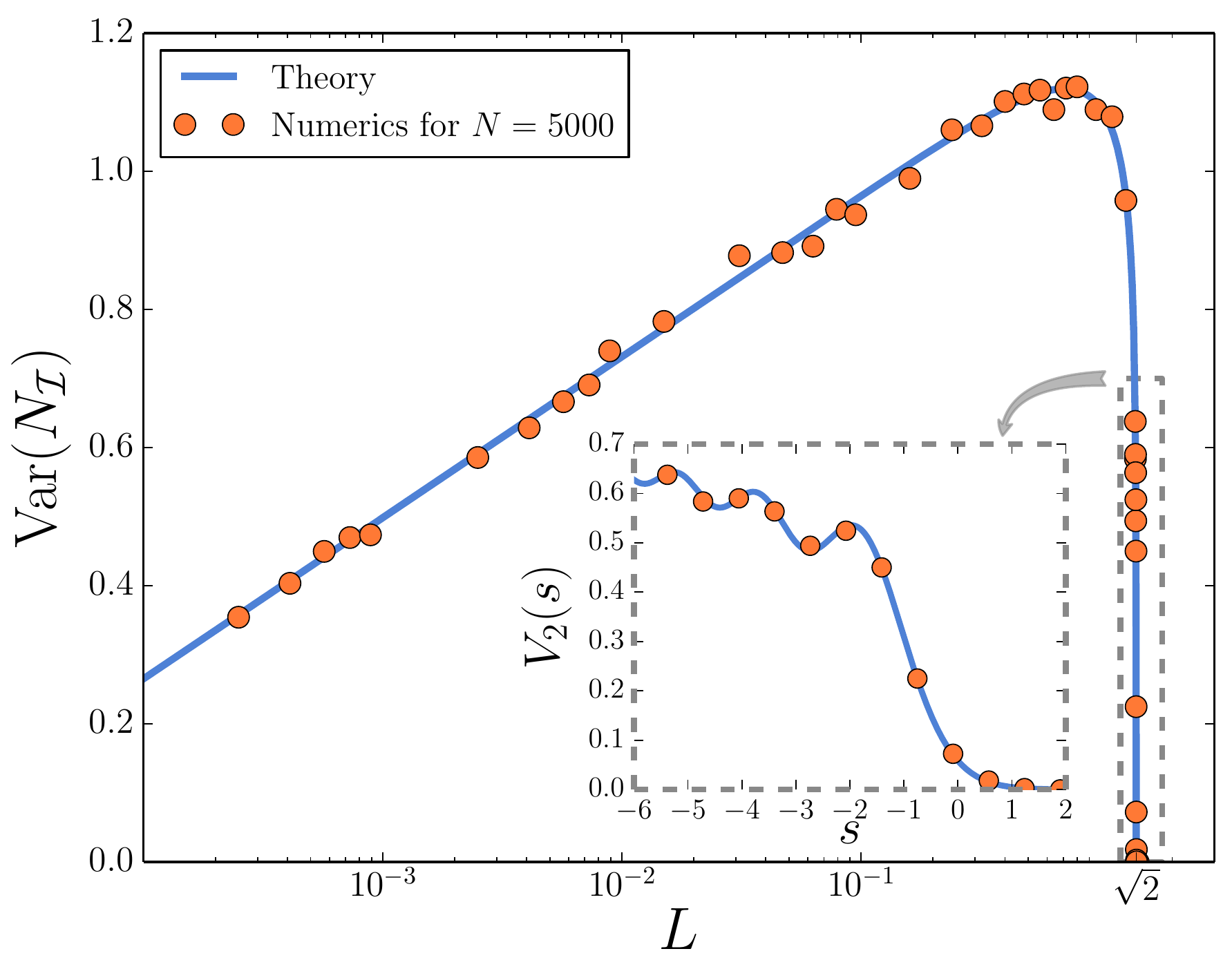}
    \caption{Results for the number variance of GUE eigenvalues when $\I=[-L,L]$. The theoretical result is in equation \eqref{eq:summary variance fermions_intro}.}
    \label{fig:variance_gaussian}
\end{figure}

However, the random variable $N_{[-L,L]}$ fluctuates around this mean value from sample to sample. It is therefore natural to study the higher cumulants of this random variable, for instance the variance ${\rm Var}(N_{[-L,L]})$, and eventually the full distribution of $N_{[-L,L]}$. In RMT, the statistics of $N_{[-L,L]}$ in Gaussian ensembles ($\beta = 1,2, 4$) is actually well known, but only in the {\it bulk} limit when $L \sim {\cal O}(1/N)$. In this case, the variance was computed by Dyson in~\cite{Dys62}  
\begin{equation}
    \Var(N_{[-L,L]})\approx \frac{2}{\beta\pi^2}\ln\left(NL\right)+C_\beta+o(1) \;, \; L \sim {\cal O}(1/N), \label{eq:dyson and mehta}
\end{equation}
where $C_\beta$ is a $\beta$-dependent constant that was computed explicitly by Dyson and Mehta \cite{DysMeh63}. For instance, for $\beta = 2$
\begin{eqnarray}\label{C2}
C_2 = \frac{1}{\pi^2}(1 +\gamma + \ln 2) = 0.230036\ldots \;,
\end{eqnarray}
where $\gamma = 0.577215\ldots$ is the Euler constant. In this bulk regime, even the full distribution of $N_{[-L,L]}$ was computed in Refs. \cite{CosLeb95} and \cite{FogShk95} and it was shown to be a simple Gaussian. However, beyond the bulk regime, {\it i.e.}, $L \gg {\cal O}(1/N)$, the behavior of the variance $\Var(N_{[-L,L]})$ (and also the full distribution of $N_{[-L,L]}$) was not studied in the RMT literature. In the context of fermions trapped in a confining potential, this number variance has found a recent renewed interest, as it was found to be closely related to the entanglement entropy at $T=0$ of the subsystem $[-L,L]$ with the rest of the system \cite{CamVic10,CalMinVic11,Vic12,Eis13,EisPes14,Vic14,CalLedMaj14}. In that context, $ \Var(N_{[-L,L]})$ was studied numerically as a function of $L$ for various confining potentials and a rather striking non-monotonous dependence on $L$ was found \cite{Vic12,Eis13} (see Fig.~\ref{fig:variance_gaussian}).

In a recent Letter \cite{MarMajSch14}, we studied $\Var(N_{[-L,L]})$ analytically for fermions in a harmonic potential. Exploiting the mapping to the GUE eigenvalues and using a Coulomb gas technique, we were able to compute $\Var(N_{[-L,L]})$ for any $L$ and large $N$. As a function of increasing $L$, we found the following behavior for the variance \cite{MarMajSch14} 
\begin{equation}
    \Var(N_{[-L,L]})\approx\begin{cases}\frac{1}{ \pi^2}\ln\left(N L (2-L^2)^{3/2}\right), & N^{-1}<L<\sqrt{2}-N^{-2/3}\\
                    V_2(s), & L=\sqrt{2}+\frac{s}{\sqrt{2}}N^{-2/3}\\
                    \exp[-2 N \phi(L)], & L>\sqrt{2}+N^{-2/3},
                    \end{cases}\label{eq:summary variance fermions_intro}
\end{equation}
where the functions $V_2(s)$ and $\phi(L)$ were computed explicitly and are given in Eqs. \eqref{V_s_gaussian} and \eqref{eq:phiL} respectively. In the bulk limit, using $L \sim {\cal O}(1/N)$ in the first line of Eq. (\ref{eq:summary variance fermions_intro}), we recover the results of Dyson in Eq. \eqref{eq:dyson and mehta} for $\beta = 2$. In Fig. \ref{fig:variance_gaussian} these theoretical results (\ref{eq:summary variance fermions_intro}) are compared to numerical simulations and one finds a very good agreement. In addition, the full distribution of $N_{[-L,L]}$ was also computed in Ref. \cite{MarMajSch14} using the Coulomb gas method. The large deviation function associated with the distribution of $N_{[-L,L]}$ was found to have a non-trivial logarithmic singularity where the distribution has its peak. This singularity was attributed to a phase transition in the associated Coulomb gas problem \cite{MarMajSch14}.

The goal of this paper is twofold. First, we present the details of the Coulomb gas calculations for $\beta$-Gaussian ensembles which were just announced in our Letter \cite{MarMajSch14}. This is a generalization of the results for the $\beta = 2$ case mentioned in Eq. (\ref{eq:summary variance fermions_intro}). Secondly, we show that our results can be generalized to compute the number variance of an interval ${\cal I}$ for two other classical random matrix ensembles: $\beta$-Wishart and $\beta$-Cauchy ensembles. The rest of the paper is organized as follows. We begin by recalling a few definitions of random matrix theory and summarizing our main results in section \ref{summary}. We then describe the details of the Coulomb gas method and the derivation of the results for the three ensembles: $\beta$-Gaussian in section \ref{sec:gaussian}, $\beta$-Wishart in section \ref{sec:wishart} and $\beta$-Cauchy in section \ref{sec:cauchy}. We present our conclusions in section \ref{sec:conclusions} and Appendix \ref{app} contains details of several calculations performed in section~\ref{sec:gaussian}.

\section{Setting and summary of results}
\label{summary}

We consider ensembles of random matrices for which the joint distribution of eigenvalues is given in Eq. (\ref{eq:eigen_jpdf_general_3}) where $V(x)$ denotes the matrix potential. In this paper we focus on three classical random matrix ensembles, $\beta$-Gaussian ($V(x)=x^2/2$), $\beta$-Wishart ($V(x)={x}/{2}-\alpha\ln x$) and $\beta$-Cauchy ($V(x)=\left((N-1)/2+1/\beta\right)\ln(1+x^2)/N$), described in details below.

Our quantity of interest is the number $N_\I$ of eigenvalues inside an interval $\I$. This is a random variable which can be expressed as $N_\I=\sum_i \mathbb{1}_\I(x_i)$, where $\mathbb{1}_\I(x_i)$ is the indicator function $=1$ when $x_i\in\I$ and zero otherwise. Note that for the Gaussian case if we choose the interval $\I = [0,+\infty)$, then $N_\I$ is just the index, i.e., the number of positive eigenvalues.
In this special case, the distribution of $N_\I$ was computed using
Coulomb gas method in Refs. \cite{MajNadSca11,MajNadSca09}. The same method was later
used to compute the distribution of the number of Wishart eigenvalues
in the interval $\I = [l,+\infty)$ \cite{MajViv12}. In all these cases, the interval $\I$ was unbounded.
Here, we show that this Coulomb gas method can also be used to compute the distribution of $N_\I$ where
the interval $\I$ is bounded, {\it i.e.}, $\I = [a,b]$. For such an interval, the probability density of $N_\I$ is given by
\begin{equation}
    \mathcal{P}_{\beta}^{(V)}(N_\I)=\int \prod_{i=1}^N\dd x_i P(\vec{x})\delta\left(N_\I-\sum_{l=1}^N\mathbb{1}_\mathcal{I}(x_l)\right),\label{prob-1}
\end{equation}
where $P(x_1,\ldots,x_N)$ is the jpdf of the eigenvalues \eqref{eq:eigen_jpdf_general_3}. The superscript $(V)$ in Eq. (\ref{prob-1}) just refers to the potential $V(x)$. The average $\langle N_{\I}\rangle$ is easily obtained in the large $N$ limit by integrating the average density over $\I$: 
\begin{equation}\label{eq:average_NI}
    \langle N_\I \rangle \approx N \, \int_\I \rho(x)\dd x \;,
\end{equation}
where $\rho(x) = \lim_{N \to \infty} \langle \rho_N(x) \rangle$ is the limiting average density of eigenvalues, which depends explicitly on the potential $V(x)$. While the computation of $\rho(x)$ is relatively straightforward and is well known for the three ensembles mentioned above, it is much harder to compute the variance, $\Var(N_\I)=\langle N_\I^2\rangle - \langle N_\I \rangle^2$, and the full distribution $\mathcal{P}_{\beta}^{(V)}(N_\mathcal{I})$. We show that for large $N$, and as long as the interval $\I$ is contained within the support of $\rho(x)$, the distribution of $N_\I$ admits a large deviation form
\begin{equation}
     \mathcal{P}_{\beta}^{(V)}(N_\mathcal{I}=k_\mathcal{I}N)\approx \e^{-\beta N^2 \psi^{(V)}(k_\I)},\label{prob_final}
\end{equation} 
where $k_\I = N_\I/N$ denotes the fraction of eigenvalues in the interval ${\cal I}$ and $\psi^{(V)}(k_\I)$ is the {\it rate function} independent of both $\beta$ and $N$ that can be derived by a general method described below. 

This rate function $\psi^{(V)}(k_\I)$ reaches its minimum at $k_\I=\overline{k_\I} = \langle N_\I\rangle/N \approx  \int_\I \rho(x)\, \dd x$ [see Eq. (\ref{eq:average_NI})]. In standard large deviation setting, usually the rate function exhibits a smooth quadratic behavior around this minimum. In our case, we find that for all three potentials the rate function, near its minimum, behaves as $\psi^{(V)}(k_\I)  \propto (k_\I - \overline{k_\I})^2/[-\ln{(|k_\I - \overline{k_\I}|/a_\I)}]$ where $a_\I$ is a  non-universal constant that depends on the interval $\I$ and on the potential. The rate function thus indeed
has a quadratic behavior close to $k_\I = \overline{k_\I}$, except that it is actually modulated by a weak logarithmic singularity. This singularity occurs when we investigate the fluctuations on a scale $N_{\I} - \langle{N_\I}\rangle \sim {\cal O}(N)$. Writing $\ln{(|k_\I - \overline{k_\I}|/a_\I)} = \ln{|N_\I - \langle N_\I\rangle|} - \ln (N \, a_\I)$, it is clear that if $|N_\I - \langle N_\I\rangle| \ll N$ one can replace $\ln{(|k_\I - \overline{k_\I}|)}$ by $-\ln (N a_\I)$. Thus on this much smaller scale, it follows from Eq. (\ref{prob_final}) that the fluctuations just become Gaussian 
\begin{equation}
\mathcal{P}_{\beta}^{(V)}(N_\mathcal{I}=k_\mathcal{I}N)\approx \exp\left[-\frac{\left(N_\I-N\overline{k_\I}\right)^2}{2\left(\mathrm{Var}(N_\I)\right)}\right]\mbox{ for }N_\I\to N\overline{k_\I}\label{Pintrovargauss} \;,
\end{equation}
with the variance growing, for large $N$, as 
\begin{equation}\label{eq:variance_logN}
    \Var(N_\I)\propto \ln (N a_\I) \;.
\end{equation}
Indeed, this quadratic behavior (modulated by logarithmic singularity) of the rate function near its minimum already appeared in the special cases of semi-infinite intervals $\I = [0,+\infty)$ (for the Gaussian case) \cite{MajNadSca11,MajNadSca09} and for $\I = [l,+\infty)$ (for the Wishart case)~\cite{MajViv12} mentioned earlier. Here we find that this logarithmic singularity is more general and holds also for arbitrary bounded intervals $\I$.

In summary, there are two scales associated with the fluctuations of $N_\I$. There is a shorter scale, $N_{\I} - \langle{N_\I}\rangle \sim {\cal O}(\sqrt{\ln (N a_\I)})$, which describes the {\it typical} fluctuations of $N_\I$ around its average and the distribution of these typical fluctuations is Gaussian with a variance $\propto \ln (N a_\I)$ for large $N$. However, the {\it atypically} large fluctuations where $N_{\I} - \langle{N_\I}\rangle \sim {\cal O}(N)$ are not described by the Gaussian distribution, rather by the large deviation form in Eq.~(\ref{prob_final}). The logarithmic singularity of $\psi^{(V)}(k_\I)$ around $\overline{k_\I}$ ensures the matching of these two scales. In this paper we compute the rate function $\psi^{(V)}(k_\I)$ for the three ensembles mentioned above. Even though the rate function $\psi^{(V)}(k_\I)$ depends on the choice of the ensemble, we will show that its logarithmic singularity near its minimum is universal. As a consequence, the variance of the typical Gaussian fluctuations around the mean grows universally as $\ln N$ for large $N$. Below we provide a summary of the results for each of these three ensembles separately.


\vspace*{0.5cm}

\emph{$\beta$-Gaussian ensemble:} $V(x)=x^2/2$. This ensemble includes real symmetric ($\beta=1$), complex Hermitian ($\beta=2$) or quaternion self-dual ($\beta=4$) matrices whose entries are Gaussian variables such as the probability density of the matrix $X$ is given by $\Pr(X)\propto \mathrm{e}^{-\beta N \Tr X^2/2}$. The average spectral density for large $N$ and for any $\beta>0$ converges to the celebrated Wigner's semi-circle law $\rho_{sc}(x)$, described in equation \eqref{eq:wigner semicircle intro}. The rate function $\psi^{(G)}(k_\I)$ in Eq. (\ref{prob_final}) is given explicitly in Eq.~\eqref{action_gauss} below.

Our results for the number variance of the Gaussian ensemble are given by
\begin{equation}
    \Var(N_{[-L,L]})\approx\begin{cases}\frac{2}{ \beta\pi^2}\ln\left(N L (2-L^2)^{3/2}\right), & N^{-1}<L<\sqrt{2}-N^{-2/3}\\
                    V_\beta(s), & L=\sqrt{2}+\frac{s}{\sqrt{2}}N^{-2/3}\\
                    \exp[-\beta N \phi(L)], & L>\sqrt{2}+N^{-2/3},
                    \end{cases}\label{eq:summary variance fermions}
\end{equation}
and were announced in the Letter \cite{MarMajSch14}. We were able to calculate $V_\beta(s)$ for $\beta=2$, equation \eqref{V_s_gaussian}, and $\phi(L)$ is given by equation \eqref{eq:phiL}. A plot of equation \eqref{eq:summary variance fermions} for $\beta=2$, which is equivalent to the statistics of the harmonically confined ideal Fermi gas when distances are scaled by $\sqrt{N}$ and $\hbar=m=\omega=1$, is presented in figure \ref{fig:variance_gaussian} compared with numerical results. 

\begin{figure}
    \centering
    \renewcommand\thesubfigure{(\roman{subfigure}) }
    \subfigure[Bulk regime.]{\includegraphics[width=0.45\textwidth]{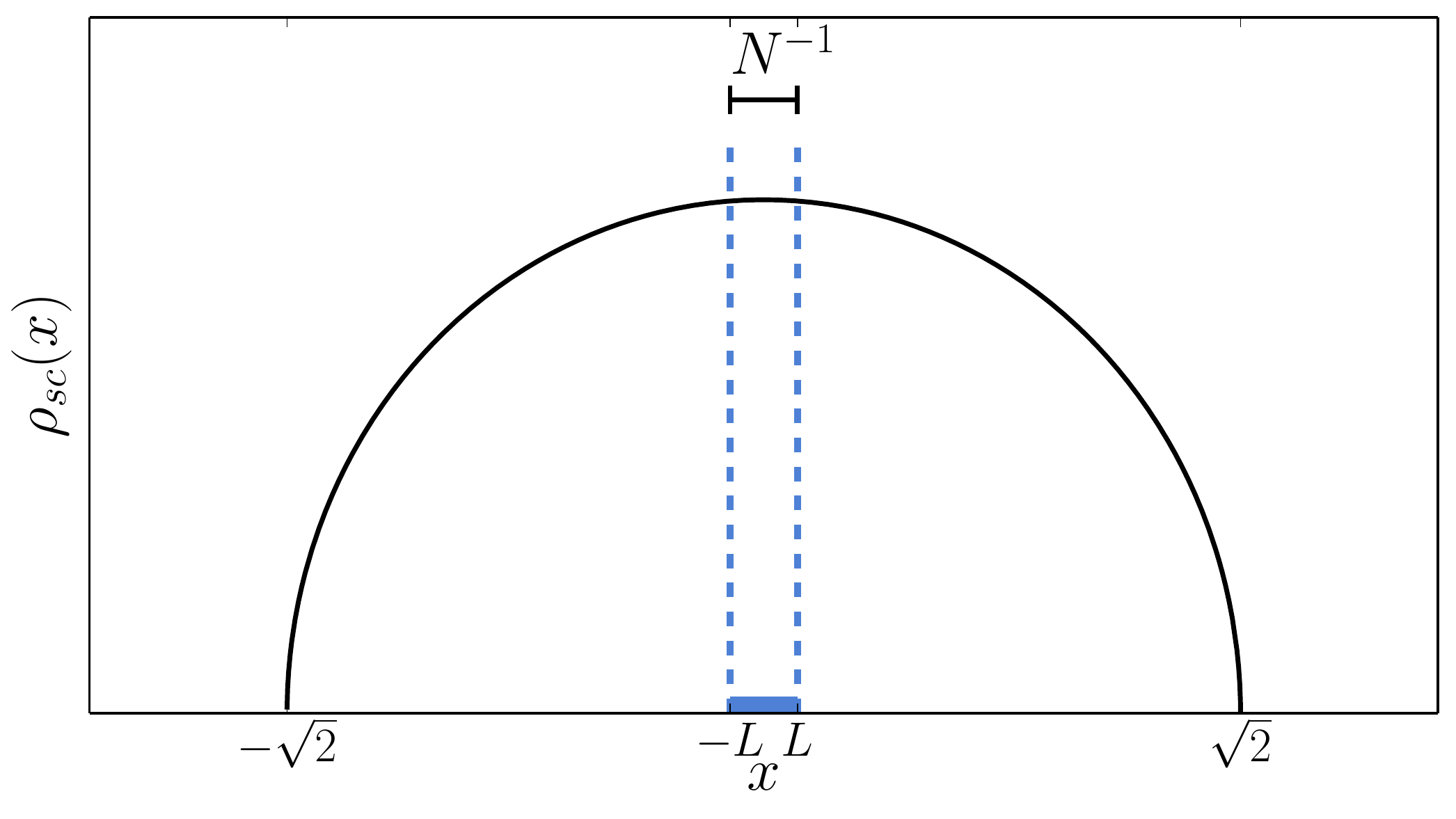}}
    \subfigure[Extended bulk regime.]{\includegraphics[width=0.45\textwidth]{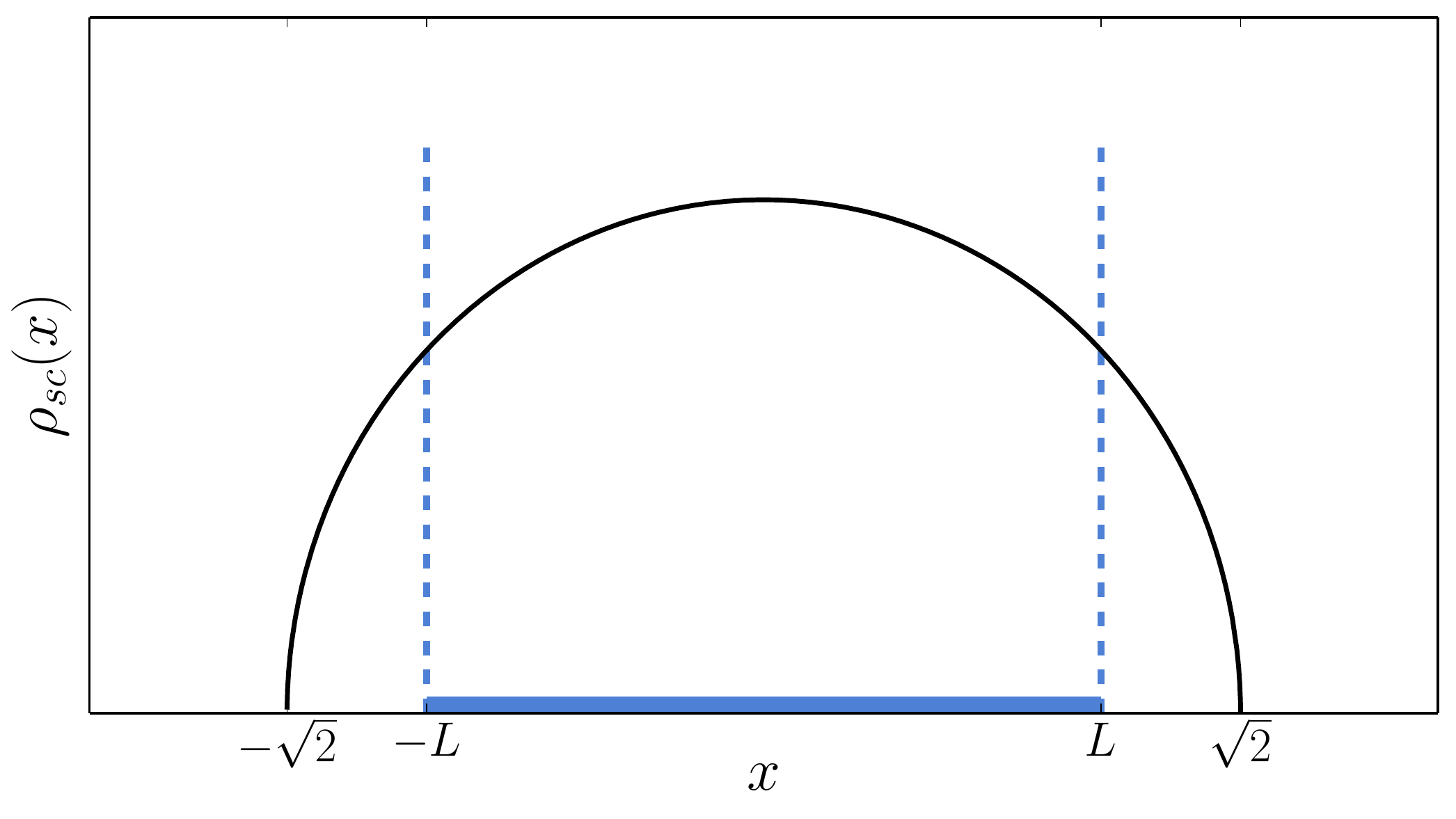}}
~
    \subfigure[Edge regime.]{\includegraphics[width=0.45\textwidth]{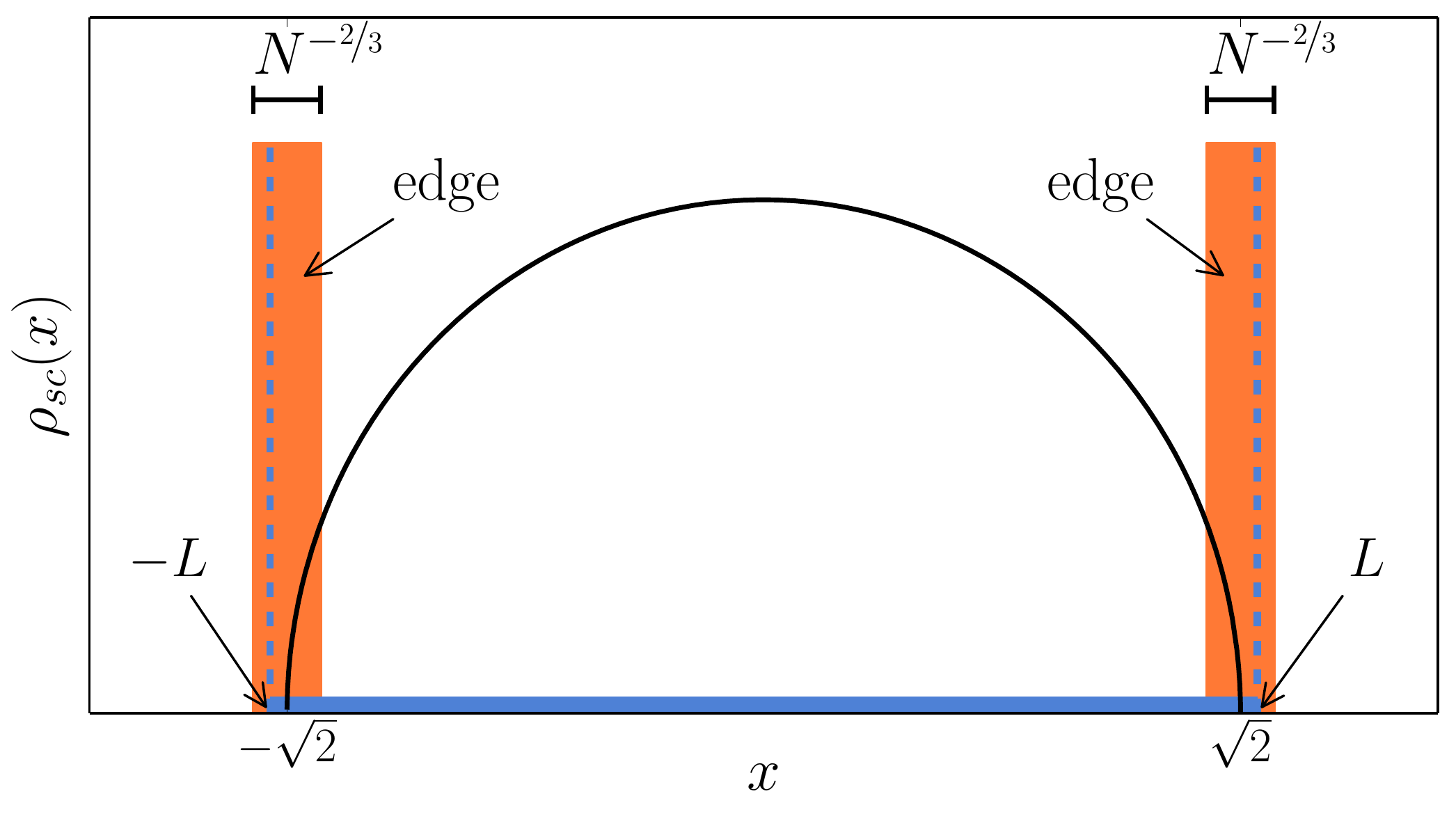}}
    \subfigure[Tail regime.]{\includegraphics[width=0.45\textwidth]{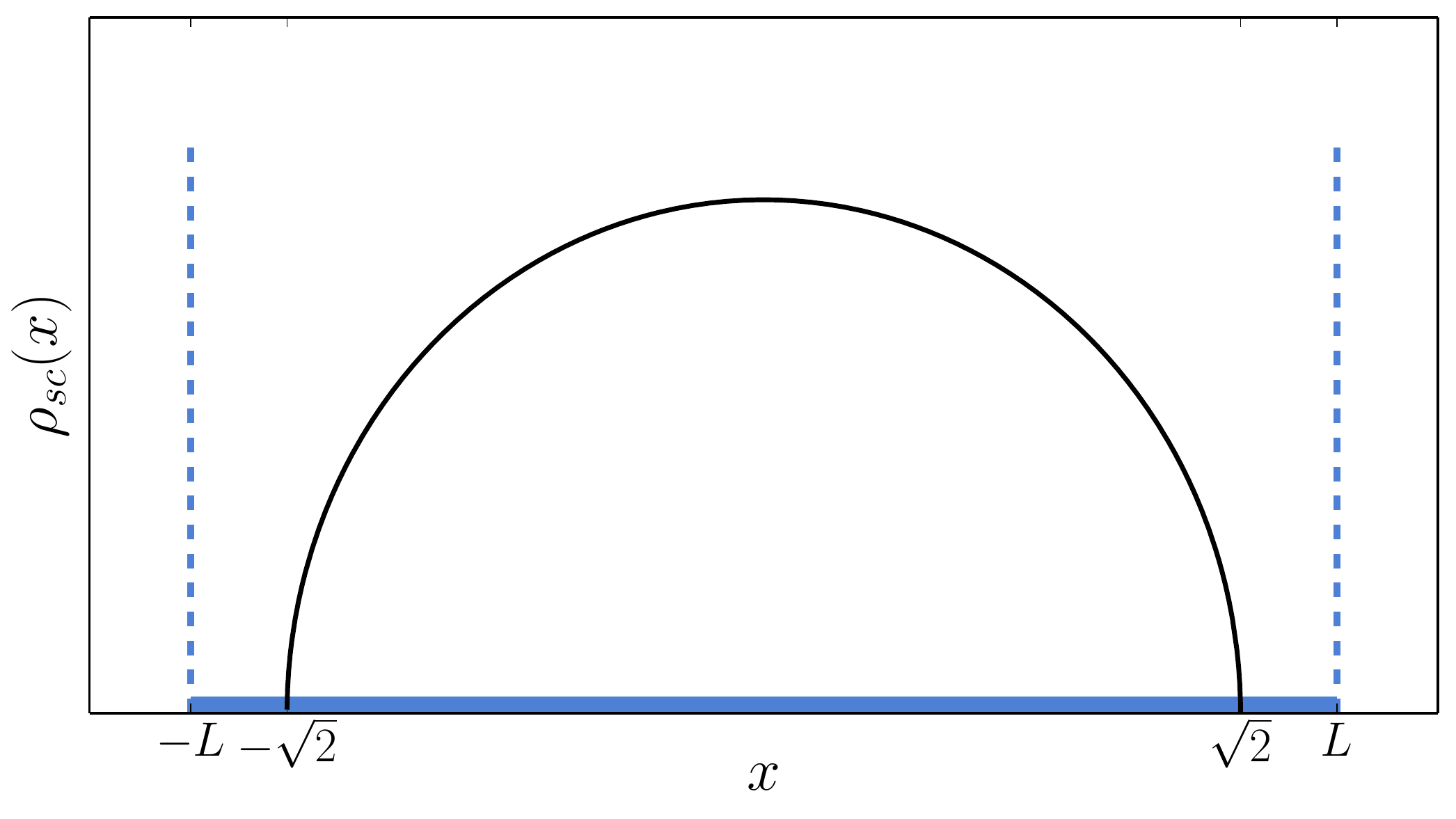}}
    \caption{Regimes of behavior of the number variance for the Gaussian ensemble. The solid blue line represents the interval $\I=[-L,L]$.}
    \label{fig:regimes_gauss_intro}
\end{figure}

As $L$ increases, the number variance for the interval $[-L,L]$ changes behavior, and we identify four regimes, highlighted in figure \ref{fig:regimes_gauss_intro}: \rm{(i)} a \emph{bulk} regime, \rm{(ii)} an \emph{extended bulk} regime, \rm{(iii)} an \emph{edge} regime and  \rm{(iv)} a \emph{tail} regime. The variance has a non-monotonic behavior, reaches its maximum at a non-trivial value ($L^\star=1/\sqrt{2}$) and drops sharply near the edge of the semicircle ($L=\sqrt{2}$). From the first line of equation \eqref{eq:summary variance fermions}, setting $L = s/N$, where $s$ is of order ${\cal O}(1)$, and taking the large $N$ limit we get ${\rm Var}(N_{[-L,L]}) \sim (2/\beta \pi^2) \ln s$, thus recovering the result of Dyson in Eq. (\ref{eq:dyson and mehta}). This is also evident in figure \ref{fig:variance_gaussian} where we see a linear growth on a logarithmic scale in the early regime. 

\begin{figure}[!htbp]
    \centering
    \subfigure[\label{fig:wishart distribution}]{\includegraphics[width=0.435\textwidth]{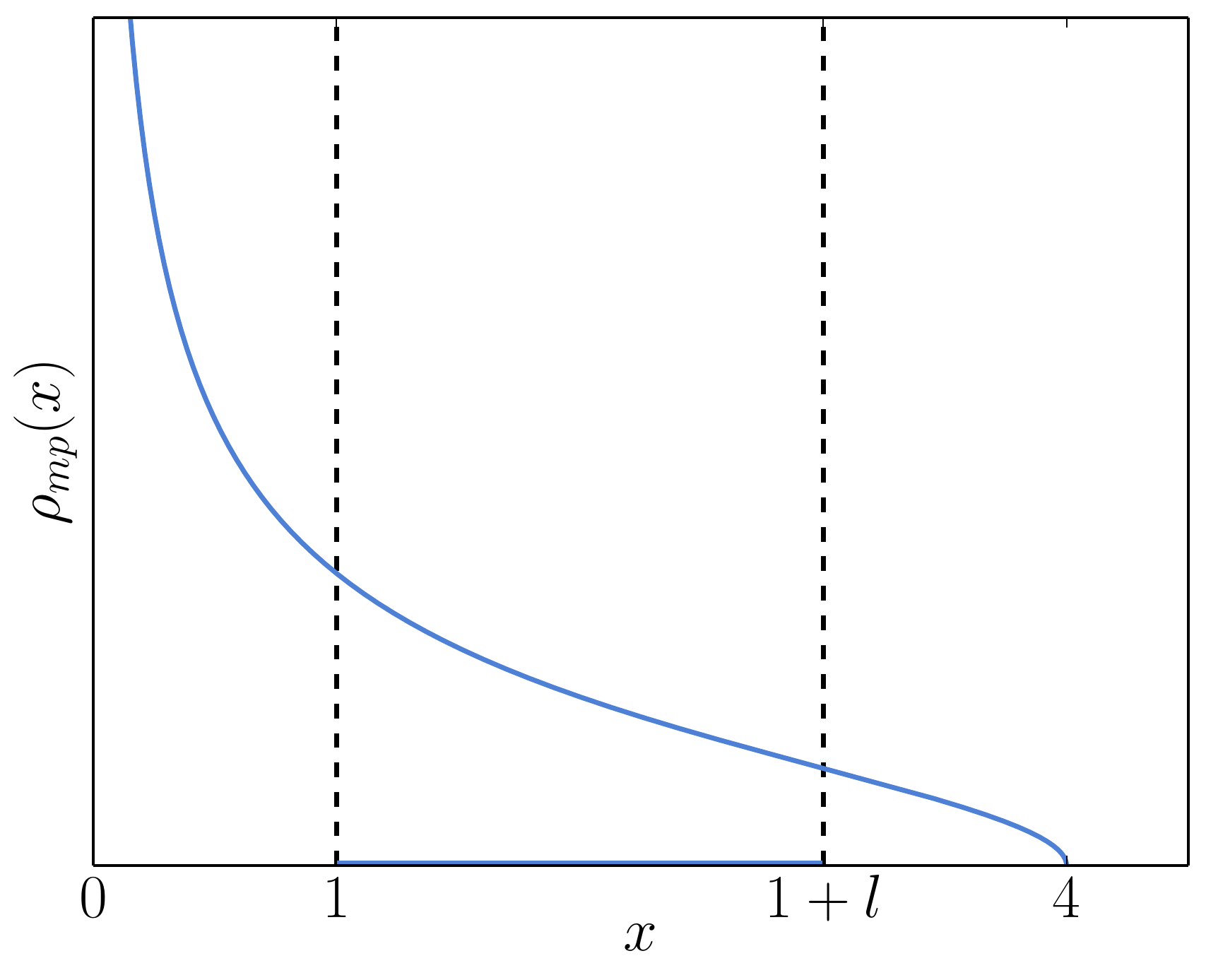}}
    \subfigure[\label{fig:var_wishart_summary}]{\includegraphics[width=0.465\textwidth]{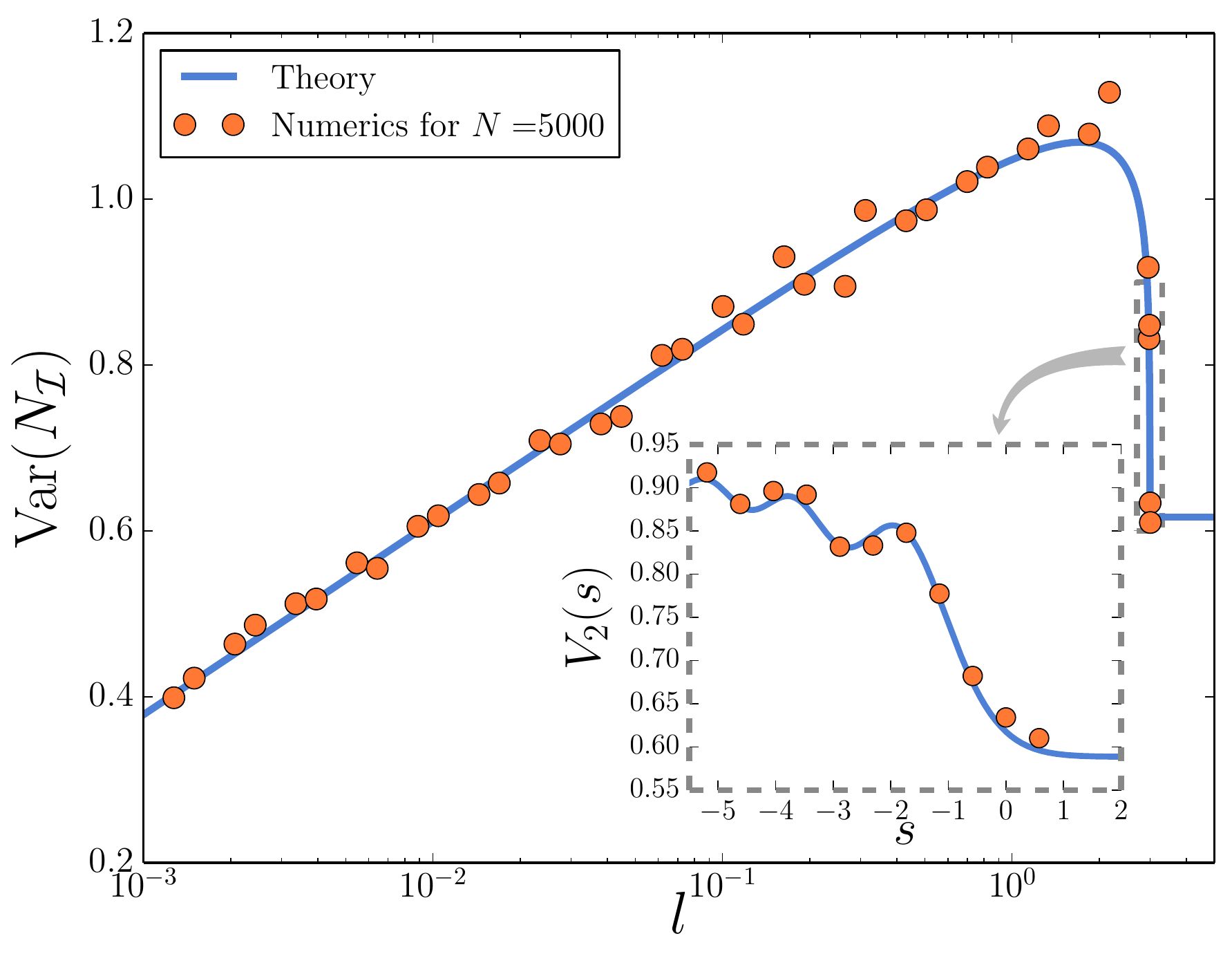}}
    \caption{(a) Average density of eigenvalues for the Wishart ensemble, which is the Mar\v{c}enko-Pastur distribution, for the $c=1$ case and (b) Results for the Wishart number variance when $\I=[1,1+l]$. The theoretical result is in equation \eqref{eq:variance wishart summary}.}
    \label{fig:wishart_ensemble}
\end{figure}

 \emph{$\beta$-Wishart ensemble:} In the Wishart ensemble \cite{Wis28} we consider a product matrix $W = X^{\dagger} X$ where $X$ is a $M \times N$ rectangular random matrix with Gaussian entries. $W$ is thus an $N \times N$ square random matrix whose entries are distributed according to $\Pr(W) \propto \e^{-\beta N {\Tr}(X^\dagger X)}$, where $\beta = 1,2$ or $4$ respectively when the entries are real, complex and quaternionic. The eigenvalues $x_i$'s of $W$ are all non-negative real numbers whose joint distribution is known to be of the form in Eq. (\ref{eq:eigen_jpdf_general_3}) with $V(x)={x}/{2}-\alpha\ln x$, where $\alpha$ is a constant given by $\alpha=\beta(1+M-N)/2N-1/N$~\cite{james}. Without any loss of generality, we consider the case where $M \geq N$. The case $N \geq M$ can be easily analyzed from the result $M \geq N$ (see for instance the discussion in \cite{VivMajBoh07}). We consider the limit $N \to \infty$ and $M \to \infty$ with the ratio $N/M=c\leq 1$ fixed. In this limit, the average spectral density is given by the Mar\v{c}enko and Pastur  law \cite{MarPas67} (see Fig. \ref{fig:wishart distribution} for the $c=1$ case) 
 \begin{equation}
 	\rho_{mp}(x)=\frac{1}{2\pi  x}\sqrt{(x-x_-)(x_+-x)}, 
    \label{wishart}
 \end{equation}
 where $x_\pm=(1\pm 1/\sqrt{c})^2$. Here we focus, for convenience, on the $c=1$ case where the density takes the form
\begin{eqnarray}
\rho_{mp}(x) = \frac{1}{2\pi} \sqrt{\frac{4-x}{x}} \label{mp_c1} \;.
\end{eqnarray}
In this case, the eigenvalues are supported over the interval $[0,4]$ with their center of mass located at $\overline{x}=1$. 

The observable of interest is again the number $N_\I$ of eigenvalues belonging to any interval $\I = [a,b]$ where
 $b>a\geq 0$. For the semi-infinite interval $[a,+\infty)$, the statistics of $N_\I$ was studied in Ref. \cite{MajViv12} using the Coulomb gas method. Here we study the statistics of $N_\I$ for any finite interval using a similar Coulomb gas method. For convenience, we provide explicit results in the case where $\I = [\overline{x}, \overline{x}+l] = [1,1+l]$, where $l$ is the size of the interval.   

In this case, the rate function $\Psi^{(W)}(k_\I)$ associated with the distribution of $N_{[1,1+l]}$ is given in equation \eqref{rate_wish} and the 
number variance is given by
\begin{equation}
    \Var(N_{[1,1+l]})\approx\begin{cases}\frac{2}{\beta \pi^2}\ln\left(N l (3-l)^{3/4}\right), & N^{-1}<l<3-N^{-2/3},\\
                    \frac{1}{\beta\pi^2}\ln N + \frac{1}{2}V_\beta(s), & l=3+\frac{s}{(N/4)^{2/3}},\\
                    \frac{1}{\beta\pi^2}\ln N + T_\beta(l), & l>3,
                    \end{cases}\label{eq:variance wishart summary}
\end{equation}
where $T_\beta(l)\approx \mathrm{e}^{-\beta N \Phi(1+l)}$, where $\Phi(w)$ is given in equation \eqref{eq:wishart_right_tail} and $V_\beta(s)$ is given for $\beta=2$ in equation \eqref{V_s_gaussian}, as in the Gaussian case. The number variance is compared to numerical results in figure \ref{fig:var_wishart_summary}. As in the Gaussian case, we notice that beyond $l=3$, where the right edge of the chosen interval coincides with the edge of the Mar\v{c}enko-Pastur distribution at $x=4$ in Eq. (\ref{mp_c1}), the variance again drops sharply as a function of $l$ and approaches a constant value $\frac{1}{\beta\pi^2}\ln N$ when $l \to \infty$, in agreement with the calculations in Ref. \cite{MajViv12} for the semi-infinite interval $[1,+\infty)$. As in the Gaussian case, the variance also has a non-monotonic behavior with a maximum at $l^\star= 12/7$.

\begin{figure}[!htbp]
    \centering
    \subfigure[\label{fig:cauchy_dist_regimes_summary}]{\includegraphics[width=0.435\textwidth]{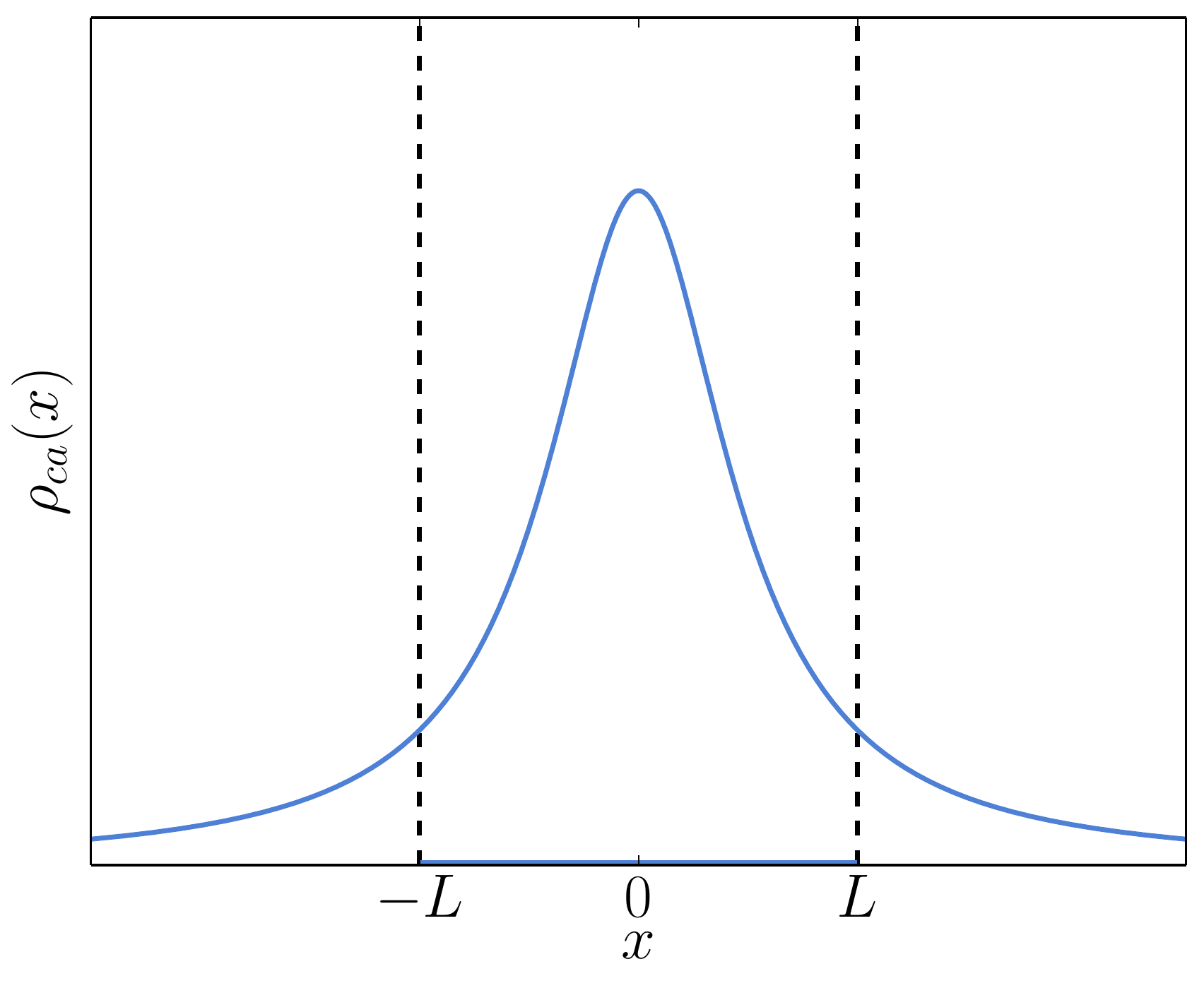}}
    \subfigure[\label{fig:var_cauchy_summary}]{\includegraphics[width=0.465\textwidth]{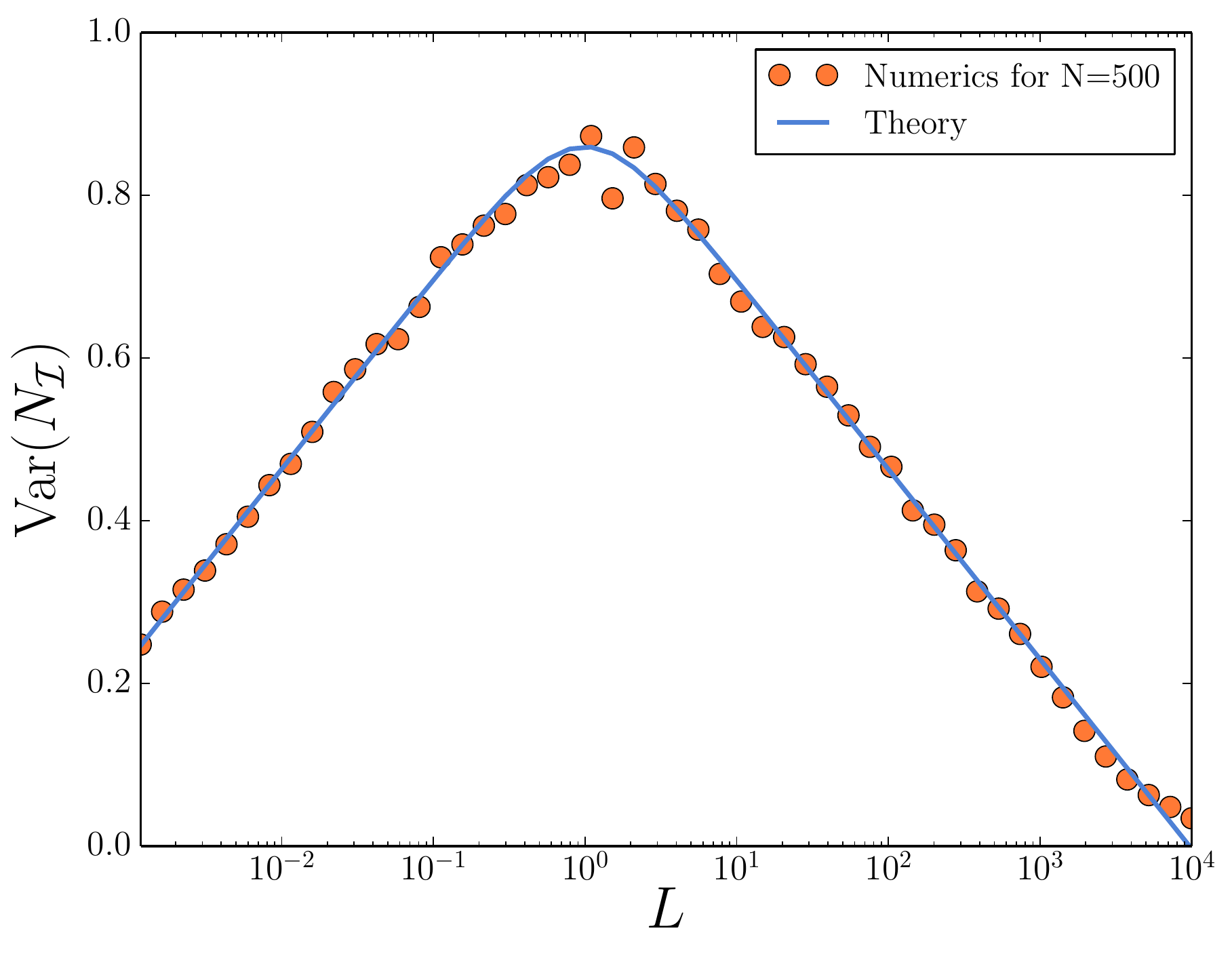}}
    \caption{(a) Average density of Cauchy ensemble, which is the Cauchy distribution and (b) Results for the Cauchy number variance when $\I=[-L,L]$. The theoretical result is equation \eqref{eq:variance cauchy summary}.}
    \label{fig:cauchy_ensemble}
\end{figure}

\emph{$\beta$-Cauchy ensemble:} 
Another well studied ensemble in RMT is the so-called $\beta$-Cauchy ensemble. In this case an $N \times N$ square matrix $X$ whose entries are distributed via $\Pr(X) \propto \det({\bf 1}_N + X^2)^{-\beta(N-1)/2-1}$, where ${\bf 1}_N$ is the $N \times N$ identity matrix and $\beta = 1,2$ or $4$ respectively for real, complex or quaternionic entries. The eigenvalues of $X$ are real and are distributed over the full real axis with a joint density given in Eq.~(\ref{eq:eigen_jpdf_general_3}) with $V(x)=\left((N-1)/(2N)+1/(\beta N)\right)\ln(1+x^2)$. The average density of eigenvalues is well known \cite{For10} and is given, for any $N$, by 
\begin{equation}
 	\langle \rho_N(x)\rangle \equiv\rho_{ca}(x)=\frac{1}{\pi}\frac{1}{1+x^2} \;.\label{cauchy}
\end{equation}
Here again, we are interested in the statistics of the number of eigenvalues $N_\I$ for an interval $\I$. For the semi-infinite interval $[0,+\infty)$, the statistics of $N_\I$ was studied for large $N$ using the Coulomb gas method \cite{MajSchVil13}. Here, we consider a finite symmetric interval $[-L,L]$. The rate function $\psi^{(C)}(k_\I)$ associated to the distribution of $N_\I$ is given by equation \eqref{rate_cauchy}. For the variance, we find
\begin{equation}
    \Var(N_{[-L,L]})\approx \frac{2}{\beta \pi^2}\ln\left(\frac{NL}{1+L^2}\right),\quad N^{-1} < L < N .\label{eq:variance cauchy summary}
\end{equation}
The variance rises to its maximum value at $L^\star=1$, and falls to zero as $L$ increases further. We note that the average position of the largest eigenvalue, $L\sim N$, behaves, for the purpose of the behavior of the number variance, as an effective edge for the Cauchy ensemble, even though its average density \eqref{cauchy} has no edge. Results are shown in figure \ref{fig:var_cauchy_summary}. 

We start by considering the $\beta$-Gaussian case in detail. The formalism based on the Coulomb gas technique combined with the resolvent method is presented for this case first and later applied to the other ensembles.

\section{$\beta$-Gaussian ensemble}
\label{sec:gaussian}

For Gaussian ensembles, the joint distribution is given by Eq. (\ref{eq:eigen_jpdf_general_3}) with $V(x) = x^2/2$ and Dyson index $\beta=1,2,4$  depending on whether the entries are real, complex or quaternionic. However, the same joint distribution with arbitrary $\beta > 0$ also appears in a class of tri-diagonal matrices found by Dumitriu and Edelman~\cite{DumEde02}. Therefore, we will consider a general $\beta > 0$. Below we first develop the Coulomb gas method for $\beta$-Gaussian ensembles (valid for all $\beta > 0$). The same method will be applicable to the other two cases.  

%

\subsection{The Coulomb gas}
\label{subsec:gaussian_coul}

 We consider the interval $\mathcal{I}=[a,b]$ for $V(x)=x^2/2$. Our goal is to derive the probability density of $N_\mathcal{I}$, the number of eigenvalues falling inside $\I$. It reads by definition
\begin{eqnarray}
    \mathcal{P}^{(G)}_\beta(N_\mathcal{I}=k_\mathcal{I}N)&=&\frac{1}{Z_{N,\beta}} \int \prod_{k=1}^N \dd x_k \mathrm{e}^{-\beta N\sum_{k=1}^N  \frac{x_k^2}{2}} \prod_{i>j}|x_i-x_j|^\beta \delta\left(k_\mathcal{I}N-\sum_{l=1}^N\mathbb{1}_\mathcal{I}(x_l)\right) \nonumber  \label{prob3}  \\
    &=&\frac{Z_{N,\beta}(k_\I)}{Z_{N,\beta}}, 
\end{eqnarray} 
where $\mathbb{1}_\mathcal{I}(x)$ is the indicator function for the interval $\mathcal{I}$, defined as 1 when $x \in \mathcal{I}$ and zero otherwise.

Using the exponential representation of the delta, it can be written as the ratio of two canonical partition functions
\begin{equation}
    Z_{N,\beta}(k_\I)=\int_{-\infty}^{\infty}\prod_{k=1}^N \dd x_k\int_{-\mathrm{i}\infty}^{\mathrm{i}\infty} \frac{\dd \mu}{2\pi} \e^{-\beta E[\bm{x},\mu,N_\mathcal{I}]},\quad\mbox{ and } Z_{N,\beta}=\int_{-\infty}^{\infty}\prod_{k=1}^N \dd x_k \e^{-\beta E[\bm{x},N_\mathcal{I}]}\ ,
\label{boltz}
\end{equation}
with energy functions
\begin{equation}\label{energy_CG_mu}
    E[\bm{x},\mu,N_\mathcal{I}]=N \sum_{i=1}^N \frac{x_i^2}{2} - \sum_{i> j}\ln|x_i-x_j| + \mu \left(\sum_{l=1}^N\mathbb{1}_\mathcal{I}(x_l)-k_\mathcal{I}N\right)\ ,  
\end{equation}
\begin{equation}\label{energy_CG}
    E[\bm{x},N_\mathcal{I}]=N \sum_{i=1}^N \frac{x_i^2}{2} - \sum_{i> j}\ln|x_i-x_j| \ .  
\end{equation}

Equation (\ref{energy_CG}) can be interpreted as the energy of a Coulomb gas of $N$ charged particles on a line, 
each subjected to an external harmonic potential $V(x)=x^2/2$ and repelling each other via a logarithmic interaction.   
Similarly, Eq. (\ref{energy_CG_mu}) represents another Coulomb gas which is subjected to an additional constraint, the $\mu$-dependent term, that prescribes the fraction of particles ({\it i.e.}, eigenvalues) inside the box $\I$. 

%

We introduce now the two-point density
\begin{equation}
    \rho_2(x,x')=\frac{1}{N(N-1)}\sum_{i\neq j}\delta(x-x_i)\delta(x'-x_j)\ .   
\end{equation} 
This function and the one-point density \eqref{densrandom} $\rho_N(x)$ (hereafter denoted simply by $\rho(x)$) allow us to write the energy in integral form, using the identities 
\begin{align}
\sum_i f(x_i) &= N\int \dd x \rho(x) f(x)\ ,\\
\sum_{i<j}f(|x_i-x_j|) &=\frac{N(N-1)}{2}\iint \dd x \dd x'\rho_2(x,x')f(|x-x'|)\ .
\end{align}
For the energy $E[\bm{x},\mu,N_\I]$ we obtain 
\begin{align}
    \nonumber E[\bm{x},\mu,N_\I]=&N^2\intinf \rho(x)\frac{x^2}{2}\dd x - \frac{N(N-1)}{2}\iint_{-\infty}^{+\infty}\!\!\! \dd x \dd x'\rho_2(x,x')\ln|x-x'| \\
        &+ \mu\left(N \int_a^b\rho(x)\dd x-N_\I\right) +\eta\left(N\intinf\rho(x)\dd x-N \right), 
\end{align}
where a supplementary Lagrange multiplier $\eta$ was introduced to enforce the normalization condition of the density $\rho(x)$. In the large-$N$ limit, we may replace the two-point density $\rho_2(x,x')$ by the product of one-point densities of $x$ and $x'$: $\rho_2(x,x')= \frac{N}{N-1}\rho(x)\rho(x')-\frac{1}{N-1}\rho(x)\delta(x-x')\sim \rho(x)\rho(x')$. Rescaling the Lagrange multipliers with $N$ yields eventually
\begin{equation}
    E[\bm{x},\mu,N_\I] \overset{N\gg 1}{\sim} N^2 S^{(G)}[\rho,\mu],
\end{equation}
where
\begin{align}
    \nonumber S^{(G)}[\rho,\mu]=&\intinf \rho(x)\frac{x^2}{2}\dd x - \frac{1}{2}\iint_{-\infty}^{+\infty}\!\!\! \dd x \dd x'\rho(x)\rho(x')\ln|x-x'| \\&+ \mu\left( \int_a^b\rho(x)\dd x-k_\I\right) +\eta\left(\intinf\rho(x)\dd x-1 \right)
    \label{action}
\end{align}
is called the \emph{action}. 

The introduction of the continuum density $\rho(x)$ converts the multiple integral \eqref{prob3} into a functional integral over all possible normalized densities, compatible with the given constraints. The Jacobian of this transformation can be shown \cite{For10} to bring in a term of order $\mathcal{O}(N)$, which is therefore subleading for large $N$. Therefore, we may write
\begin{equation}
  \mathcal{P}^{(G)}_\beta(N_\mathcal{I}= k_\mathcal{I}N)= \frac{1}{Z_{N,\beta}}\int \mathcal{D}[\rho] \dd \mu \dd \eta\, \e^{-\beta N^2 S^{(G)}[\rho,\mu]+\mathcal{O}(N)}\ , \label{action1}
\end{equation}
where the resulting functional integral may be eventually evaluated by the saddle-point method. 

While the original Coulomb gas approach was introduced by Dyson \cite{Dys62}, it was realized recently that the probability distribution 
of various linear statistics of eigenvalues can also be determined explicitly by adapting the same Coulomb gas approach. 
Depending on the linear statistics, the effective external potential of the Coulomb gas gets modified. For instance, in Eq. (\ref{energy_CG_mu}) the effective potential $V_{\rm eff}(x) = x^2/2 + \mu \mathbb{1}_\mathcal{I}(x)$. To solve this, we need to find the saddle point
density of the Coulomb gas in this effective potential. This method of calculating the distribution of
linear statistics of eigenvalues has found numerous recent applications: {\it e.g.} in finding the large deviations properties 
of the largest eigenvalue of a random matrix \cite{DeaMaj06,DeaMaj08,VivMajBoh07,MajVer09,MajSchVil13}, the evaluation of the conductance and shot noise power in chaotic mesoscopic cavities \cite{VivMajBoh07,VivMajBoh10}, the study of mutual information and data transmission in multiple input multiple output (MIMO) channels \cite{KazMerMou11}, bipartite entanglement of quantum systems \cite{CalLedMaj14,FacMarPar08,NadMajVer10,NadMajVer11}, non-intersecting Brownian motions and Yang-Mills gauge theory in $2$-dimensions \cite{reunion_proba} and many other problems~\cite{PerKatViv14,CunFacViv15,MarMajSch14_2} -- for a review see~\cite{MajSch14}.  
 
Note that the normalization constant $Z_{N,\beta}$ can also be written, for large $N$, as a functional integral without the additional constraint imposed by $\mu$
\begin{eqnarray}
Z_{N,\beta} = \int \mathcal{D}[\rho] \dd \mu \dd \eta\, \e^{-\beta N^2 S^{(G)}[\rho,\mu=0]+\mathcal{O}(N)} \;. \label{action0}
\end{eqnarray}
To evaluate the integrals in \eqref{action1} and (\ref{action0}) by the saddle point method, we need to find the density that minimizes the action. Consider first the denominator $Z_{N,\beta}$ in Eq. (\ref{action0}). The saddle point density in this case is given by the semi-circular law $\rho_{sc}(x) = \sqrt{2-x^2}/\pi$. If $\rho^\star(x)$ denotes the saddle-point density in presence of $\mu$ then we may express Eq. (\ref{action1}) as 
\begin{equation}
   \mathcal{P}^{(G)}_\beta(N_\mathcal{I}= k_\mathcal{I}N)= \frac{1}{Z_{N,\beta}}\int \mathcal{D}[\rho] \dd \mu \dd \eta\, \e^{-\beta N^2 S^{(G)}[\rho]} \approx \e^{-\beta N^2 \left(S^{(G)}[\rho^\star]-S^{(G)}[\rho_{sc}]\right)}=\e^{-\beta N^2 \psi^{(G)}(k_\I)}\ .
\label{saddle} 
\end{equation}
This then naturally provides the large deviation estimate for the probability that the fraction $k_\I$ of eigenvalues contained within $\I$ is atypically far from its average value $\overline{k_\I}$.
The rate function is $\psi^{(G)}(k_\I)=S^{(G)}[\rho^\star]-S^{(G)}[\rho_{sc}]$.
To obtain $\rho^\star(x)$, we differentiate $S^{(G)}[\rho]$ functionally with respect to $\rho$
\begin{equation} 
    \left. \frac{\delta S^{(G)}}{\delta \rho}\right|_{\rho^\star}=0=\frac{x^2}{2} - \int\dd x' \rho^\star(x') \ln|x-x'| + \mu\mathbb{1}_{\I}(x) + \eta, \,\,\,\,\,\,\, x\in\text{supp }\rho^\star.
    \label{funct_der}
\end{equation}
Differentiating it once more with respect to $x$ we obtain
\begin{equation} 
    x + \mu\left( \delta(x-b) - \delta(x-a) \right) = {\rm PV} \int\frac{\rho^\star(x')}{x-x'}\dd x', \,\,\,\,\,\,\, x\in\text{supp }\rho^\star,
    \label{int_eq_gaussian}
\end{equation}
where PV denotes Cauchy's principal value.
This is the singular integral equation that, combined with the constraints $\int\rho^\star(x)\dd x = 1$ and $\int_\I\rho^\star(x)\dd x = k_\I$, defines $\rho^\star$. In the next section, we show how to convert this singular integral equation into an algebraic equation for the \emph{resolvent} (or Green's function), which is much simpler to handle.


\subsection{Resolvent method}

Let $G(z)$ be a function of a complex variable $z$, called the \emph{resolvent}, defined as
\begin{equation} 
    G(z) = \int \frac{\rho^\star(x)}{z-x} \dd x, \,\,\,\, z\in\mathbb{C}\setminus\text{supp }\rho^\star.
    \label{resolvent_def}
\end{equation} 
Normalization of $\rho^\star$ to 1 implies that $G(z)$ behaves as $1/z$ when $|z|$ is large. Using the following Sochocki-Plemelj identity
\begin{align}
&    \lim_{\epsilon\to 0^+}G(x+\mathrm{i}\epsilon) =\lim_{\epsilon\to 0^+}\int \frac{\rho^\star(y)}{x+\mathrm{i}\epsilon -y}\dd y =
    \lim_{\epsilon\to 0^+}\int \frac{\rho^\star(y)(x-y-\mathrm{i}\epsilon)}{(x+\mathrm{i}\epsilon -y)(x-y-\mathrm{i}\epsilon)}\dd y \\
    &    \lim_{\epsilon\to 0^+}\left[\int \frac{\rho^\star(y)(x-y)}{(x-y)^2+\epsilon^2}\dd y-\mathrm{i}\int \frac{\rho^\star(y)\epsilon}{(x-y)^2+\epsilon^2}\dd y\right]={\rm PV} \int \frac{\rho^\star(y)}{x-y}\dd y -\mathrm{i}\pi\rho^\star(x),
\end{align}
where one uses $\delta(x)=\lim_{\epsilon\to 0^+}(1/\pi)\epsilon/(x^2+\epsilon^2)$, the density $\rho^\star(x)$ can be reconstructed from the imaginary part of the resolvent 
\begin{equation} 
    -\frac{1}{\pi}\lim_{\epsilon\to 0^+}\Im G(x+\mathrm{i}\epsilon)=\rho^\star(x)\ .
    \label{S-P-identity}
\end{equation}
To find an equation for $G(z)$, we multiply equation \eqref{int_eq_gaussian} by $\frac{\rho^\star(x)}{z-x}$ and integrate over $x$. If either $a$ or $b$ are outside the support of $\rho^\star$, their contribution is zero. If $a$ and $b$ belong to the support of $\rho^\star$, this procedure yields
\begin{equation} 
    \int x\frac{\rho^\star(x)}{z-x} \dd x + \int \frac{\rho^\star (x)}{z-x}\mu\left( \delta(x-b) - \delta(x-a) \right) \dd x   = \iint \frac{\rho^\star(x)}{z-x}\frac{\rho^\star(y)}{x-y}\dd y \dd x\ .
    \label{int_eq_1.5}
\end{equation}

\begin{figure}[!htbp]
    \subfigure[{Sketch of the expected behavior of $\rho^\star(x)$ for the Gaussian ensemble when $\I=[a,b]$, $-\sqrt{2}<a<b<\sqrt{2}$ and $k_\I<\overline{k_\I}$.}\label{fig:gaussian_density_smaller}]{\includegraphics[width=0.45\textwidth]{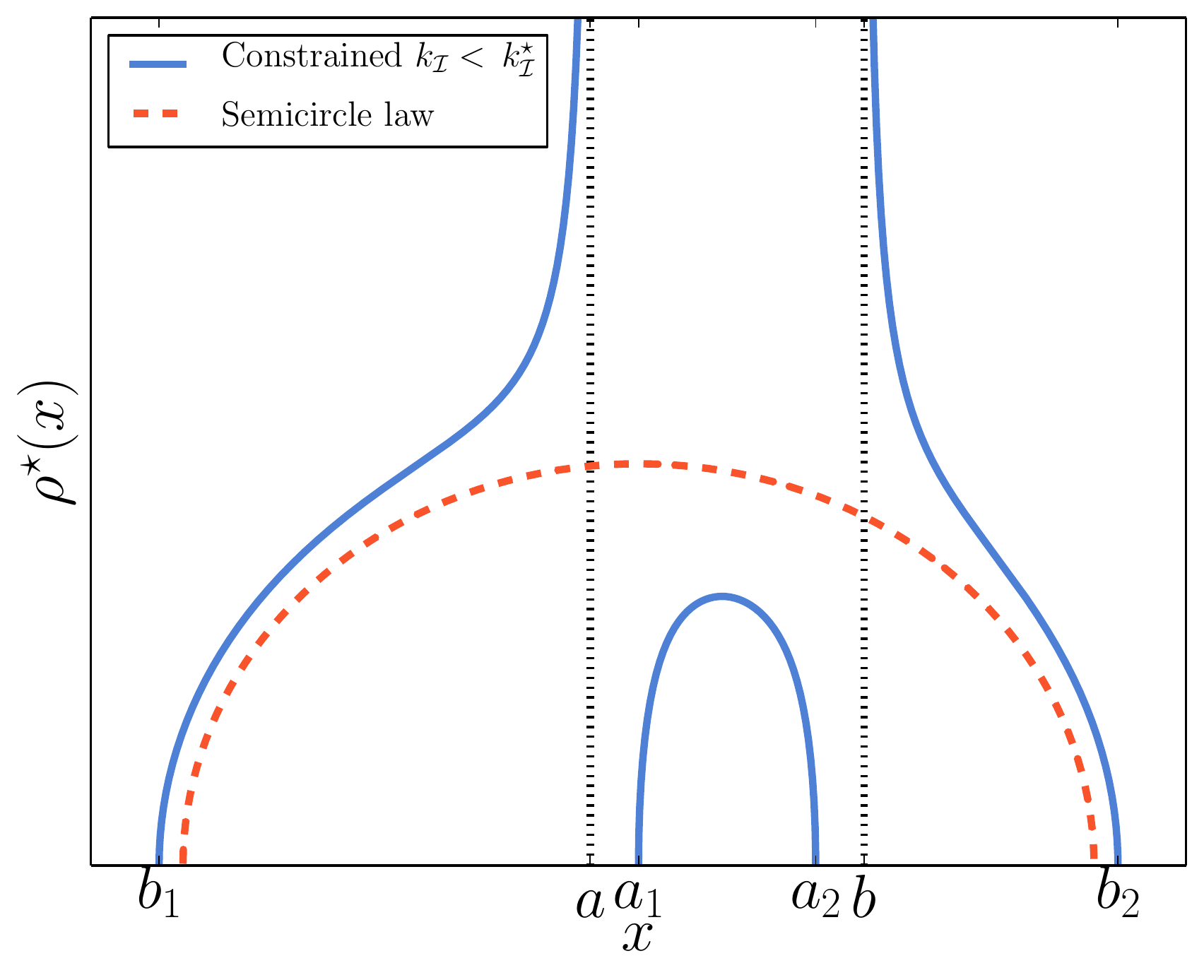}}\quad
    \subfigure[{Sketch of the expected behavior of $\rho^{\star}(x)$ for the Gaussian ensemble when $\I=[a,b]$, $-\sqrt{2}<a<b<\sqrt{2}$ and $k_\I>\overline{k_\I}$.}\label{fig:gaussian_density_larger}]{\includegraphics[width=0.45\textwidth]{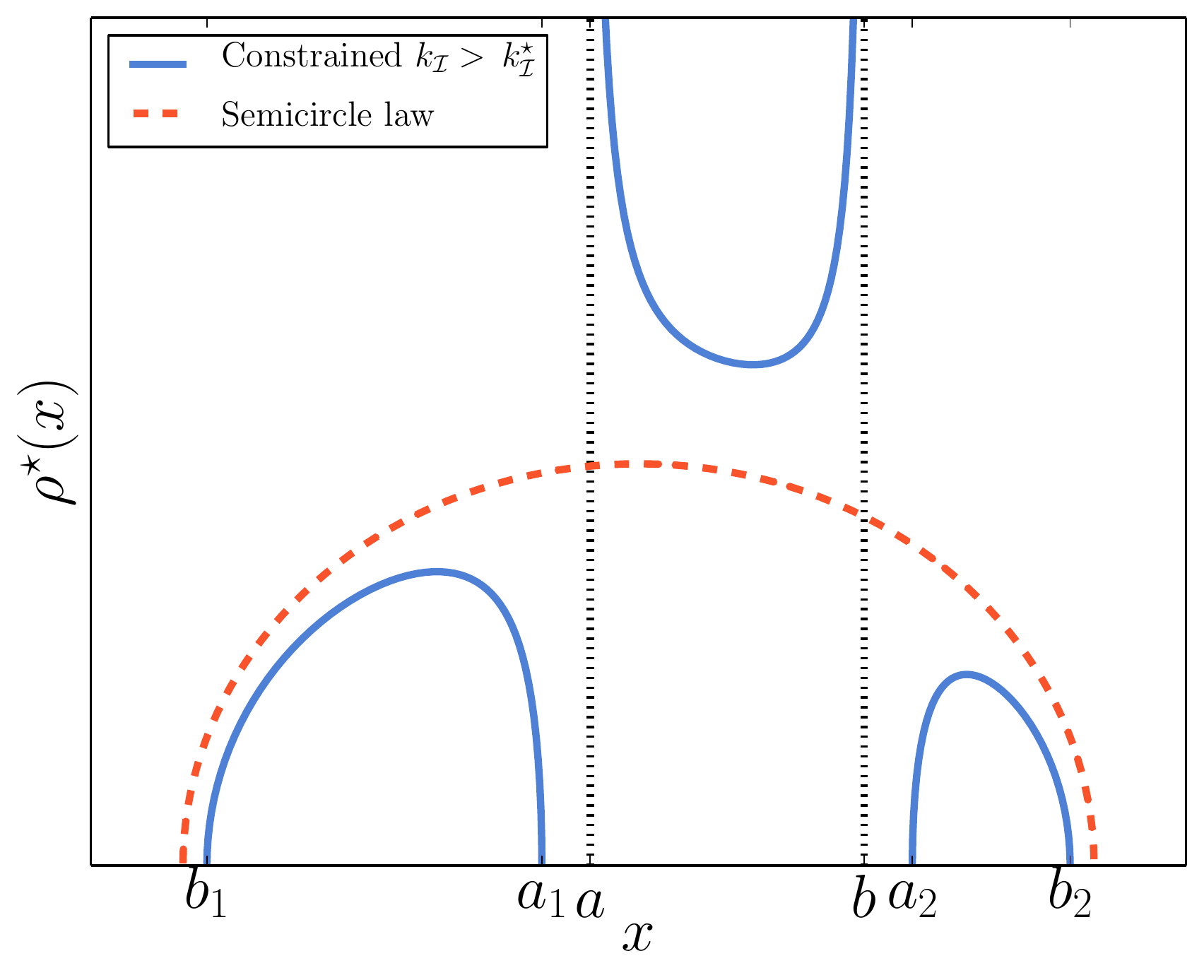}}
    \caption{}
    \label{fig:gaussian_density}
\end{figure}

Some regularization is required to perform the integrals involving deltas, as the density $\rho^\star$ may diverge at $x=a$ or $x=b$. We set $a\to a\mp \varepsilon$ and $b\to b\pm \varepsilon$, and we take $\varepsilon \to 0$. The choice of signs depends on whether the possible divergence happens inside or outside the interval $\I=[a,b]$ (see figures \ref{fig:gaussian_density_smaller} and \ref{fig:gaussian_density_larger}). This procedure yields
\begin{equation}
    \int \frac{\rho^\star (x)}{z-x}\mu\left( \delta(x-b\mp \epsilon) - \delta(x-a\pm \varepsilon) \right) \dd x = \frac{B(\epsilon)}{z-b}+\frac{A(\epsilon)}{z-a},
\end{equation}
where $A(\epsilon)$ and $B(\epsilon)$ are functions of $\epsilon$ that may individually diverge when $\epsilon\to 0$. However, the normalization condition of the average density implies that the resolvent $G(z)$ behaves as $1/z$ when $|z|$ is large. This condition requires that $A(\epsilon\to 0)=A$ and $B(\epsilon\to 0)=B$ should be well-defined constants that we can determine prescribing the fraction of eigenvalues $k_\I$ inside the interval and an extra condition, called the chemical equilibrium condition, which will be described below. The equation for the resolvent thus reads
\begin{equation} 
    \int x\frac{\rho^\star(x)}{z-x} \dd x + \frac{A}{z-a}+\frac{B}{z-b}   = \iint \frac{\rho^\star(x)}{z-x}\frac{\rho^\star(y)}{x-y}\dd y \dd x\ .
    \label{int_eq_2}
\end{equation}
It is worth noticing that the presence of hard walls in the potential confining the charged particles always yields extra terms of the form $A/(z-a)$ in the equation for the resolvent. 

The RHS of equation \eqref{int_eq_2} can be written in terms of $G(z)$ with the following manipulations. We use the identity
\begin{equation} 
    \frac{1}{(z-x)(x-y)}=\left(\frac{1}{z-x}+\frac{1}{x-y}\right)\frac{1}{z-y},
\end{equation}
to write the RHS as
\begin{equation} 
    \iint \frac{\rho^\star(x)}{z-x}\frac{\rho^\star(y)}{x-y} \dd x \dd y=\iint \frac{\rho^\star(x)\rho^\star(y)}{(z-x)(z-y)} \dd x \dd y-\iint \frac{\rho^\star(x)}{z-x}\frac{\rho^\star(y)}{x-y} \dd x \dd y\ ,
    \label{RHS}
\end{equation}
where in the last term we have exchanged $x\to y$ and flipped the sign in front accordingly. Equation \eqref{RHS} then read $RHS = G^2(z)-RHS$,  implying that the original RHS of equation \eqref{int_eq_2} is $(1/2)G^2(z)$. 

The manipulations required to express the LHS in terms of $G(z)$ depend on the specific form of the potential, but are straightforward in the Gaussian case. The integral over the derivative of the potential in the LHS reads 
\begin{equation}
    \int x\frac{\rho^\star(x)}{z-x} \dd x = \int (x-z+z)\frac{\rho^\star(x)}{z-x} \dd x =-1+zG(z).
    \label{LHS_gaussian}
\end{equation}
The final equation for the resolvent thus reads 
\begin{equation}
	-1+zG(z)+ \frac{A}{z-a} + \frac{B}{z-b} = \frac{1}{2}G^2(z)\ .
	\label{algebr_1}
\end{equation}
Its solution is given by
\begin{equation} 
    G_{\pm}(z)=z\pm\sqrt{z^2-2+\frac{2A}{z-a}+\frac{2B}{z-b}}=z\pm\sqrt{\frac{(z-a_1)(z-a_2)(z-b_1)(z-b_2)}{(z-a)(z-b)}}\ ,\label{resolv_gauss}
\end{equation}
where $a_1$, $a_2$, $b_1$ and $b_2$ are the roots of the polynomial $(z^2-2)(z-a)(z-b)+2A(z-b)+2B(z-a)$ (we chose $b_1<a_1<a_2<b_2$) and the sign of the resolvent is chosen according to the normalization condition $G(z)\to 1/z$ when $|z|\to\infty$. We note that the resolvent is well defined on the real line outside on the support of $\rho^\star(x)$, and the sign convention for points on the real line is given by
\begin{equation}
    G(x)=\begin{cases} 
                    G_-(x) \text{ for } x>b_2\\
                    G_+(x) \text{ for } x<b_1\\
                    G_+(x) \text{ for } k_\I>\overline{k_\I}\text{ and }b<x<a_2 \\ 
                    G_+(x) \text{ for } k_\I<\overline{k_\I}\text{ and }a<x<a_1 \\ 
                    G_-(x) \text{ otherwise}\ ,
                \end{cases}\label{Goutside}
\end{equation}
these regions being either ``gaps'' in the spectrum or semi-infinite intervals $(-\infty,b_1]$, $[b_2,\infty)$. They are clearly illustrated in figure \ref{fig:gaussian_density}. 

The average density can then be extracted straightforwardly from the resolvent
\begin{equation} 
    \rho^\star(x)=-\frac{1}{\pi}\lim_{\epsilon\to 0^+}\Im G(x+\mathrm{i}\epsilon)=\frac{1}{\pi}\sqrt{\frac{(a_1-x)(x-a_2)(x-b_1)(x-b_2)}{(x-a)(x-b)}}.\label{gaussian_density}
\end{equation}
We note that the constants $a_1$, $a_2$, $b_1$ and $b_2$ define the edges of the support of $\rho^\star$ (see figure \ref{fig:gaussian_density}) and we need four equations to fix them. Two equations are provided by the normalization constraint, which is equivalent to identifying the edges as the roots of the polynomial $(z^2-2)(z-a)(z-b)+2A(z-b)+2B(z-a)$. A simple matching of coefficients of this polynomials yields
\begin{align}
    a_1+a_2+b_1+b_2=&\;a+b\label{norm_cond_gauss_1}\ ,\\
    a_1a_2+a_1b_1+a_1b_2+a_2b_1+a_2b_2+b_1b_2=&\;-2+ab\ .
    \label{norm_cond_gauss_2}
\end{align}
The remaining two equations are obtained prescribing the fraction of eigenvalues inside the interval, $\int_a^b \rho^\star(x)\dd x = k_\I$, and a supplementary condition described below (equation \eqref{chemical_equilibrium}).

The average density of eigenvalues $\rho^\star(x)$ depends only on the fraction $k_\I$ of eigenvalues constrained to be inside the interval. We see from figure \ref{fig:gaussian_density} that $\rho^\star(x)$ changes shape abruptly when $k_\I=\overline{k_\I}$, where 
\begin{equation}\label{kI_bar}
\overline{k_\I}=\int_a^b \rho_{sc}(x)\dd x
\end{equation}
is the average number of eigenvalues inside $\I$ for the Gaussian ensemble. The shape of the average density shown in figures \ref{fig:gaussian_density} diverges at the edges of the interval in either case. As an example, let us take $k_\I>\overline{k_\I}$ (figure \ref{fig:gaussian_density_larger}). This means that the number of eigenvalues (particles) stacked inside the interval is larger than their naturally expected value. Physically they will then tend to repel and accumulate towards the box walls, while the charges \emph{outside} the box will be pushed further out by this excess of charge. The reverse case $k_\I<\overline{k_\I}$ (figure \ref{fig:gaussian_density_smaller}) causes more charges to be outside the box than their expected value, and they accumulate towards the box walls on the external side, pushing those inside to agglomerate in a compact blob. When $k_\I=\overline{k_\I}$, there are as many charges in the interval as naturally expected in the absence of any constraint, and therefore the equilibrium density reconstructs Wigner's semicircle \eqref{eq:wigner semicircle intro}.


\subsection{Calculation of the rate function $\psi^{(G)}(k_\I)$}

With the average density $\rho^\star(x)$, we may now proceed to calculate the rate function $\psi^{(G)}(k_\I)$. We must insert $\rho^\star$ into equation \eqref{action} and perform the necessary integrals to compute the full probability density of the random variable $N_\I$. It is first convenient to rewrite the action using the saddle point condition \eqref{funct_der}.


We first multiply the equation \eqref{funct_der} by $\rho^\star(x)$ and integrate it with respect to $x$ to obtain
\begin{equation} 
    \int \rho^\star(x)\frac{x^2}{2}\dd x + \mu\underbrace{\int_a^b\rho^\star(x) \dd x}_{=k_\I} + \eta\underbrace{\int\rho^\star(x)\dd x}_{=1}= \iint \dd x \dd x' \rho^\star(x)\rho^\star(x') \ln|x-x'|\ .
\end{equation}
The RHS is exactly the double integral present in the original action \eqref{action}. Therefore we can replace it with single integrals in the action $S^{(G)}$ \eqref{action} evaluated at the saddle point $\rho^\star$, obtaining eventually
\begin{equation} 
    S^{(G)}[\rho^\star]=\frac{1}{2}\int \frac{x^2}{2}\rho^\star(x)\dd x - \frac{\mu}{2}k_\I - \frac{\eta}{2}\ .
    \label{action_simple}
\end{equation}
What remains to be determined are the multipliers $\mu$ and $\eta$. Equations for these parameters will arise from \eqref{funct_der}, once specialized to certain points within the domain of $\rho^\star(x)$.

Let us consider the case $k_\I>\overline{k_{\I}}$ (figure \ref{fig:gaussian_density_larger}).  We call now $H(x)=\int \rho^\star(x')\ln|x-x'| \dd x'$. Evaluating \eqref{funct_der} at $x=b$ and $x=a_2$ we obtain
\begin{align}
    H(a_2)=&\,\frac{a_2^2}{2}+\eta\label{Heta0}\ ,\\
    H(b)=&\,\frac{b^2}{2}+\mu+\eta,\label{Heta}
\end{align}
which allows us to write $\mu=H(b)-H(a_2)+\frac{a_2^2}{2}-\frac{b^2}{2}$. Using the definition of the resolvent in equation \eqref{resolvent_def}, we note that $H(b)-H(a_2)=\int_{a_2}^{b} G(x) \dd x$. The resolvent is well-defined between $a_2$ and $b$, since this interval does not belong to the support of $\rho^\star(x)$. The same reasoning may be applied to the other edge of the box, and we would obtain a similar equation with $a$ and $a_1$ (see figure \ref{fig:gaussian_density_larger}). These equalities allow us to calculate $\mu$ in terms of the edges and an integral of the resolvent:
\begin{equation} 
    \mu=-\int_{b}^{a_2} G(x) \dd x+\frac{a_2^2}{2}-\frac{b^2}{2} = \int_{a_1}^a G(x) \dd x+\frac{a_1^2}{2}-\frac{a^2}{2}.
    \label{chemical_equilibrium}
\end{equation}

We name equation \eqref{chemical_equilibrium} the \emph{chemical equilibrium condition} \cite{Jur90}. Its physical interpretation in the context of the Coulomb gas is the following: the amount of energy required to transport a charge from outside to inside the interval must be the same on both sides of the interval. The integral of the resolvent from $b$ to $a_2$ (see figure \ref{fig:gaussian_density}) represents the variation of the intensity of the repulsion felt by a test charge at positions $b$ and $a_2$, while $\mu$ represents the balance between repulsion and potential, {\it i.e.} the amount of energy required to move one charge from inside to outside the box. The imposition of boundaries and constraints creates a pressure on the edges of the interval and equation \eqref{chemical_equilibrium} expresses the fact that such pressure is the same at both sides of the interval. This equation is equivalent to applying equation \eqref{funct_der} at a point in the support of the average density. 

To obtain $\eta$, we use \eqref{funct_der} for values of $x$ inside the support of $\rho^\star$, but outside the box: $H(x)=\frac{x^2}{2}+\eta$. One can show that, if $b_2$ is the upper edge of the support $\rho^\star$ (see figure \ref{fig:gaussian_density}), we may expand $H(x)$ around $b_2$ to obtain
\begin{equation} 
    H(b_2)=\ln b_2 - \int_{b_2}^\infty \left(G(x)-\frac{1}{x}\right)\dd x=\frac{b_2^2}{2}+\eta\ .
    \label{border_condition}
\end{equation}
We then obtain the following expression for $\eta$
\begin{equation} 
	\eta = \ln b_2 -\frac{b_2^2}{2}- \int_{b_2}^\infty \left(G(x)-\frac{1}{x}\right)\dd x\ .
	\label{eta}
\end{equation}

\begin{figure}[!htbp]
    \centering
    \includegraphics[width=0.6\linewidth]{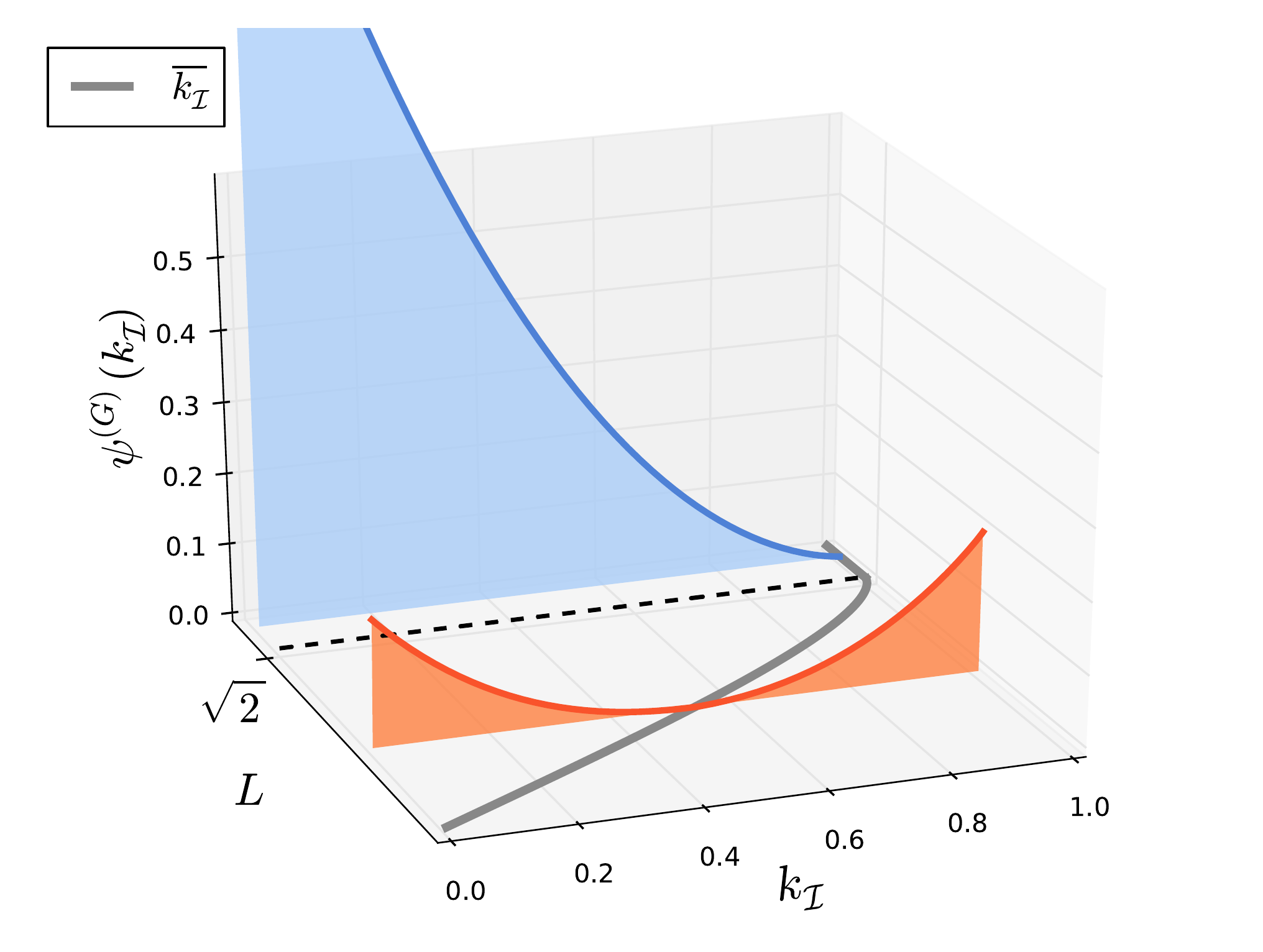}
    \caption{Behavior of the rate function $\psi^{(G)}(k_\I)$ as a function of $k_\I\in[0,1]$ for two different values of $L$: $L=0.6<\sqrt{2}$ and $L=1.6>\sqrt{2}$. $k_\I$ is the fraction of eigenvalues inside the interval $\I=[-L,L]$. The solid gray line in the plane $(L,k_\I)$ is the critical line $\overline{k_\I}$ given in Eq. (\ref{k_Ibar2}), where $\psi^{(G)}(k_\I)$ has a minimum (zero).}
    \label{fig:rate function gaussian}
\end{figure}

It remains only to calculate the normalization constant, given by the action $S^{(G)}$ calculated with Wigner's semi circle \eqref{eq:wigner semicircle intro}. All integrals simplify greatly and we easily compute $S^{(G)}[\rho_{sc}]=\frac{3}{8}+\frac{\ln 2}{4}$. The final formula for the rate function of the probability of finding $N_\I=Nk_\I$ eigenvalues inside interval $\I=[a,b]$ is eventually
\begin{align}
\nonumber	\psi^{(G)}(k_\I)=&\frac{1}{2}\int \frac{x^2}{2}\rho^\star(x)\dd x - \frac{\mu}{2}k_\I - \frac{\eta}{2}-S^{(G)}[\rho_{sc}]\\
\nonumber	=&\,\frac{1}{2}\intinf \rho^{\star}(x)\frac{x^2}{2}\dd x - \frac{1}{2}\left(\int_{b}^{a_2} G(x) \dd x-\frac{a_2^2}{2}+\frac{b^2}{2}\right)k_\I \\
    &- \frac{1}{2}\left[\ln b_2- \frac{b_2^2}{2} - \int_{b_2}^\infty \left(G(x)-\frac{1}{x}\right)\dd x\right]-\frac{3}{8}-\frac{\ln 2}{4}\ .\label{action_gauss}
\end{align}
The resolvent $G(x)$ is given by equation \eqref{resolv_gauss} with the choice of signs given by equation \eqref{Goutside}, the average density $\rho^\star(x)$ is in  equation \eqref{gaussian_density} and the edges of its support are given by equations \eqref{norm_cond_gauss_1}, \eqref{norm_cond_gauss_2}, \eqref{chemical_equilibrium} and $\int_a^b \rho^\star(x)\dd x=k_\I$. This expression can be easily evaluated numerically, and results for the interval $\I=[-L,L]$ are shown in figure \ref{fig:rate function gaussian}. As we can see, the minimum of the rate function is reached at the average value $k_\I = \overline{k_\I}$ where 
\begin{equation}\label{k_Ibar2}
\overline{k_\I} = \int_{-L}^{+L} \rho_{sc}(x) \dd x = \frac{1}{\pi} \left(L\,\sqrt{2-L^2} +2 \sin ^{-1}\left(\frac{L}{\sqrt{2}}\right)\right) \;, \; 0 \leq L \leq \sqrt{2},
\end{equation}
indeed the average value of $k_\I$.

\subsection{Number variance}

Formula \eqref{action_gauss} is, however, too general to be analyzed directly. We turn our attention to a more specific problem: calculate the fluctuations of eigenvalues inside an interval, the number variance for the Gaussian ensemble. For simplicity, we take the interval $\I=[-L,L]$ and we determine the variance of the variable $N_{[-L,L]}$.

\begin{figure}[!htbp]
    \centering
    \includegraphics[width=0.6\linewidth]{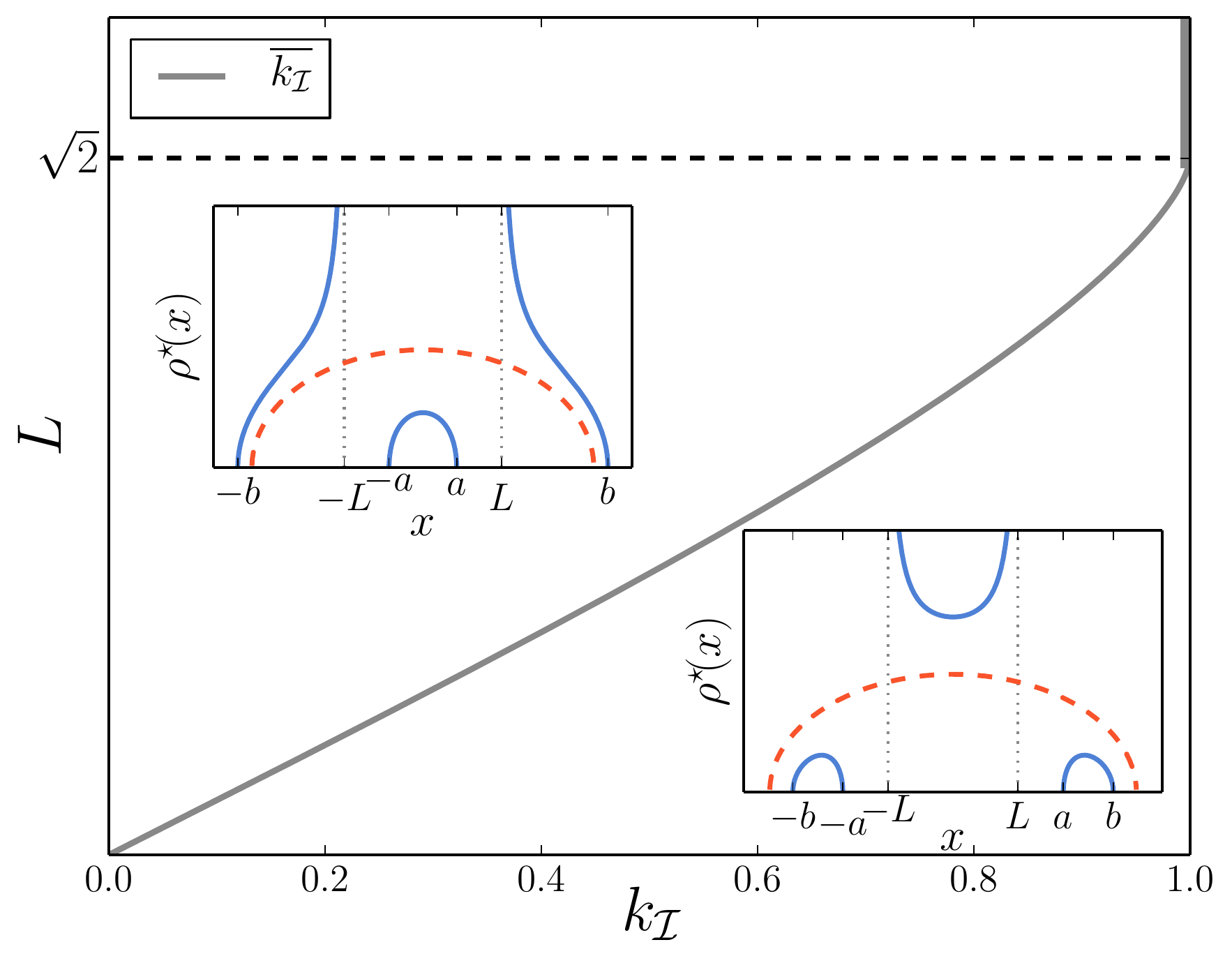}
    \caption{Phase transition of the average density $\rho^\star(x)$, considering the interval $\I=[-L,L]$. When $k_\I=\overline{k_\I}$, the average density becomes the Wigner semicircle law given in Eq. (\ref{k_Ibar2}), represented as a dashed orange line.}
    \label{fig:phase transition kstar}
\end{figure}

Usually one can read off the variance of a random variable satisfying a large deviation principle directly from the quadratic behavior of the rate function around its minimum: this implies indeed that small fluctuations around the minimum are Gaussian. In our case, though, the rate function is not analytic and its expansion is not simply quadratic around the minimum \cite{MajNadSca11,MajNadSca09}, so more effort is needed to extract the variance.

The origin of this non-analytic behavior is the phase transition in the average density when $k_\I$ crosses the value $\overline{k_\I}$, the average fraction of eigenvalues inside the interval $\I=[-L,L]$. This phase transition is illustrated in figure \ref{fig:phase transition kstar}.

In order to find the number variance, we perturb the parameters of \eqref{gaussian_density} and expand the rate function $\psi^{(G)}(k_\I)$ around the unconstrained case $k_\I=\overline{k_\I}+\delta$, where $\overline{k_\I}$ is the location of its minimum. We keep only the leading terms in the expansion and read off the variance from the quadratic term that remains. Depending on the value of $L$, however, we must consider three different regimes: ({\it i}) an extended bulk $N^{-1}<L<\sqrt{2}$, ({\it ii}) an edge regime $|L-\sqrt{2}|\sim N^{-2/3}$ and ({\it iii}) a tail regime $L>\sqrt{2}$ (see figure \ref{fig:regimes_gauss_intro}). We provide the analysis of each regime separately.

\subsubsection{Extended bulk regime}

\

\begin{figure}[!htbp]
  \centering
  \includegraphics[width=0.55\linewidth]{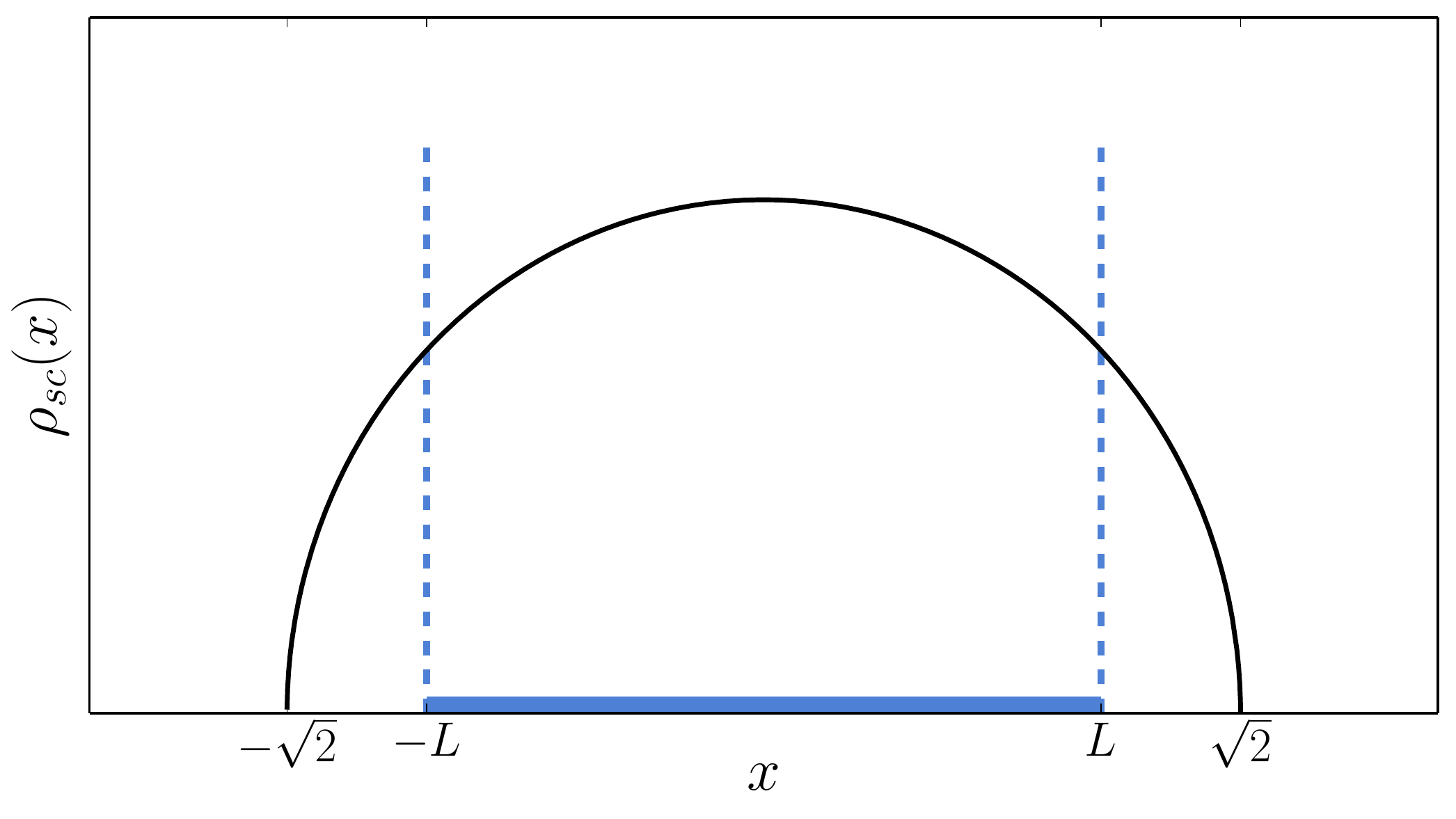}
  \caption{Extended bulk regime for the Gaussian ensemble.}
  \label{fig:extended gaussian}
\end{figure}

The Gaussian potential and the interval $\I=[-L,L]$ are symmetrical, so the equilibrium density will also be symmetrical: $a_1=-a_2$ and $b_1=-b_2$. This reduces the number of parameters that we must determine. Note that the chemical equilibrium condition \eqref{chemical_equilibrium} is trivially satisfied in this case.

The unconstrained case, when $\mu=0$ and the number of eigenvalues inside the interval is equal to their naturally expected value, corresponds to $a_1=-L$, $a_2=L$, $b_1=-\sqrt{2}$, $b_2=\sqrt{2}$, which turns $\rho^\star$ into Wigner's semicircle. We perturb the parameter $a_1$ by a small shift $\epsilon$: $a_1=-L-\epsilon$. The perturbation is negative because we expect a modification on the edges according to figure \ref{fig:phase transition kstar}, and we compute the resulting perturbation applied to the other parameters using equation \eqref{norm_cond_gauss_2}. We write
\begin{align}
	a_2=L+\epsilon, & & b_1 = -\sqrt{2}+\epsilon_2, & & b_2 = \sqrt{2}-\epsilon_2\ .
\end{align}
From the normalization condition \eqref{norm_cond_gauss_2} we find to leading order in $\epsilon$
\begin{equation}
	\sqrt{2}\epsilon_2=L\epsilon\ .
\end{equation}
Having computed the effect of the perturbation on each parameter appearing in the resolvent and average density, we proceed to expand each of the integrals in the rate function \eqref{action_gauss} around its minimum value. This procedure is rather lengthy and we leave the details of calculations to Appendix \ref{app}. We eventually obtain

\begin{equation} 
	\psi^{(G)}(k_\I)=\frac{1}{4}\int x^2\rho^\star(x)\dd x - \frac{\eta}{2}-\frac{\mu}{2}k_\I -\frac{3}{8}-\frac{\ln 2}{4} =  \frac{\pi}{4}\sqrt{2-L^2}\epsilon\delta + o(\epsilon^2\ln\epsilon),\label{rate_2}
\end{equation}
where we denote $k_\I=\overline{k_\I}+\delta$. The value of $\delta$ as a function of $\epsilon$ is given by \eqref{delta_app}
\begin{equation} 
	\delta \approx \epsilon\frac{\sqrt{2-L^2}}{ \pi }\left[\ln \left( \left(2-L^2\right)L\right)- \ln \epsilon \right]\ . \label{delta}
\end{equation}
	
We now have to find $\epsilon$ as a function of $\delta$ by inverting \eqref{delta}. We propose the following ansatz, correct to leading orders in $\delta$ and $\epsilon$
\begin{equation}
 	\epsilon\approx \frac{\pi}{\sqrt{2-L^2}} \frac{\delta}{\ln  \left(\left(2-L^2\right)^{\frac{3}{2}}L\right)- \ln \delta }\ .
\end{equation}
We conclude that, for small fluctuations of $k_\I$ around its average value $\overline{k_\I}$ we have
\begin{equation}
	\psi^{(G)}(k_\I=\overline{k_\I}+\delta)\approx \frac{\pi^2}{4}\frac{\delta^2}{\ln  \left(\left(2-L^2\right)^{\frac{3}{2}}L\right)- \ln \delta}\ .
\end{equation}

Replacing the asymptotic expansion of the rate function around its minimum we obtain, for small $\delta>0$
\begin{equation}
	\mathcal{P}^{(G)}_\beta(N_{[-L,L]}= (\overline{k_\I} +\delta)N)\approx\exp\left[-\beta N^2 \frac{\pi^2}{4}\frac{\delta^2}{\ln  \left(\left(2-L^2\right)^{\frac{3}{2}}L\right)- \ln \delta}\right].
\end{equation}
We replace $\delta\to (N_{[-L,L]}-\overline{N_{[-L,L]}})/N$, obtaining for small deviations of $N_{[-L,L]}$ from $\overline{N_{[-L,L]}}$
\begin{equation}
	\mathcal{P}^{(G)}_\beta(N_{[-L,L]})\approx\exp\left[{-\beta \frac{\pi^2}{4}\frac{(N_{[-L,L]}-\overline{N_{[-L,L]}})^2}{\ln  \left[\left(2-L^2\right)^{\frac{3}{2}}LN\right]  }}\right].\label{Pnlasymp}
\end{equation}
Hence for small fluctuations around the average, $N_{[-L,L]}$ has a Gaussian distribution modulated by a logarithmic dependence on $N$. From \eqref{Pnlasymp}, we may directly read off the number variance $\Var(N_{[-L,L]})$
\begin{equation}
	\Var(N_{[-L,L]})=\frac{2}{\beta\pi^2}\ln  \left[NL\left(2-L^2\right)^{3/2}\right] + \mathcal{O}(1). \label{var_gauss_index}
\end{equation}
The variance for the number of eigenvalues initially grows logarithmically with the size of the box, reaches a peak at $1/\sqrt{2}$ and decreases due to the presence of the edge of the semicircle at $\sqrt{2}$ (see fig \ref{fig:variance_gaussian}). The critical value $L^\star=1/\sqrt{2}$ for the $[-L,L]$ box, where the fluctuations are maximal, is a new result found by this method. Notably, when $L$ is small, we retrieve Dyson's result, equation \eqref{eq:dyson and mehta}
\begin{equation}
    \Var(N_{[-L,L]})\sim \frac{2}{\beta\pi^2}\ln  \left[NL\left(2-L^2\right)^{3/2}\right] \xrightarrow{L\sim 1/N} \frac{2}{\beta \pi^2}\ln (NL)+\mathcal{O}(1).
\end{equation}

\subsubsection{Edge regime}

\

\begin{figure}[!htbp]
  \centering
  \includegraphics[width=0.55\linewidth]{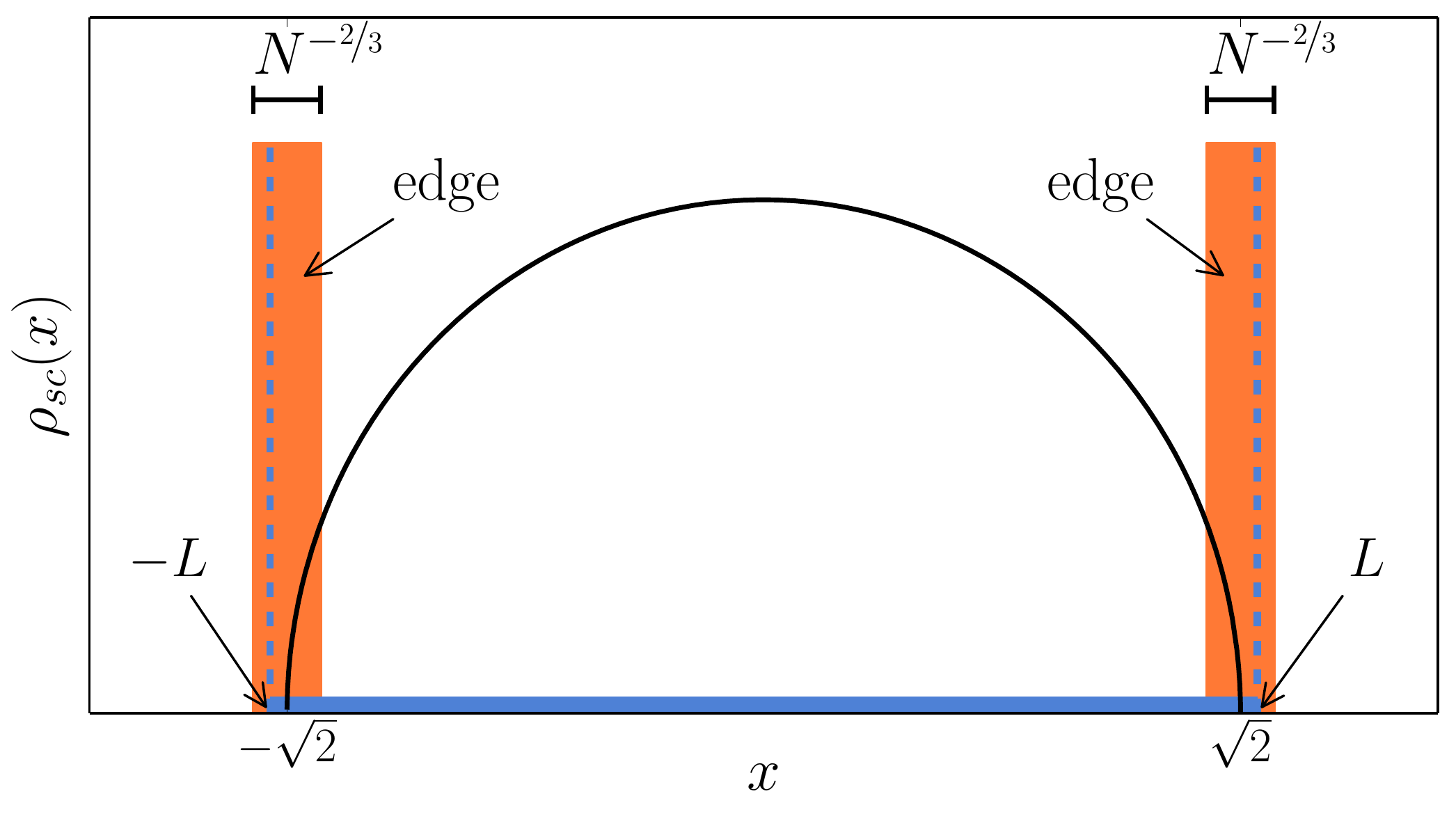}
  \caption{Edge regime for the Gaussian ensemble.}
  \label{fig:edge gaussian}
\end{figure}

When the interval reaches the limit of the support, most of the approximations used in the last section are no longer valid. We then resort to the technique of orthogonal polynomials to investigate the scaling region around the semicircle edge. For Gaussian unitary matrices, with $\beta=2$, one can show that the number variance of an interval $\I$ is given by \cite{Meh91}
\begin{equation}
      \Var (N_\I)=\int_\I K_N(x,x)\dd x- \int_\I\!\int_\I K_N(x,y)^2\dd x \dd y, \label{gustavsson}
\end{equation}  
where $K_N(x,y)$ is the Christoffel-Darboux kernel \cite{Meh91}
\begin{equation}
    K_N(x,y)=\sqrt{\frac{N}{\pi}}\frac{1}{2^N(N-1)!}\e^{-N\frac{x^2+y^2}{2}}\frac{H_N(x\sqrt{N})H_{N-1}(y\sqrt{N})-H_N(y\sqrt{N})H_{N-1}(x\sqrt{N})}{\sqrt{N}(x-y)},
\end{equation}
and $H_N(x)$ are Hermite polynomials. The kernel satisfies the ``reproducing" property
\begin{equation}\label{identity}
\int_{-\infty}^{\infty}  K_N(x,z) K_N(z,y)\dd z = K_N(x,y)
\end{equation}
as well as the symmetry relation
\begin{equation}\label{inversion}
K_N(-x,-y) = K_N(x,y)\ .
\end{equation}
The average spectral density may be written as $K_N(x,x)=N\langle\rho_N(x)\rangle$.

In the case $\I=[-L,L]$, we split the integral \eqref{gustavsson} in three contributions
\begin{align}
    \nonumber   \int_\I K_N(x,x)\dd x- \int_\I\!\int_\I K_N(x,y)^2\dd x \dd y &= \int_{-L}^L\!\!\!\dd x\left(\int_{-\infty}^{-L}\!\!\!\dd y  + \int_{L}^\infty\!\!\!\dd y \right)K_N(x,y)^2 \\ 
    \nonumber=& \int_{-L}^\infty \!\!\!\dd x\int_{-\infty}^{-L}\!\!\!\dd yK_N(x,y)^2 + \int_L^\infty\!\!\!\dd x\int_{-\infty}^L\!\!\!\dd y K_N(x,y)^2\\
    &- 2\int_L^\infty \!\!\!\dd x\int_{-\infty}^{-L}\!\!\!\dd y K_N(x,y)^2.\label{split}
\end{align}
We notice that the first two terms of the RHS of equation \eqref{split} are $\Var(N_{(-\infty,L]})$ and $\Var(N_{[L,\infty)})$, respectively. It remains to show that, when $L$ is near $\sqrt{2}$, the last term of the RHS vanish when $N$ is large. We place ourselves in the \emph{edge} regime, where the value of $L$ is close to $\sqrt{2}$ and we define the scaled variable $s=(L-\sqrt{2})\sqrt{2}N^{2/3}$. This variable defines the width of the edge, $\mathcal{O}(N^{-2/3})$, and is the region where the kernel behaves asymptotically as the Airy kernel \cite{TraWid94,TraWid96,For10}
\begin{equation}
    K_N\left(\sqrt{2} + \frac{u}{\sqrt{2}N^{2/3}},  \sqrt{2} + \frac{v}{\sqrt{2}N^{2/3}}\right) \overset{N\gg 1}{\sim} \sqrt{2} N^{2/3} K_{\rm Ai}(u,v) \;,\label{airy_k}
\end{equation}
\begin{equation} \; 
    K_{\rm Ai}(u,v) = \frac{{\rm Ai}(u) {\rm Ai}'(v)-{\rm Ai}(v) {\rm Ai}'(u)}{u - v} \;,
\end{equation}
where ${\rm Ai}(x)$ is the Airy function. We study the last term of the RHS of \eqref{split} in this regime. By changing variables $x=\sqrt{2} + \frac{u}{\sqrt{2}N^{2/3}}$, $y=\sqrt{2} + \frac{v}{\sqrt{2}N^{2/3}}$ and using the behavior for large $N$ we find
\begin{align}
     2\int_L^\infty \!\!\!\dd x\int_{-\infty}^{-L}\!\!\!\dd y K_N(x,y)^2 \overset{N\gg 1}{\longrightarrow} 2\int_s^\infty \!\!\!\dd u\int_{-\infty}^{-4N^{2/3}-s}\!\!\!\!\!\!\dd v K_{\rm Ai}(u,v)^2 \to 0 .\label{airy_convergence}
\end{align} 
Since the integrand is finite and the domain on integration shrinks to zero as $N$ increases, this term is negligible for any $s$ in this scaling limit and, using the box symmetry, we may write
\begin{equation}
    \Var(N_{[-L,L]})\overset{N\gg 1}{\sim} \Var(N_{(-\infty,-L]})+\Var(N_{[L,\infty)})=2\Var(N_{[L,\infty)})\text{, for }L\sim\sqrt{2}\ .\label{independent}
\end{equation}
This is expected, since for large $N$ and when $L$ is around the edge, the intervals $(-\infty,-L]$ and $[L,\infty)$ are sufficiently far apart that the random variables $N_{(-\infty,-L]}$ and $N_{[L,\infty)}$ can be treated as independent when fluctuating around their average values.

Writing $\Var(N_{[-L,L]})$ as \eqref{independent} is convenient because $\Var(N_{[L,\infty)})$ is known: it was obtained in \cite{Gus05} using asymptotics of the Airy kernel. Using the scaling $s=(L-\sqrt{2})\sqrt{2}N^{2/3}$ we find
\begin{equation}
    2\Var(N_{[L,\infty)})=V_2(s)=2\int_s^\infty \! \dd u\int_{-\infty}^s \! \dd v K_{\rm Ai}^2(u,v)\ . \label{V_s_gaussian}
\end{equation}
%
%
The variance of the number of eigenvalues inside the $L$-box is then given, in the edge regime for $\beta=2$, by
\begin{equation}
  \Var(N_{[-L,L]})= V_2(s),~\text{ for }L=\sqrt{2} + \frac{s}{\sqrt{2}N^{2/3}}\ .\label{edge_formula}
\end{equation}
The asymptotic behavior of $V_2(s)$ is given by \cite{Gus05}
\begin{align}
    V_2(s)\sim\begin{cases} \frac{3}{2\pi^2}\ln|s|, & \text{ for }s\to -\infty \\ \frac{1}{8\pi s^{3/2}}\exp\left(-\frac{4}{3}s^{3/2}\right), & \text{ for }s\to \infty \end{cases}.\label{edge_cases}
\end{align}

We expect a matching between the limit $L\to\sqrt{2}$ from the extended bulk regime $N^{-1}<L<\sqrt{2}$ and the limit $s\to-\infty$ in the edge regime. Indeed, replacing $L=\sqrt{2} + \frac{s}{\sqrt{2}N^{2/3}}$ in equation \eqref{var_gauss_index} for $\beta=2$ and taking the large $N$ limit yields
\begin{equation}
    \frac{1}{\pi^2}\ln\left(NL(2-L^2)^{3/2}\right)\overset{N\gg 1}{\longrightarrow}  \frac{3}{2\pi^2}\ln(-s),~\text{ for }L=\sqrt{2} + \frac{s}{\sqrt{2}N^{2/3}}\ ,
\end{equation}
in agreement with the limit in equation \eqref{edge_cases} when $s\to -\infty$. Assuming this matching holds for all values of $\beta$, we expect the following asymptotic behaviors
\begin{align}
    V_\beta(s)\sim\begin{cases} \frac{3}{\beta\pi^2}\ln|s|, & \text{ for }s\to -\infty \\ \frac{1}{C_\beta(s)}\exp\left(-\frac{2\beta}{3}s^{3/2}\right), & \text{ for }s\to \infty \end{cases},\label{edge_cases_full}
\end{align}
where $C_\beta(s)$ is a power of $s$ dependent on $\beta$ whose value for $\beta=2$ is given by $C_2(s)=8\pi s^{3/2}$.

\subsubsection{Tail regime}

\

\begin{figure}[!htbp]
  \centering
  \includegraphics[width=0.55\linewidth]{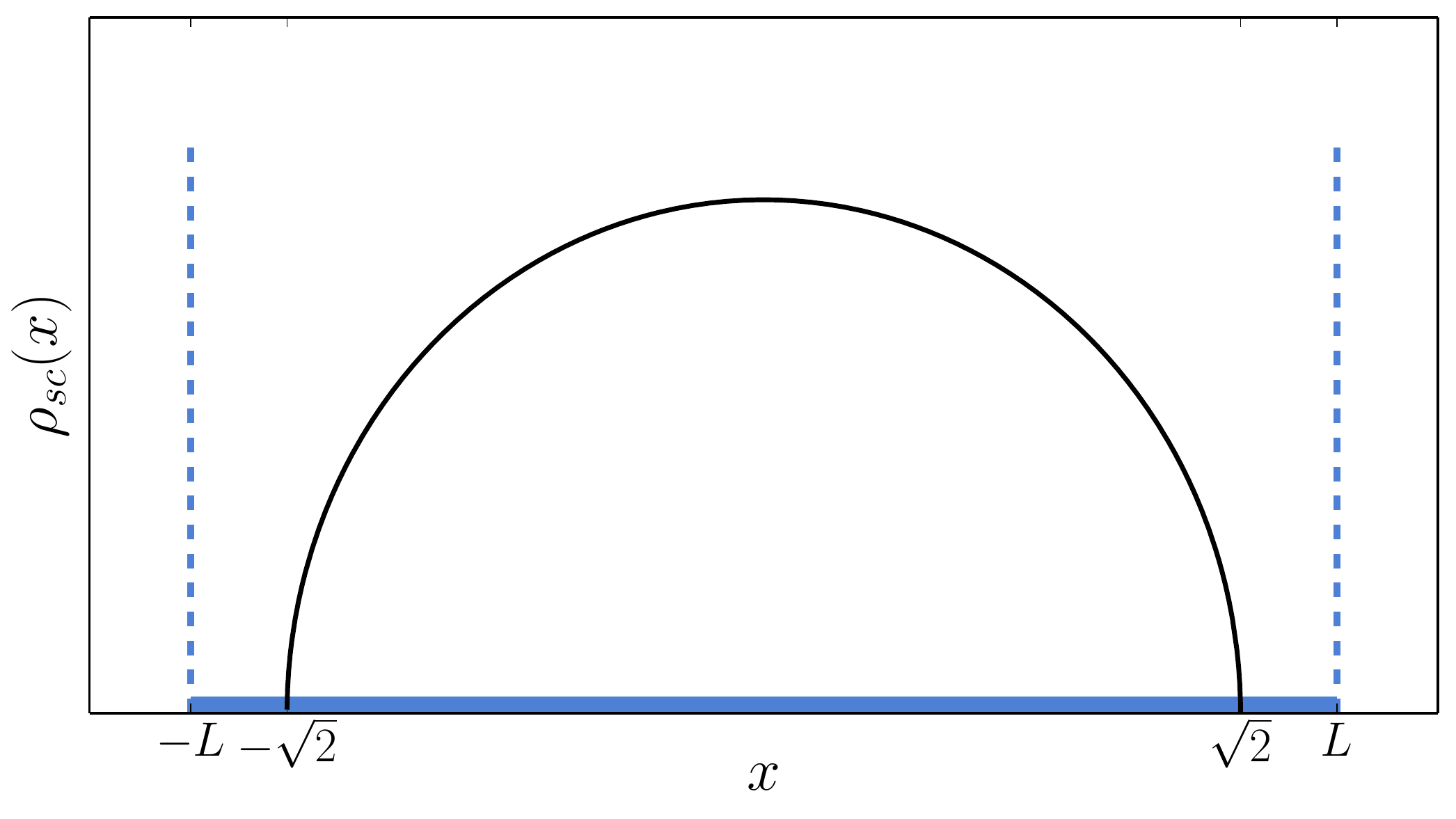}
  \caption{Tail regime for the Gaussian ensemble.}
  \label{fig:tail gaussian}
\end{figure}

While it is certainly possible to use the Coulomb gas technique for the \emph{tail} regime, where $|L-\sqrt{2}|\gg N^{-2/3}$, it is more convenient to obtain the number variance $\Var(N_{[-L,L]})$ there by recalling a few properties of the statistics of atypical fluctuations of the spectrum. 

We first note that equation \eqref{independent} may be also applied for any value of $L$ larger than $\sqrt{2}$. We thus need only to determine $\Var(N_{[L,\infty)})$ when $L>\sqrt{2}$. For values of $L$ larger than the edge of the semi-circle, the probability of finding one eigenvalue in the interval $[L,\infty)$ is equivalent to the probability of having the largest eigenvalue located at $x\geq L$. This is an atypical configuration whose statistics was first computed in \cite{MajVer09}. Keeping only leading order terms for large $N$, the pdf of the largest eigenvalue $x_{\text{max}}$ for the normalized Gaussian ensemble is given by $P(x_{\text{max}}=L)\approx \exp\left( -\beta N \phi(L)\right)$, where
\begin{equation}
    \phi(L)=\frac{1}{2}L\sqrt{L^2-2}+\ln\left[\frac{L-\sqrt{L^2-2}}{\sqrt{2}}\right]\ . \label{eq:phiL}
\end{equation}

\begin{figure}[!htbp]
	\centering
	\includegraphics[width=0.5\linewidth]{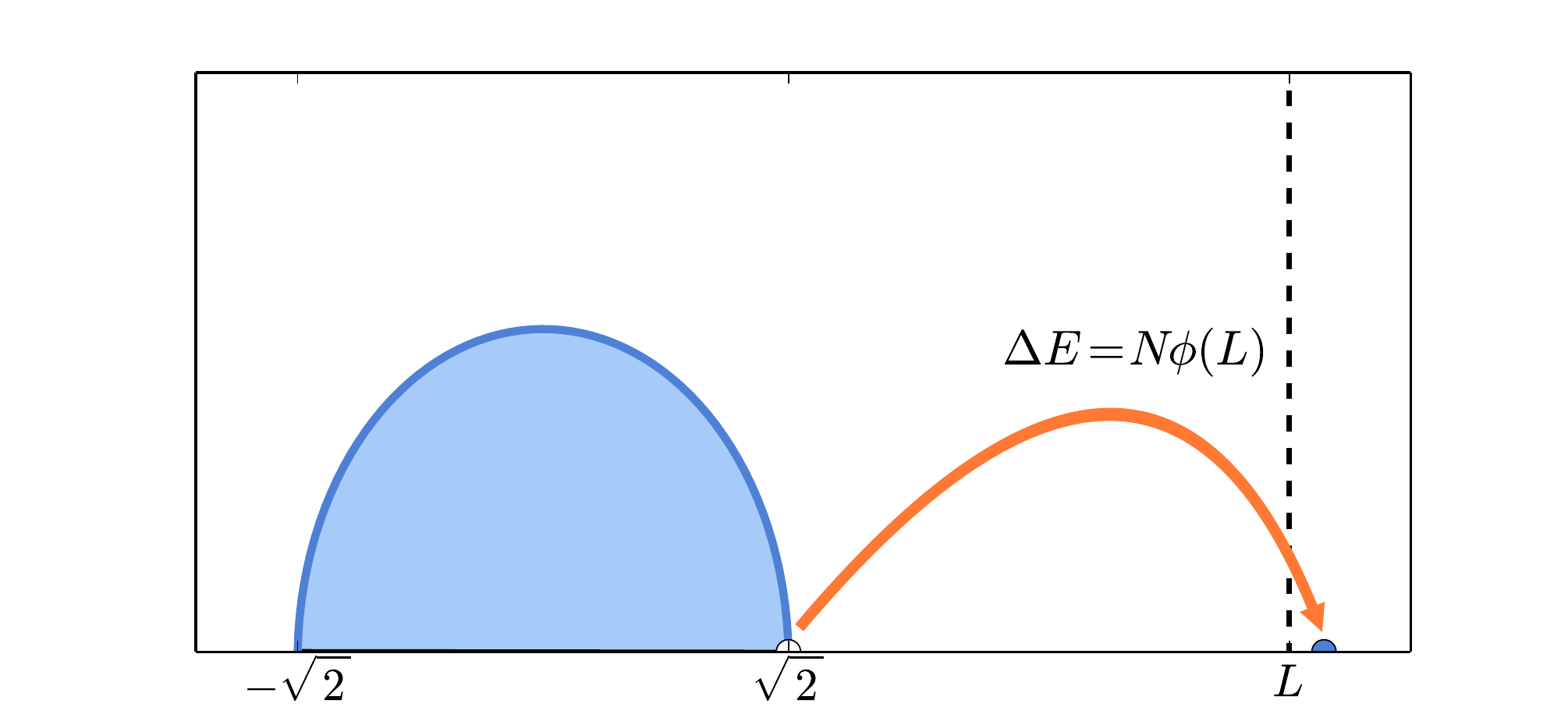}
	\caption{Energy required to pull one eigenvalue from the semi-circle sea and place it at $L>\sqrt{2}$.}
	\label{fig:shift_larger}
\end{figure}

In the Coulomb gas framework, we identify $N\phi(L)$ as the energy required to pull one charge from the Wigner's sea and bring it to $L$ (see figure \ref{fig:shift_larger}). For small values of $k$, the probability of finding $k$ eigenvalues in the interval $[L,\infty)$ can be approximated by the product of their individual chance of falling outside the interval. This becomes $P(k,L) \propto \exp\left(-\beta N k \phi(L) \right)$, where $A$ is a normalization constant. For large values of $N$, $A$ may be approximated by $\left(\sum_{k=0}^\infty P(k,L)\right)^{-1} = 1-\exp\left(-\beta N \phi(L)\right)$. The number variance on $[L,\infty)$ may then be calculated directly as
 \begin{equation}
     \Var(N_{[L,\infty)}) = \la k^2 \ra-\la k \ra^2= \sum_{k=1}^\infty k^2 P(k,L)-\left(\sum_{k=1}^\infty k P(k,L)\right)^2 \approx\frac{\mathrm{e}^{-\beta N \phi(L)}}{(1-\mathrm{e}^{-\beta N \phi(L)})^2}\approx \mathrm{e}^{-\beta N \phi(L)}\ ,\label{tail_result}
\end{equation}
to leading order in $N$. Therefore, the variance for the number of eigenvalues between $-L$ and $L$, when $L>\sqrt{2}$ and $N$ is large, reads
\begin{equation}
    \Var(N_{[-L,L]})= 2\Var(N_{[L,\infty)}) \approx \mathrm{e}^{-\beta N \phi(L)}\ ,
    \label{tail_final_result}
\end{equation}
which means that fluctuations of eigenvalues inside an interval larger than their equilibrium support in the $N\to\infty$ limit decay exponentially with the size of the interval.

Replacing $L=\sqrt{2} + \frac{s}{\sqrt{2}N^{2/3}}$ and taking the large $N$ limit on equation \eqref{tail_final_result} will retrieve
\begin{equation}
   \Var\left(N_{\left[-L,L\right]}\right)\approx \e^{-\beta N \phi\left(\sqrt{2} + \frac{s}{\sqrt{2}N^{2/3}}\right)} \xrightarrow{N\gg 1} \e^{-\frac{2\beta}{3}  s^{3/2}},~\text{ for }L=\sqrt{2} + \frac{s}{\sqrt{2}N^{2/3}}\ ,
\end{equation}
which is the leading order in $s$ of equation \eqref{edge_formula} for $s\to \infty$. This confirms the matching between the right limit of the edge regime and the tail regime.

\subsection{Comparison with numerics}

Eigenvalues of $\beta$-Gaussian random matrices can be simulated with little computational effort using the tridiagonal representation found by Dumitriu and Edelman \cite{DumEde02}. Results for the extended bulk and edge regimes are shown in figure \ref{fig:variance_gaussian}. We are not able to perform simulations for the tail regime due to the extreme small probability of finding an eigenvalue above $\sqrt{2}$.

\section{$\beta$-Wishart ensemble}
\label{sec:wishart}


In this section we apply the formalism based on the Coulomb gas technique combined with the resolvent method to another classical ensemble of random matrices, the $\beta$-Wishart. The jpdf of Wishart eigenvalues is given by equation \eqref{eq:eigen_jpdf_general_3} with potential $V(x)=\frac{x}{2}-\alpha \ln x$. Again, for general $\beta$, there exist tri-diagonal ensembles that give rise to this jpdf \cite{DumEde02}. We consider $N\sim M$, which implies $\alpha \sim N^{-1}$ and hence we neglect the logarithmic contribution to the potential in the large $N$ limit. Following the same route as in the Gaussian case,  
the pdf of $N_\I$ can be written as
\begin{equation}
    \mathcal{P}^{(W)}_\beta(N_\mathcal{I}=k_\mathcal{I}N)=\frac{1}{Z_{N,\beta}} \int \prod_{k=1}^N \dd x_k \mathrm{e}^{-\beta N\sum_{k=1}^N  \frac{x_k}{2}} \prod_{i>j}|x_i-x_j|^\beta \delta\left(k_\mathcal{I}N-\sum_{l=1}^N\mathbb{1}_\mathcal{I}(x_l)\right).
    \label{prob_wishart}    
\end{equation} 
Replacing once again the multiple integral by a functional integral over the normalized density of charges, we find for large $N$
\begin{align}
   \mathcal{P}^{(W)}_\beta(N_\mathcal{I}= k_\mathcal{I}N)=& \; \frac{1}{Z_N}\int \mathcal{D}[\rho] \dd \mu \dd \eta \e^{-\beta N^2 S^{(W)}[\rho]+\mathcal{O}(N)}\\
    \approx &\; \e^{-\beta N^2 \left(S^{(W)}[\rho^\star]-S^{(W)}[\rho_{mp}]\right)}=\e^{-\beta N^2 \psi^{(W)}(k_\I)},
\label{saddle_wish} 
\end{align}
where
\begin{align}
    \nonumber S^{(W)}[\rho]=&\int \frac{x}{2}\rho(x)\dd x - \frac{1}{2}\iint \dd x \dd x'\rho(x)\rho(x')\ln|x-x'| \\&+ \mu\left( \int_a^b\rho(x)\dd x-k_\I\right) +\eta\left(\int_0^\infty\rho(x)\dd x-1 \right).
    \label{action_wish}
\end{align}


For large $N$ we apply the saddle-point method to obtain the constrained average density of eigenvalues $\rho^\star(x)$. The functional derivative of the action reads
\begin{equation} 
    \left. \frac{\delta S^{(W)}}{\delta \rho}\right|_{\rho^\star}=0=\frac{x}{2} - \int\dd x' \rho^\star(x') \ln|x-x'| + \mu\mathbb{1}_{\I}(x) + \eta\mathbb{1}_{[0,\infty)}(x), \,\,\,\,\,\,\, x\in\text{supp }\rho^\star.
    \label{funct_der_wish}
\end{equation}
Differentiating with respect to $x$ yields
\begin{equation}
    \frac{1}{2}+ \eta\delta(x)+ \mu\left( \delta(x-b) - \delta(x-a) \right) = {\rm PV} \int\frac{\rho^\star(x')}{x-x'}\dd x',\qquad x\in\text{supp }\rho^\star .\label{deriv_integ_eq_wish}
\end{equation}
We notice how the hard wall at $x=0$ is responsible for an extra $\eta$-dependent term in the integral equation \eqref{deriv_integ_eq_wish}. As we did for the Gaussian ensemble, we now wish to convert the singular integral equation \eqref{deriv_integ_eq_wish} into an algebraic equation for the resolvent $G(z)=\int \dd x\rho^\star (x)/(z-x)$. We then first multiply both sides by $\frac{\rho^\star(x)}{z-x}$ and integrate over $x$. If either $a$ or $b$ are outside the support of $\rho^\star$, their contribution is zero. If $a$ and $b$ belong to the support of $\rho^\star$, this procedure yields
\begin{equation} 
    \frac{1}{2}\int \frac{\rho^\star(x)}{z-x} \dd x + \frac{A}{z-a} + \frac{B}{z-b} + \frac{C}{z} = \iint \frac{\rho^\star(x)}{z-x}\frac{\rho^\star(y)}{x-y}\dd y \dd x.
    \label{int_eq_2_wish}
\end{equation}
The RHS of integral equation \eqref{int_eq_2_wish} is again $G(z)^2/2$, while the LHS can be trivially written in terms of $G(z)$. We obtain
\begin{equation} 
    \frac{1}{2}G(z)+\frac{A}{z-a}+\frac{B}{z-b}+\frac{C}{z} = \frac{1}{2}G^2(z), \label{resolvent_eq_wish}
\end{equation}
with solution
\begin{equation} 
    G(z)=\frac{1}{2}\pm\sqrt{\frac{1}{4}+\frac{2A}{z-a}+\frac{2B}{z-b}+\frac{2C}{z}}=\frac{1}{2}\pm\frac{1}{2}\sqrt{\frac{(z-a_1)(z-a_2)(z-b_2)}{z(z-a)(z-b)}}\ .
    \label{resolvent_wish}
\end{equation}

\begin{figure}[!htbp]
    \centering
    \subfigure[{Sketch of the expected behavior of $\rho^\star(x)$ for the Wishart ensemble when $\I=[1,L]$ and $k_\I<\overline{k_\I}$.}\label{fig:wishart_density_smaller}]{\includegraphics[width=0.45\textwidth]{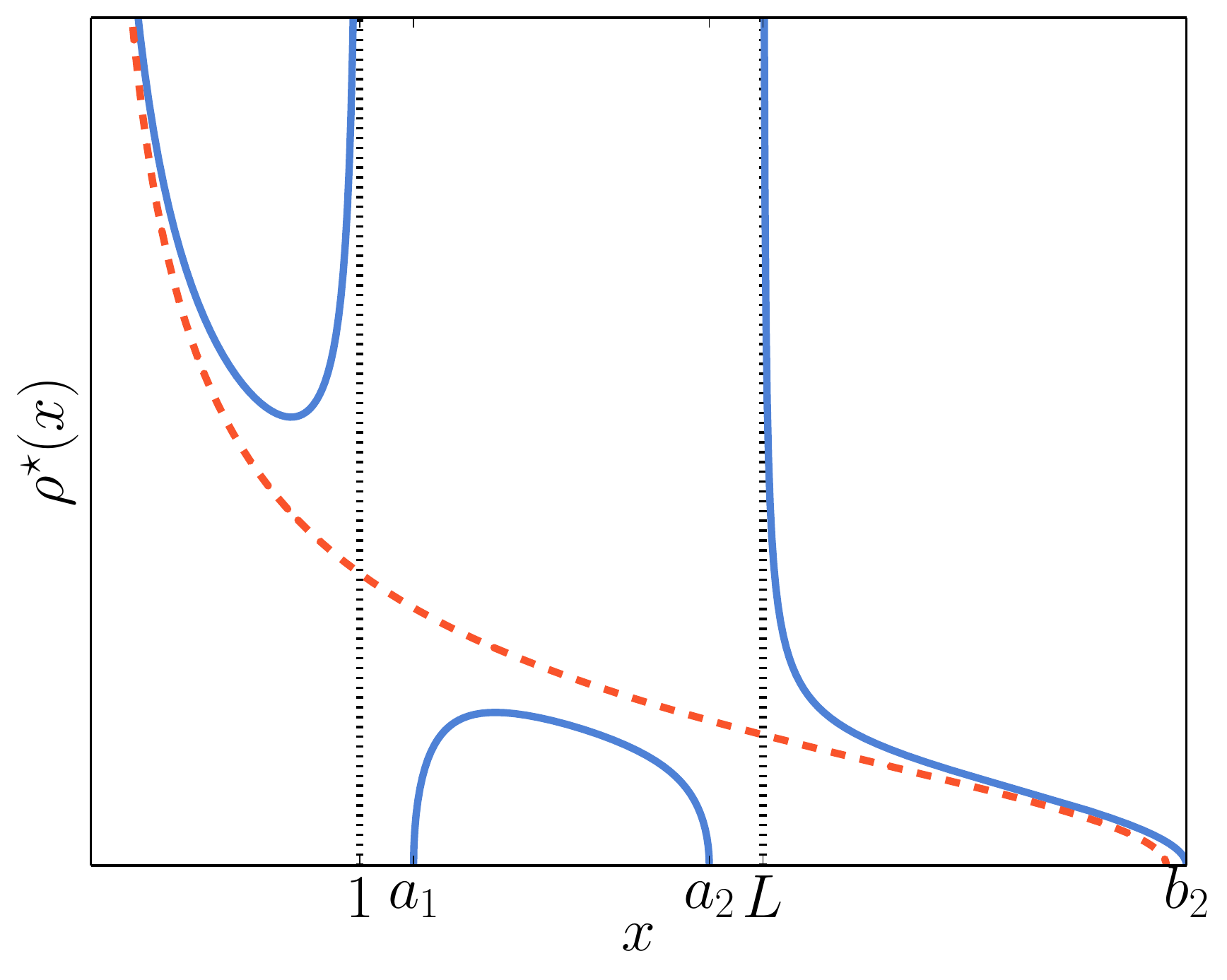}}\quad
    \subfigure[{Sketch of the expected behavior of $\rho^\star(x)$ for the Wishart ensemble when $\I=[1,L]$ and $k_\I>\overline{k_\I}$.}\label{fig:wishart_density_larger}]{\includegraphics[width=0.45\textwidth]{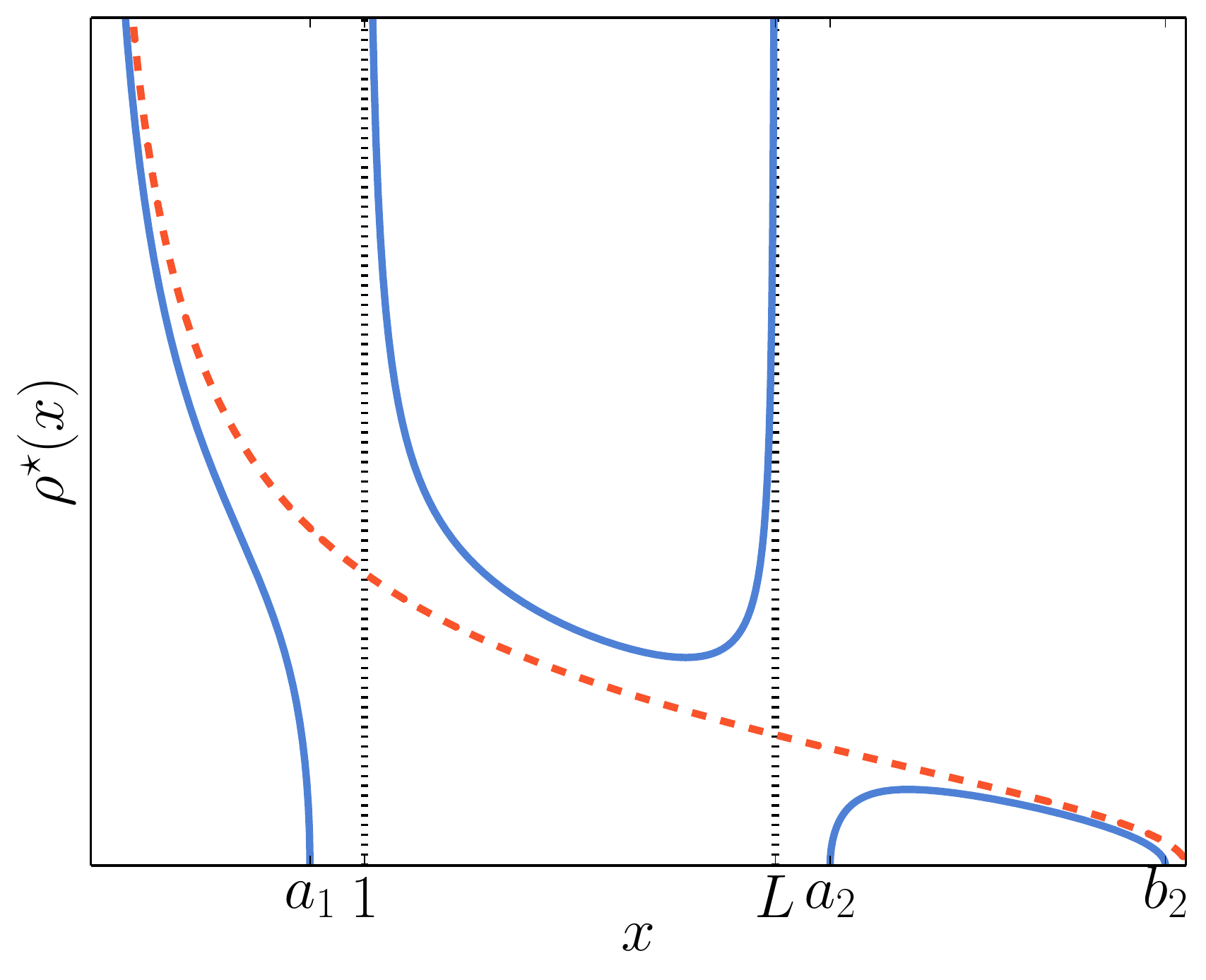}}
    \caption{}
    \label{fig:wishart_density}
\end{figure}


The constrained spectral density (when a prescribed fraction of eigenvalues is assigned to the interval $\mathcal{I}$) can be directly read off from $G(z)$

\begin{equation} 
    \rho^\star(x)=\frac{1}{2\pi}\sqrt{\frac{(a_1-x)(a_2-x)(b_2-x)}{x(a-x)(b-x)}}\label{support_wish} ,
\end{equation}
where the constants $a_1$, $a_2$ and $b_2$ are the edges of the support (see figure \ref{fig:wishart_density}) and are determined by the normalization condition $G(z)\to 1/z$ for $|z|\to\infty$ and the fraction of eigenvalues stacked inside $\I$. They imply
\begin{align}
   a_1+a_2+b_2=4+a+b & & \int_a^b\rho^\star(x)\dd x=k_\I.
\end{align}
Together with the chemical equilibrium condition \eqref{chemical_equilibrium}
\begin{equation} 
    -\int_{b}^{a_2} G(x) \dd x+\frac{a_2}{2}-\frac{b}{2} = \int_{a_1}^a G(x) \dd x+\frac{a_1}{2}-\frac{a}{2},
    \label{chemical_equilibrium_wish}
\end{equation}
 they uniquely determine the edges $a_1$, $a_2$ and $b_2$ of the support \eqref{support_wish}.

\subsection{Calculation of the rate function $\psi^{(W)}(k_\I)$}

To calculate the rate function, we notice that all steps taken in the Gaussian case may be applied if we replace the Gaussian potential by the Wishart potential. Using \eqref{funct_der_wish}, we may write the action as
\begin{equation} 
    S^{(W)}[\rho^\star]=\frac{1}{2}\int \frac{x}{2}\rho^\star(x)\dd x - \frac{\mu}{2}k_\I - \frac{\eta}{2}.
    \label{action_simple_wish}
\end{equation}
The Lagrange multipliers $\mu$ and $\eta$ will be calculated in precisely the same way as the Gaussian case, their formulas will only differ on the potential:
\begin{align}
	\mu=-\int_{b}^{a_2} G(x) \dd x+\frac{a_2}{2}-\frac{b}{2}, & & \eta = \ln b_2 -\frac{b_2}{2}- \int_{b_2}^\infty \left(G(x)-\frac{1}{x}\right)\dd x.
\end{align}
The action for the unconstrained case can also be easily computed, when $k_\I=\overline{k_\I}$ and the average density is the Mar\v{c}enko-Pastur distribution \eqref{wishart}. We find $S^{(W)}[\rho_{mp}]=3/4$.

The final formula for the rate function of $N_\I$ in the Wishart ensemble reads
\begin{align}
\nonumber	\psi^{(W)}(k_\I)=&\frac{1}{2}\int_0^\infty \frac{x}{2}\rho^\star(x)\dd x - \frac{\mu}{2}k_\I - \frac{\eta}{2}-S^{(W)}[\rho_{mp}]\\
\nonumber	=&\,\frac{1}{2}\int_0^\infty \frac{x}{2}\rho^{\star}(x)\dd x - \frac{1}{2}\left(\int_{b}^{a_2} G(x) \dd x-\frac{a_2^2}{2}+\frac{b^2}{2}\right)k_\I \\
    &- \frac{1}{2}\left[\ln b_2- \frac{b_2^2}{2} - \int_{b_2}^\infty \left(G(x)-\frac{1}{x}\right)\dd x\right]-\frac{3}{4}\ ,\label{rate_wish}
\end{align}
where $G(x)$ is equation \eqref{resolvent_wish} with the following choice of sign
\begin{equation}
    G(x)=\begin{cases} 
                    G_-(x) \text{ for } x>b_2\\
                    G_+(x) \text{ for } x<0\\
                    G_+(x) \text{ for } k_\I>\overline{k_\I}\text{ and }b<x<a_2 \\ 
                    G_+(x) \text{ for } k_\I<\overline{k_\I}\text{ and }a<x<a_1 \\ 
                    G_-(x) \text{ otherwise}\ ,
                \end{cases}\label{Goutside_Wish}
\end{equation}
and $\rho^\star(x)$ is \eqref{support_wish}.

\subsection{Number variance}

As in the Gaussian case, equation \eqref{rate_wish} is too general to be analyzed directly. We turn our attention to a more specific case, the interval $\I=[1,L]=[1,1+l]$, where $l$ is the length of the interval. 
We wish to perturb the rate function around its minimum to read off the variance of $N_\I$ from the quadratic behavior, modulated by a logarithmic singularity as in the Gaussian case. 

\begin{figure}
    \centering
    \renewcommand\thesubfigure{(\roman{subfigure}) }
    \subfigure[Bulk regime.\label{fig:wishart bulk regime}]{\includegraphics[width=0.45\textwidth]{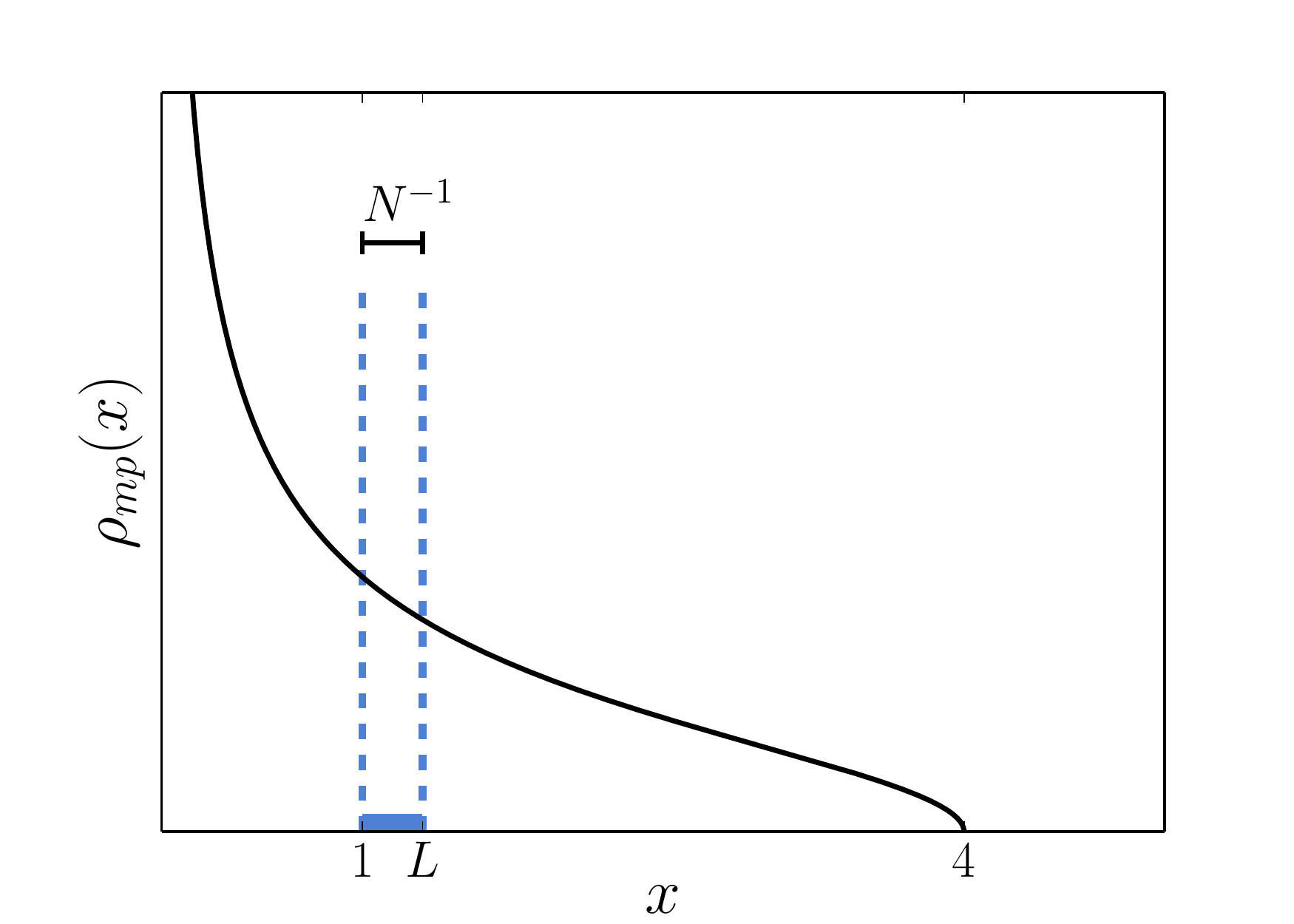}}
    \subfigure[Extended bulk regime.\label{fig:wishart extended regime}]{\includegraphics[width=0.45\textwidth]{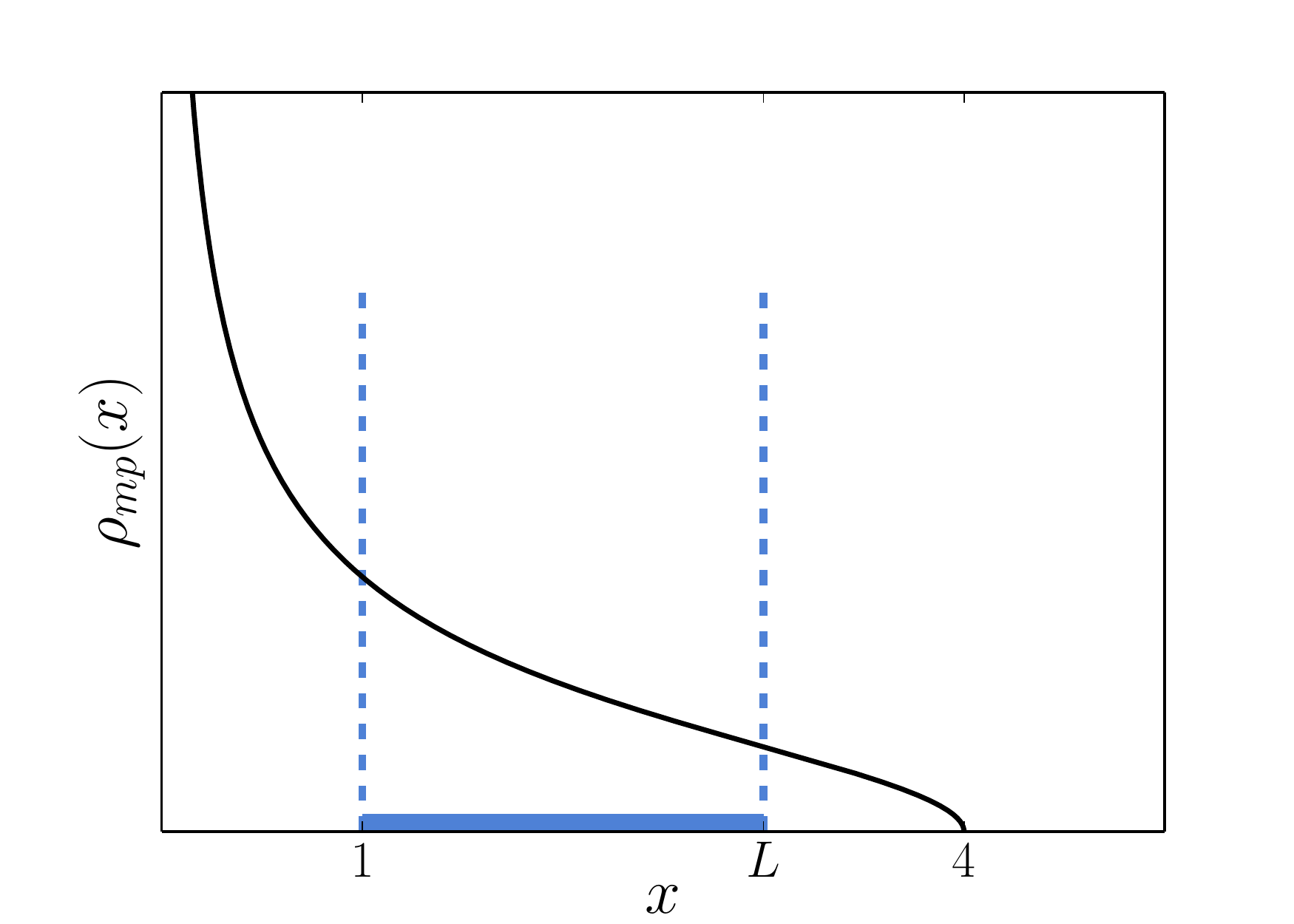}}
~
    \subfigure[Edge regime.\label{fig:wishart edge regime}]{\includegraphics[width=0.45\textwidth]{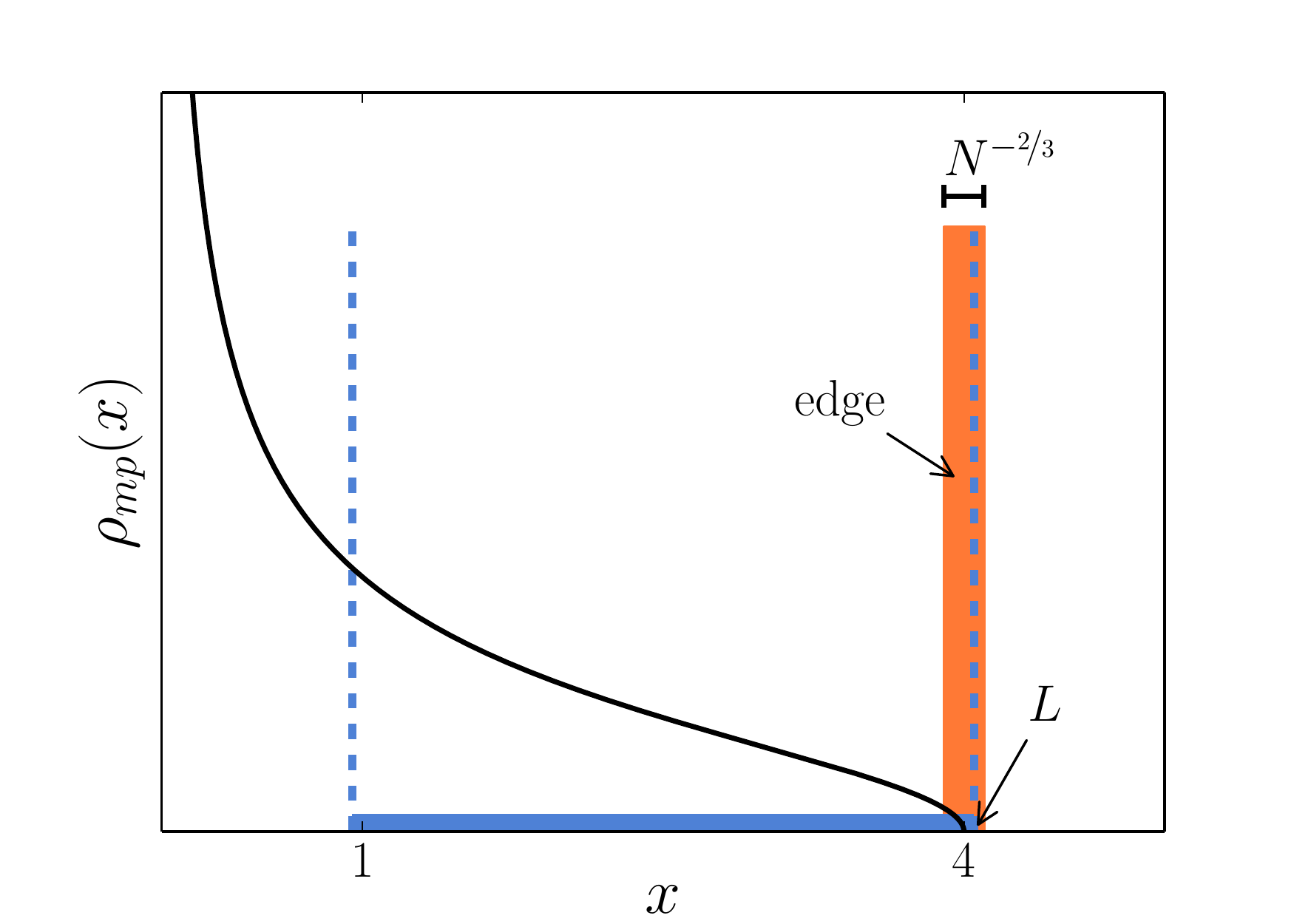}}
    \subfigure[Tail regime.\label{fig:wishart tail regime}]{\includegraphics[width=0.45\textwidth]{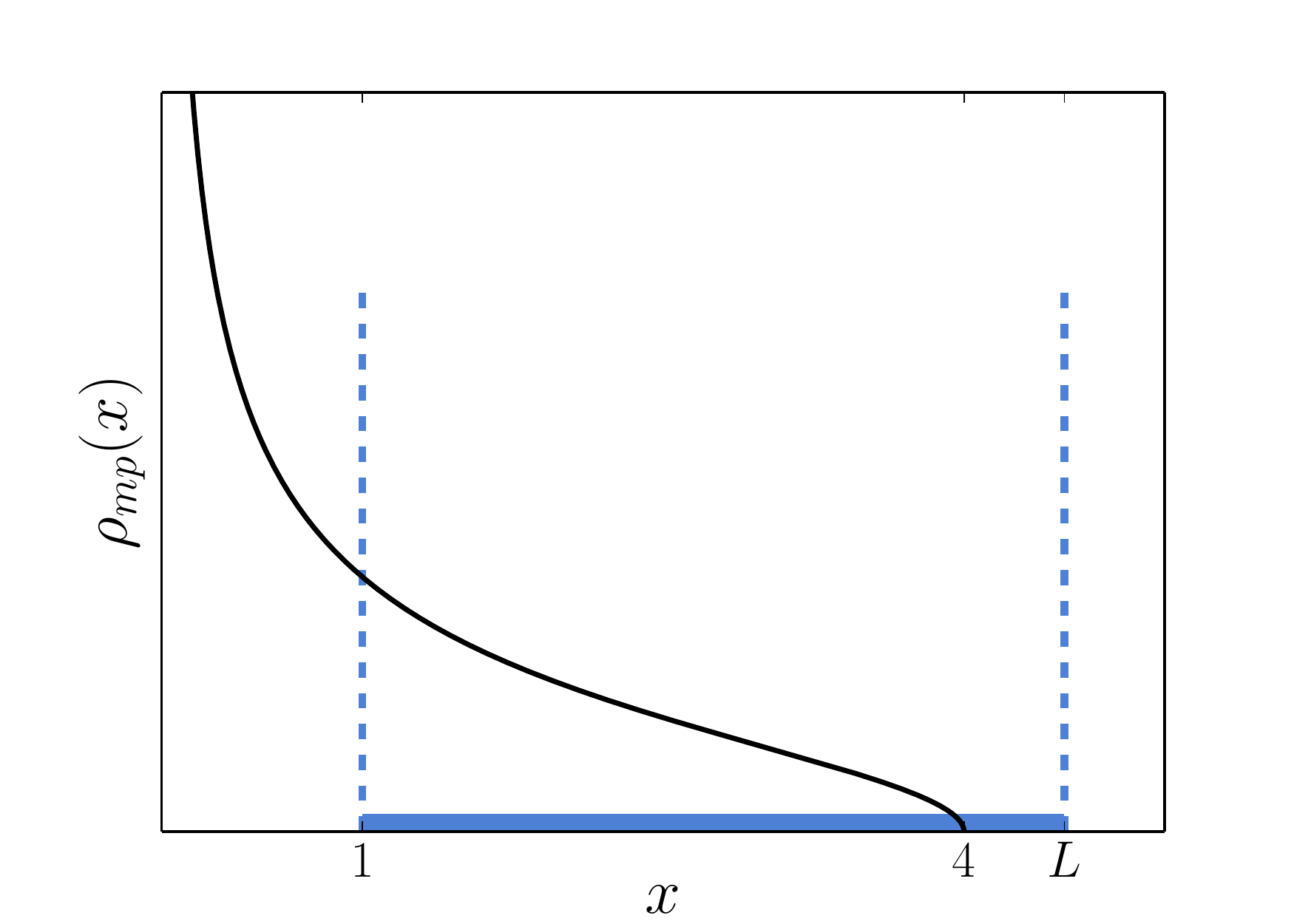}}
    \caption{Regimes of behavior of the number variance for the Wishart ensemble. The solid blue line represents the interval $\I=[1,1+l]$.}
    \label{fig:regimes wishart}
\end{figure}

The perturbative approach is similar, and we once more find three separate regimes for the number variance: an extended bulk, an edge and a tail regime, described in figure \ref{fig:regimes wishart}. We deal with these regimes separately.

\subsubsection{Extended bulk regime}

The calculation is very similar to the Gaussian case, therefore we refer to section \ref{sec:gaussian} and to Appendix \ref{app} for the details of the calculations of $\mu$ and $\eta$. Denoting the number of eigenvalues inside $[1,1+l]$ as $N_\I$, we obtain its variance, to leading orders of $N$ and for $1/N < l < 3$,
\begin{equation}
    \Var (N_{[1,1+l]}) = \frac{2}{\beta \pi^2}\ln\left(Nl(3-l\right)^{3/4})+\mathcal{O}(1),
    \label{var_wishart_index}
\end{equation} 
where $l=3$ is a special point, corresponding to $L=4$, the soft edge. 

As expected, the behavior on the extended bulk is extremely similar to the Gaussian case. The number variance has a non-monotonic behavior as a function of $l$, whose maximum is reached at the critical point $l^\star=12/7$ for the Wishart ensemble.

\subsubsection{Edge regime}

Results in the edge regime are similar to the Gaussian case, since both ensembles are governed by the Airy kernel near their soft edges. For the Wishart ensemble, we may write an equivalent of equation \eqref{independent},
\begin{equation}
    \Var(N_{[1,L]})\overset{\text{large }N}{\sim} \Var(N_{[0,1]})+\Var(N_{[L,\infty)})\text{, for }L\sim 4\ .\label{independent_wish}
\end{equation}

Since $[0,1]$ and $[1,\infty)$ are complementary sets on the positive semi-axis, $\Var(N_{[0,1]})=\Var(N_{[1,\infty)})$ and the first term of the RHS of equation \eqref{independent_wish} is in fact the variance of the index of the Wishart ensemble. This result was obtained in \cite{MajViv12}. 
\begin{equation}
    \Var(N_{[0,1]})=\Var(N_{[1,\infty)})=\frac{1}{\beta\pi^2}\ln N.
\end{equation}
To obtain $\Var(N_{[L,\infty)})$, we consider the $\beta=2$ case. Equation \eqref{airy_convergence} applies if we use the scaling $u=(N/4)^{2/3}(4-x)$ \cite{Sos02,Joh01}
\begin{equation}\label{edge_limit}
    K_N\left(4 + \frac{u}{(N/4)^{2/3}},  4 + \frac{v}{(N/4)^{2/3}}\right) \overset{N\gg 1}{\sim} (N/4)^{2/3} K_{\rm Ai}(u,v) \;,
\end{equation}
\begin{equation} \; 
    K_{\rm Ai}(u,v) = \frac{{\rm Ai}(u) {\rm Ai}'(v)-{\rm Ai}(v) {\rm Ai}'(u)}{u - v} \ .\label{airy_k_wish}
\end{equation}
Using the scaling variable $s=(L-4)(N/4)^{2/3}$, we find 
\begin{equation}
    \Var(N_{[L,\infty)})=\frac{1}{2}V_2(s)=\int_s^\infty \! \dd u\int_{-\infty}^s \! \dd v K_{\rm Ai}^2(u,v). \label{V_s_wishart}
\end{equation}
The variance becomes, for $\beta=2$ 
\begin{equation}
  \Var(N_{[1,L]})= \frac{1}{2\pi^2}\ln N + \frac{1}{2}V_2(s)\text{, for }L=4 + \frac{s}{(N/4)^{2/3}}.\label{edge_formula_wish}
\end{equation}
Asymptotics of $V_2(s)$ for $s\to\pm\infty$ were studied by \cite{Gus05} and are given by equation \eqref{edge_cases}. We notice a fundamental difference between this case and the Gaussian case, the presence of the $\ln N$ term. This contribution arises from fluctuations on the left side on the interval, on the $x=1$ point. The other contribution, $V_2(s)$, represents fluctuations on the right side on the interval, and increasing $L$ decreases the probability of having an eigenvalues to the right of the interval and hence decreases fluctuations on this side.

We notice that, as in the Gaussian case, we can explore the limits $s\to\pm\infty$ to obtain the matching between the edge regime and the extended bulk and tail regimes. For $\beta=2$ we confirm that this matching holds, and we can conjecture that it holds for all values of $\beta$. If we take $L=4+ \frac{s}{(N/4)^{2/3}}$ in equation \eqref{var_wishart_index}, we obtain, for large values of $N$:
\begin{equation}
    \frac{2}{\beta \pi^2}\ln\left[Nl(3-l)^{3/4}\right]\xrightarrow{N\gg 1}\frac{1}{\beta\pi^2}\ln N + \frac{3}{2\beta \pi^2}\ln |s| + \mathcal{O}(1)\text{, for }L=4 + \frac{s}{(N/4)^{2/3}}.
\end{equation}

If we assume that this matching holds for all values of $\beta$, we can write a general expression for the variance of the edge regime
\begin{equation}
  \Var(N_{[1,L]})= \frac{1}{\beta\pi^2}\ln N + \frac{1}{2}V_\beta(s)\text{, for }L=4 + \frac{s}{(N/4)^{2/3}}.\label{edge_formula_wish_beta}
\end{equation}
and its asymptotic behavior, to match its neighboring regions, should be
\begin{align}
    V_\beta(s)\sim\begin{cases} \frac{3}{\beta\pi^2}\ln|s|, & \text{ for }s\to -\infty \\ \frac{1}{C_\beta(s)}\exp\left(-\frac{2\beta}{3}s^{3/2}\right), & \text{ for }s\to \infty \end{cases},\label{edge_cases_full_wish}
\end{align}
where $C_\beta(s)$ is a power of $s$ dependent on $\beta$ whose value for $\beta=2$ is given by $C_2(s)=8\pi s^{3/2}$. The asymptotic behavior for $s\to\infty$ for general $\beta$ is obtained using the matching with the tail regime, presented in the next section.

\subsubsection{Tail regime}

\begin{figure}[!htpb]
    \centering
    \includegraphics[width=0.5\textwidth]{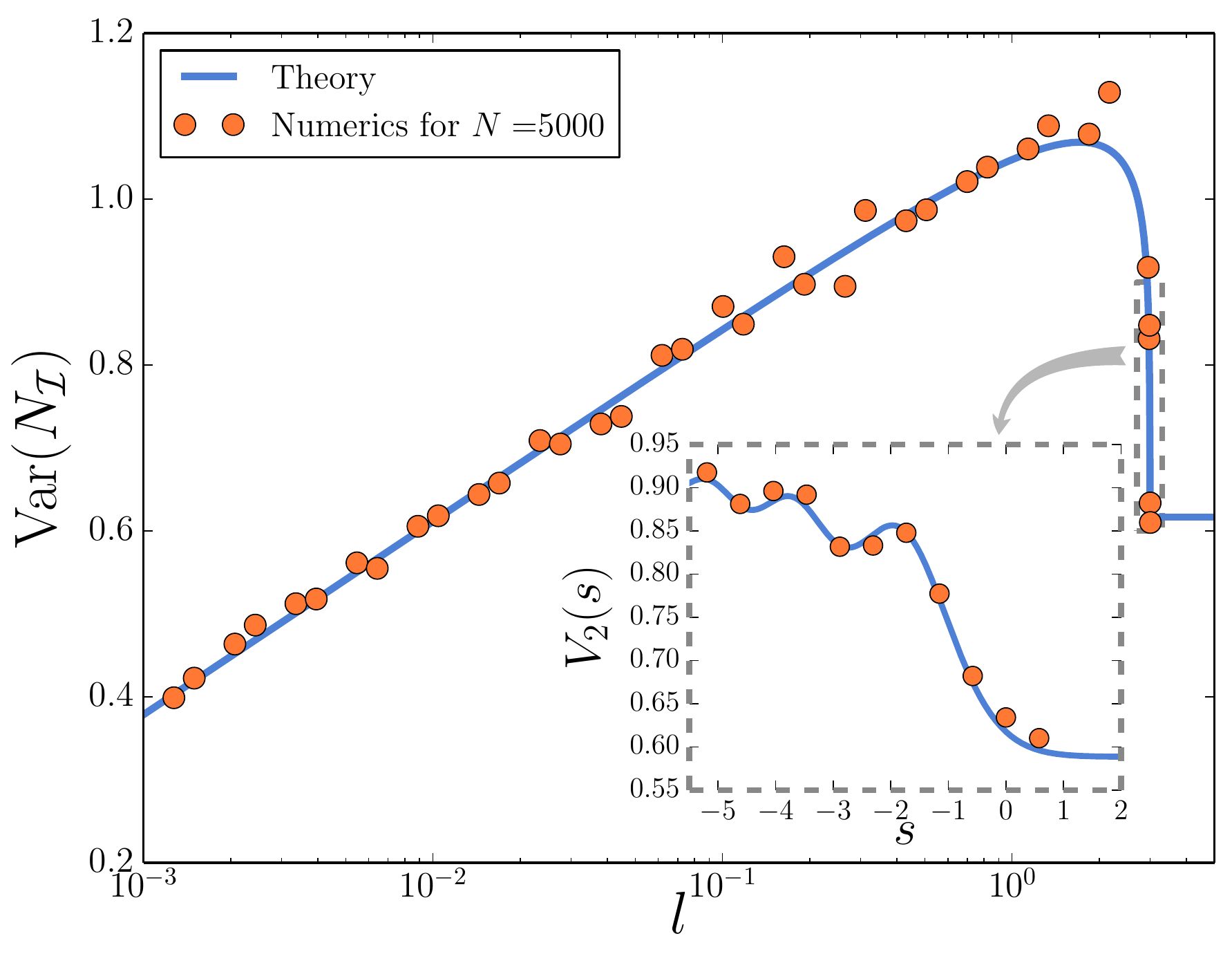}
    \caption{Results for the variance of $N_\I$ for the Wishart ensemble when $\I=[1,1+l]$ and $l<3$. Theory is equation \eqref{var_wishart_index}. Inset: results for the edge regime, where $s=(l-3)(N/4)^{2/3}$ and theory is equation \eqref{edge_formula}.}
    \label{fig:wishart_results}
\end{figure}

As the kernel $K_N(x,y)$ of both Gaussian and Wishart ensembles is described by the Airy kernel \eqref{airy_k} on the correct scaling limit, the statistics of their largest eigenvalue is equally described by the Tracy-Widom distribution. This implies that the probability of finding the largest eigenvalue far away from the support edge can be derived in an equivalent way for both ensembles, and we find a similar result for the variance of $N_\I$ when $L$ is larger than the soft edge $x=4$. The only difference is the form of the large deviation function $\phi$. For the Wishart ensemble \cite{MajVer09,NadMajVer11}, the probability of finding the largest eigenvalue to be much larger than its average value is given by
\begin{equation}
    P(\lambda_{\text{max}}>w)\sim \e^{-\beta N \Phi(w)},\quad w>4,
\end{equation}
where
\begin{equation}
    \Phi(w)=\sqrt{\frac{w(w-4)}{4}}+\ln\left[\frac{w-2-\sqrt{w(w-4)}}{2}\right].\label{eq:wishart_right_tail}
\end{equation}

We denote by $P(k,L)$ the probability of finding $k$ eigenvalues at positions larger than $L$, and $L>4$. Since the energy required to move one eigenvalue to position $L$ is $N\Phi(L)$, we obtain
\begin{equation}
     \Var(N_{[L,\infty)}) = \la k^2 \ra-\la k \ra^2= \sum_{k=1}^\infty k^2 P(k,L)-\left(\sum_{k=1}^\infty k P(k,L)\right)^2 \approx \e^{-\beta N \Phi(L)}\ ,\label{tail_result_wish}
\end{equation}
to leading order in $N$. Since equation \eqref{independent} applies also to the Wishart ensemble, we find, for $L>4$
\begin{equation}
    \Var(N_{[1,L]})= \frac{1}{\beta\pi^2}\ln N + T_\beta(L),
\end{equation}
where $T_\beta(L)\approx \e^{-\beta N \Phi(L)}$.

As above, replacing $L=4 + \frac{s}{(N/4)^{2/3}}$ and taking the leading order in $N$ on equation \eqref{tail_result_wish} will retrieve
\begin{equation}
   T_\beta\left(4+\frac{s}{(N/4)^{2/3}}\right)= \e^{-\beta N \Phi\left(4+\frac{s}{(N/4)^{2/3}}\right)} \xrightarrow{N\gg 1} \e^{-\frac{2\beta}{3}  s^{3/2}},
\end{equation}
which is the leading order in $s$ of equation \eqref{edge_formula_wish_beta} for $s\to \infty$. This confirms, as in the Gaussian case, the matching between the right limit of the edge regime and the tail regime.

\subsection{Comparison with numerics}

Numerical calculations in the Wishart ensemble are similar to the Gaussian ensemble. A tridiagonal ensemble with equivalent statistics was also found by Dumitriu and Edelman \cite{DumEde02} and its diagonalization is less costly than taking Gaussian matrices, multiplying them by their conjugate and diagonalizing the Wishart matrix. Results obtained are in figure \ref{fig:wishart_results}.

\section{$\beta$-Cauchy ensemble}
\label{sec:cauchy}


For $\beta=1,2,4$, this ensemble contains $N \times N$ matrices which might be real symmetric ($\beta=1$), complex hermitian ($\beta=2$) or quaternion self-dual ($\beta=4$) drawn from the distribution $P(X)\propto\left[\det\left({\bf 1}_N+X^2)\right)\right]^{-\beta (N-1)/2-1}$, where ${\bf 1}_N$ is the identity matrix $N\times N$. The Cauchy ensemble has found applications in the context of quantum transport \cite{Bro95} and free probability \cite{BurJanJur02,BurJur11,FyoKhoNoc13}. It is also one of the few exactly solvable ensembles whose average spectral density has fat tails extending over the full real axis, and given by (see Fig. \ref{fig:cauchy_dist_regimes_summary})
\begin{equation}
    \rho_{ca}(x)=\frac{1}{\pi}\frac{1}{1+x^2}.\label{cauchy_2}
\end{equation}

We recall its associated potential $V(x)=\left((N-1)/(2N)+1/(\beta N)\right)\ln(1+x^2)$. Since the potential is of the same order as the repulsion between charges in the Coulomb gas, $V(x)$ is not strong enough to confine the eigenvalues into a compact region and the average density extends to infinity. We define the interval $\I=[a,b]$ and we repeat the previous derivation of the Coulomb gas method.

%
In the large-$N$ limit, the Cauchy potential can be simply written as $V(x)= \ln (1+x^2)/2$. Using this potential, we write the pdf of $N_\I$ as
\begin{equation}
    \mathcal{P}^{(C)}_\beta(N_\mathcal{I}=k_\mathcal{I}N)=\frac{1}{Z_{N,\beta}} \int \prod_{k=1}^N \dd x_k \mathrm{e}^{-\beta N\sum_{i=1}^N  \frac{1}{2}\ln(1+x_i^2)} \prod_{i>j}|x_i-x_j|^\beta \delta\left(k_\mathcal{I}N-\sum_{l=1}^N\mathbb{1}_\mathcal{I}(x_l)\right).
    \label{prob_cauchy}    
\end{equation} 
Again, we write this pdf in a continuum approximation as 
\begin{equation}
   \mathcal{P}^{(C)}_\beta(N_\mathcal{I}= k_\mathcal{I}N)= \frac{1}{Z_N}\int \mathcal{D}[\rho] \dd \mu \dd \eta \e^{-\beta N^2 S^{(C)}[\rho]+\mathcal{O}(N)} \approx \e^{-\beta N^2 \left(S^{(C)}[\rho^\star]-S^{(C)}[\rho_{ca}]\right)}=\e^{-\beta N^2 \psi^{(C)}(k_\I)},
\label{saddle_cauchy} 
\end{equation}
where
\begin{align}
    \nonumber S^{(C)}[\rho]=&\frac{1}{2}\intinf \ln(1+x^2) \rho(x)\dd x - \frac{1}{2}\iint_{-\infty}^{+\infty}\!\! \dd x \dd x'\rho(x)\rho(x')\ln|x-x'| \\&+ \mu\left( \int_a^b\rho(x)\dd x-k_\I\right) +\eta\left(\intinf \rho(x)\dd x-1 \right),
    \label{action_cauchy}
\end{align}
and $\rho_{ca}(x)$ is the equilibrium Cauchy density \eqref{cauchy_2} in the unconstrained case.


We obtain the average density $\rho^\star(x)$ by differentiating \eqref{action_cauchy} functionally with respect to the density $\rho$
\begin{equation} 
    \left. \frac{\delta S^{(C)}}{\delta \rho}\right|_{\rho^\star}=0=\frac{\ln(1+x^2)}{2} - \int\dd x' \rho^\star(x') \ln|x-x'| + \mu\mathbb{1}_{\I}(x) + \eta, \,\,\,\,\,\,\, x\in\text{supp }\rho^\star\ .
    \label{funct_der_cauchy}
\end{equation}
Its derivative with respect to $x$ yields
\begin{equation}
	\frac{x}{1+x^2}+\mu\left(\delta(x-b)-\delta(x-a)\right)={\rm PV} \int \frac{\rho^\star(y)}{x-y}\dd y\ . \label{int_eq_cauchy}
\end{equation}

Once again, we multiply both sides of equation \eqref{int_eq_cauchy} by $\rho^\star(x)/(z-x)$ and integrate with respect to $x$. While the RHS turns out to be the same as the previous examples, the LHS is not polynomial and the technique used for the Gaussian and Wishart ensembles needs to be modified. This procedure is analogous to the case described in \cite{MarMajSch14_2}. The algebraic equation for the resolvent eventually reads
\begin{equation}
	aA + bB -\frac{1}{2}+z(A+B)+zG(z)=(1+z^2)\left(\frac{1}{2}G^2(z)+\frac{A}{z-a}+\frac{B}{z-b}\right), \label{algebraic_cauchy}
\end{equation}
where $A$ and $B$ are constants to be determined by the fraction of eigenvalues $k_\I$ stacked inside $\I=[a,b]$ and the chemical equilibrium condition \eqref{chemical_equilibrium}. The solution of \eqref{algebraic_cauchy} is lengthy, but straightforward
\begin{equation}
    G(z)=\frac{z}{1+z^2}\pm\frac{1}{1+z^2}\sqrt{\frac{P(A,B,a,b,z)}{(z-a)(z-b)}},
\end{equation}
where $P(A,B,a,b,z)=a b - 2 A b - 2 a^2 A b - 2 a B - 
 2 a b^2 B + (-a + 2 A + 2 a^2 A + 2 B + b (-1 + 2 b B)) z + (1 - 2 A b - 2 a^2 A b - 2 a B - 2 a b^2 B) z^2 + (2 A(1+a^2) + 2 (1 + b^2) B) z^3$ is a third-degree polynomial in $z$. We can write the resolvent in the more appealing form
\begin{equation}
    G(z)=\frac{z}{1+z^2}\pm\frac{1}{1+z^2}\sqrt{\frac{(z-a_1)(z-a_2)(z-b_2)}{K(z-a)(z-b)}},\label{cauchy_resolvent}
\end{equation}
where $K$, $a_1$, $a_2$ and $b_2$ are determined by equating $(z-a_1)(z-a_2)(z-b_2)/K=P(A,B,a,b,z)$. The average density is obtained directly
\begin{equation}
    \rho^\star(x)=\frac{1}{\pi}\frac{1}{1+x^2}\sqrt{\frac{(a_1-x)(x-a_2)(x-b_2)}{K(x-a)(x-b)}}.
\end{equation}

\subsection{Analysis for $[-L,L]$}

We restrict ourselves for definiteness to the symmetric case $\I=[-L,L]$. Equation \eqref{algebraic_cauchy} greatly simplifies, as $a=-b=L$ and $A=-B$. The resolvent thus becomes
\begin{equation}
    G(z)=\frac{z}{1+z^2}\pm\frac{1}{1+z^2}\sqrt{\frac{a_1^2-z^2}{z^2-L^2}}\ .
 \end{equation} 
\begin{figure}[!htbp]
    \centering
    \subfigure[{Sketch of the expected behavior of $\rho^\star(x)$ for the Cauchy ensemble when $\I=[-L,L]$ and $k_\I<\overline{k_\I}$.}\label{fig:cauchy_density_smaller}]{\includegraphics[width=0.45\textwidth]{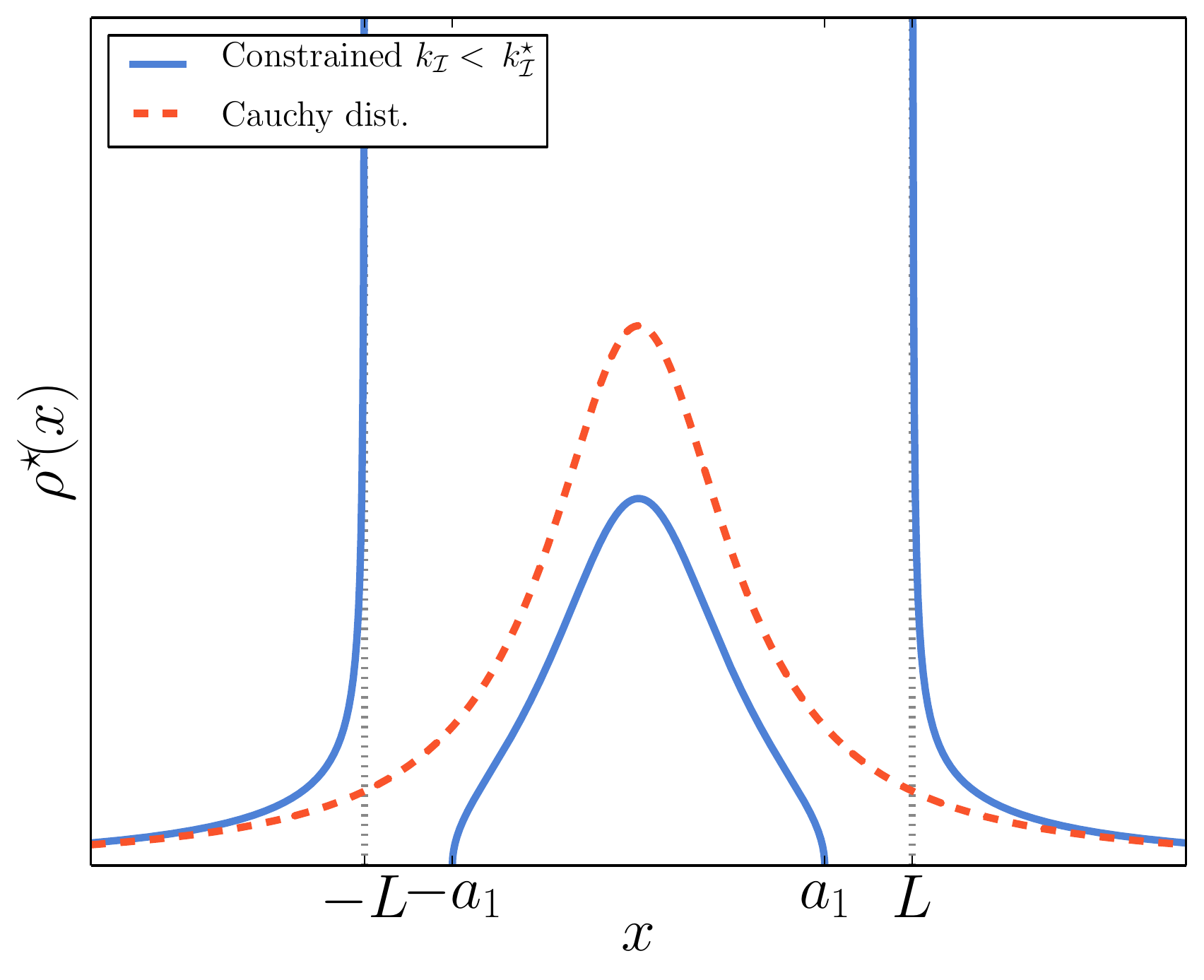}}\quad
    \subfigure[{Sketch of the expected behavior of $\rho^\star(x)$ for the Cauchy ensemble when $\I=[-L,L]$ and $k_\I>\overline{k_\I}$.}\label{fig:cauchy_density_larger}]{\includegraphics[width=0.45\textwidth]{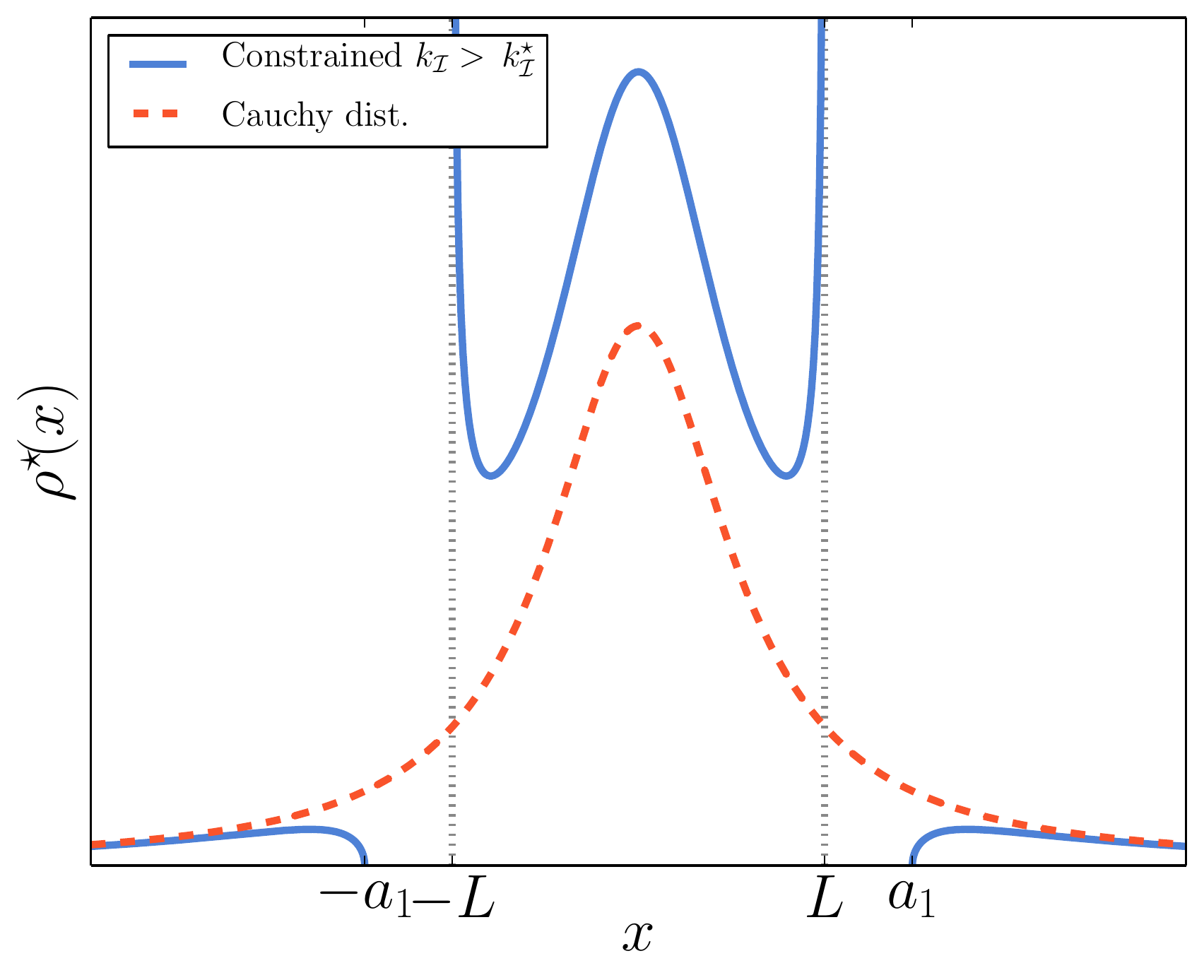}}
    \caption{}
    \label{fig:cauchy_density}
\end{figure}
%
The constrained density $\rho^\star(x)$ reads (see figure \ref{fig:cauchy_density})
\begin{equation}
    \rho^\star(x)=\frac{1}{\pi}\frac{1}{1+x^2}\sqrt{\frac{x^2-a_1^2}{x^2-L^2}}\ ,\label{cauchy_dist}
\end{equation}
where the only constant to be determined is $a_1$ by imposing $\int_{-L}^L \rho^\star(x)\dd x=k_{[-L,L]}$. 

We recall once more the simplified equation for the action \eqref{action_simple}, modified for the Cauchy potential
\begin{equation} 
    S^{(C)}[\rho^\star]=\frac{1}{2}\int \frac{\ln(1+x^2)}{2}\rho^\star(x)\dd x - \frac{\mu}{2}k_\I - \frac{\eta}{2}.
    \label{action_simple_cauchy}
\end{equation}
The Lagrange multipliers are obtained as in the previous two cases, with a small difference. The formula for $\mu$ is precisely the same as \eqref{chemical_equilibrium}, adapted to the Cauchy potential, but for $\eta$ we cannot apply the simplification \eqref{border_condition}, as there is no ``upper edge'' for the support of $\rho^\star$ in the Cauchy case. This forces us to write $\eta$ as a slightly more complicated integral compared to previous cases. Applying equation \eqref{funct_der_cauchy} at a point $p$ inside the support of $\rho^\star$ but outside the interval $[-L,L]$ yields an expression for $\eta$ in terms of integrals over $\rho^\star$. For the Lagrange multipliers we thus find
\begin{align}
    \mu=-\int_{L}^{a_1} \!\!G(x) \dd x+\frac{1}{2}\ln\left(\frac{1+a_1^2}{1+L^2}\right), & & \eta = \int \rho^\star(x)\ln|p-x| \dd x-\frac{1}{2}\ln(1+p^2),
\end{align}
where $p$ is a point inside the support of $\rho^\star$ but outside the interval $[-L,L]$.

Calculating the action for the unconstrained case is, once again, straightforward and yields $S^{(C)}[\rho_{ca}]=\ln 2$. The final formula for the rate function for the number statistics of the Cauchy ensemble when the interval is $\I=[-L,L]$ is
\begin{align}
 \nonumber \psi^{(C)}(k_\I)=&\,\frac{1}{2}\intinf \rho^\star(x)\frac{\ln(1+x^2)}{2}\dd x - \frac{1}{2}\left[\int_{L}^{a_1} G(x) \dd x-\frac{1}{2}\ln\left(\frac{1+a_1^2}{1+L^2}\right)\right]k_{[-L,L]} \\
    &- \frac{1}{2}\left(H(p)-\frac{\ln(1+p^2)}{2}\right)-\frac{\ln 2}{2}\ ,\label{rate_cauchy}
\end{align}
where $H(x)=\int \rho^\star(x')\ln|x-x'| \dd x'$ and $p\in\text{supp }\rho^\star\setminus [-L,L]$. 
\subsection{Number variance}

To obtain the variance of $N_{[-L,L]}=Nk_{[-L,L]}$ for typical fluctuations, we repeat the asymptotic method applied above for Gaussian and Wishart ensembles. Since the Cauchy distribution has no edge, we expect no sharp decline on the variance. However, our method is not capable of describing the variance for the full range of $L$. Again, we find four different regimes: a {\it (i)} bulk regime, {\it (ii)} an extended bulk regime $1/N < L < N$, {\it (iii)} an  effective edge regime $L \sim N$ and {\it (iv)} a tail regime $L>N$. This is surprising, since the Cauchy distribution has no edge, but indeed in terms of the behavior of the variance we find that the average position of the largest eigenvalue (see \cite{MajSchVil13}), which is of order $N$, behaves as an effective edge for the Cauchy distribution.

\begin{figure}
    \centering
    \renewcommand\thesubfigure{(\roman{subfigure}) }
    \subfigure[Bulk regime.\label{fig:Cauchy bulk regime}]{\includegraphics[width=0.45\textwidth]{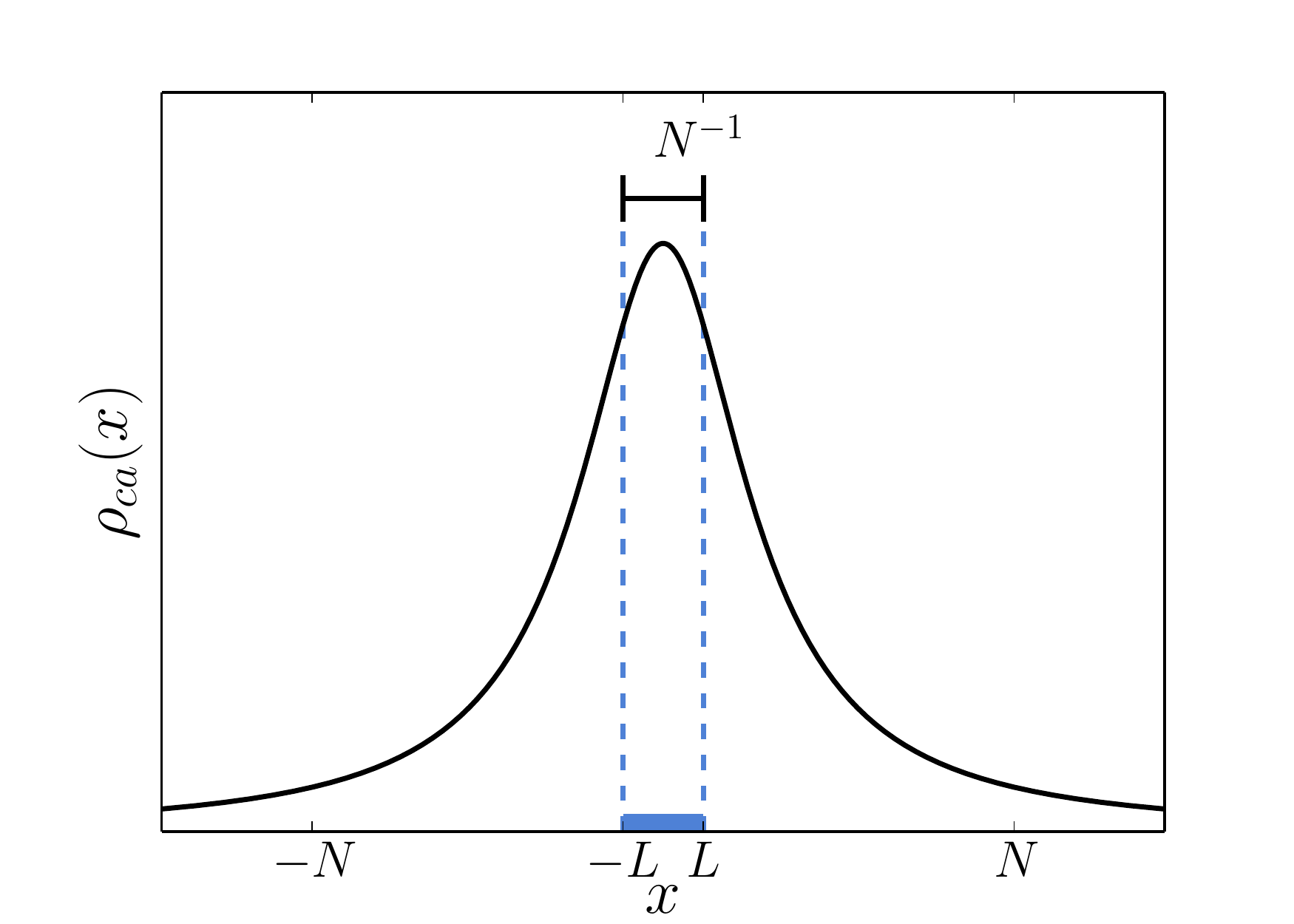}}
    \subfigure[Extended bulk regime.\label{fig:Cauchy extended regime}]{\includegraphics[width=0.45\textwidth]{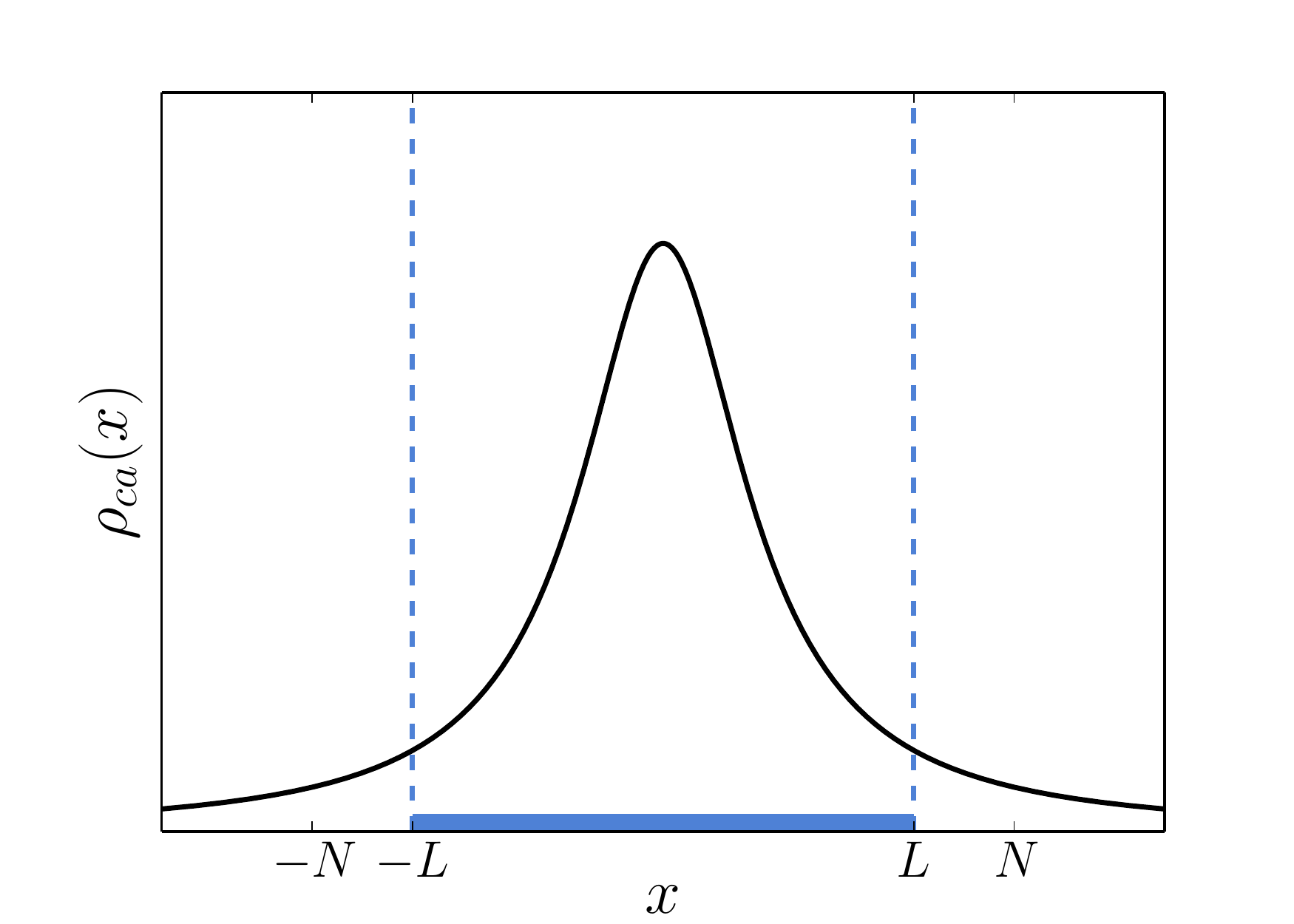}}
~
    \subfigure[Edge regime.\label{fig:Cauchy edge regime}]{\includegraphics[width=0.45\textwidth]{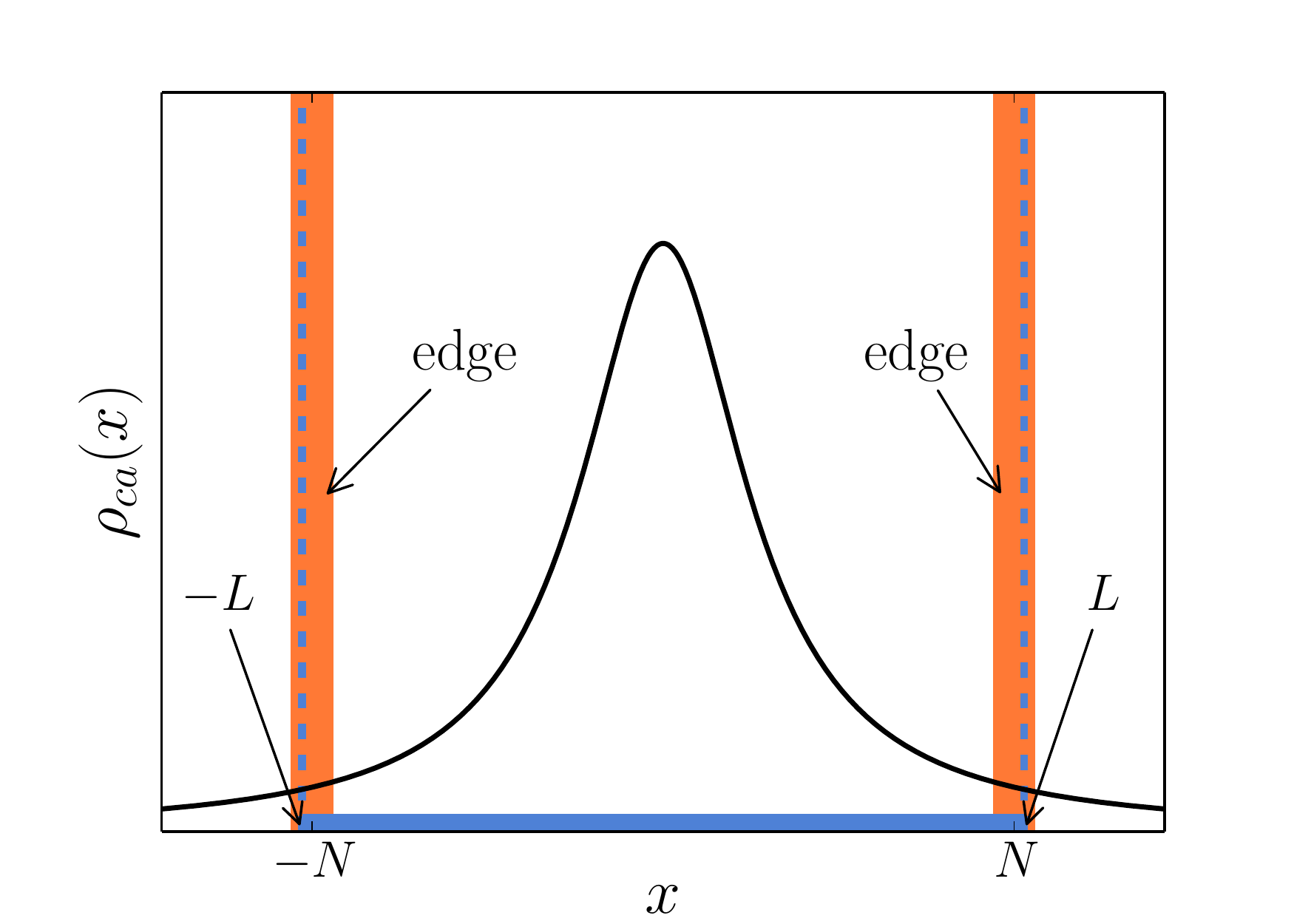}}
    \subfigure[Tail regime.\label{fig:Cauchy tail regime}]{\includegraphics[width=0.45\textwidth]{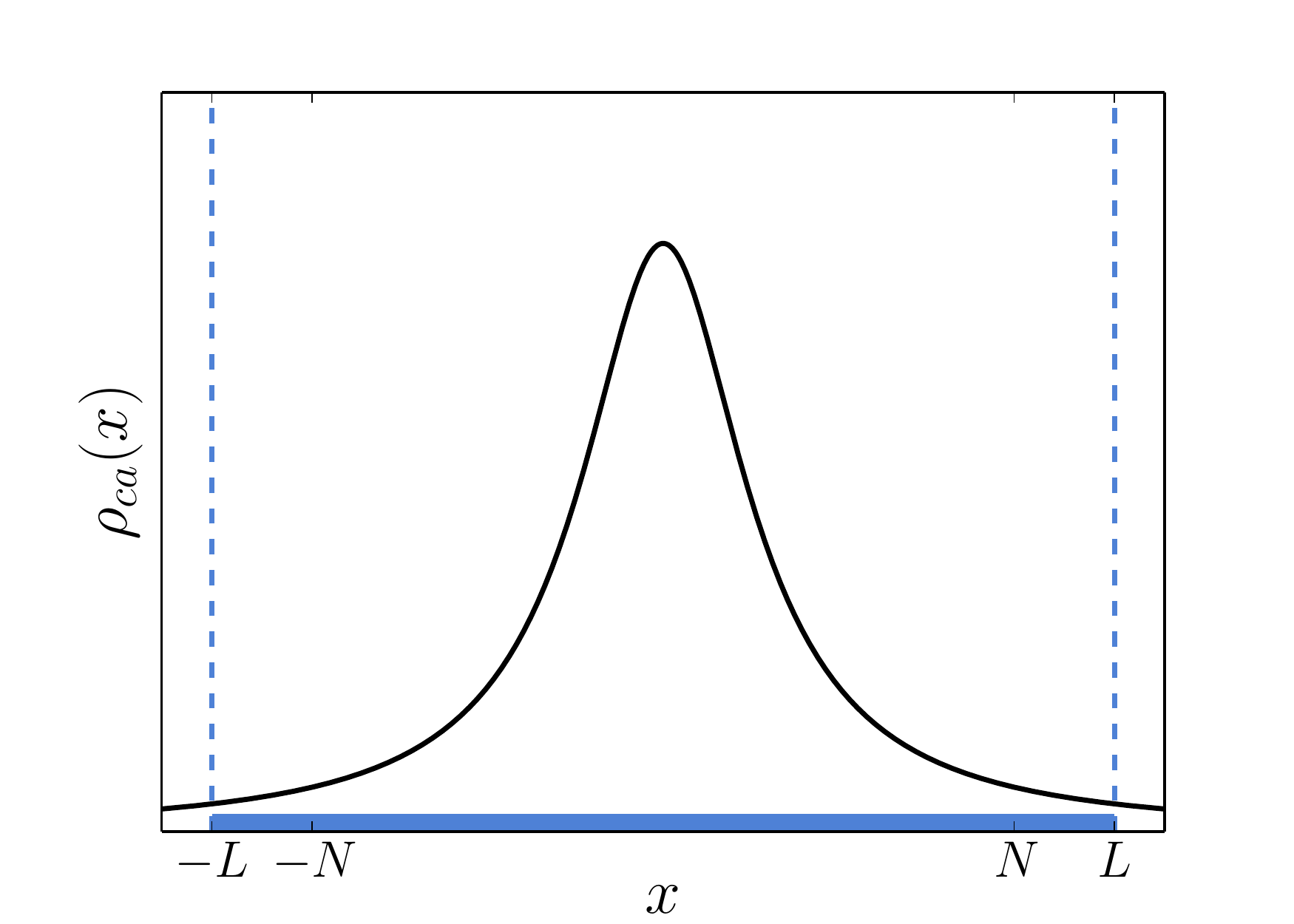}}
    \caption{Regimes of behavior of the number variance for the Cauchy ensemble. The solid blue line represents the interval $\I=[-L,L]$.}
    \label{fig:regimes Cauchy}
\end{figure}


Even in this simplified symmetric case, the number variance for the Cauchy regime is extremely difficult to obtain for all identified regimes, and we are only able to compute it for the extended bulk regime. The effective edge regime is problematic because the kernel for the Cauchy ensemble is built out of Jacobi polynomials at imaginary argument (see {\it e.g.} \cite{MajSchVil13}), whose scaled limit is complicated \cite{WitFor00}. The tail regime also presents problems, since a calculation for a finite number of eigenvalues beyond the effective edge $N$, similar to \eqref{tail_result} and \eqref{tail_result_wish}, exhibits individual divergences in $\langle k^ 2 \rangle$ and $\langle k \rangle^2$. Although their difference is finite, it is hard to extract with this method. 


The perturbative calculation of the rate function \eqref{rate_cauchy} around its minimum is  performed in a very similar way to the previous ensembles, and we omit it for brevity. For values of $L$ such as $1/N<L<N$ we find
\begin{equation}
    \Var(N_{[-L,L]})=\frac{2}{\beta\pi^2}\ln\left(\frac{NL}{1+L^2}\right)+\mathcal{O}(1).\label{var_cauchy_eq}
  \end{equation}  
The variance for the extended bulk rises until reaching its maximum value at $L^\star=1$ and falls towards zero as the interval size increases. Our calculations required approximations that are no longer valid if $L\sim N$, therefore we identify the regime of validity of this result as $1/N<L<N$. In this sense, the value $L\sim N$ behaves just as $L\sim \sqrt{2}$ for the Gaussian ensemble or $L\sim 4$ for the Wishart ensemble. It defines an ``effective edge'' that changes the behavior of the number variance when considering intervals whose edges lie in the vicinity of the average position of the large Cauchy eigenvalue.

\begin{figure}[!htbp]
    \centering
    \includegraphics[width=0.5\linewidth]{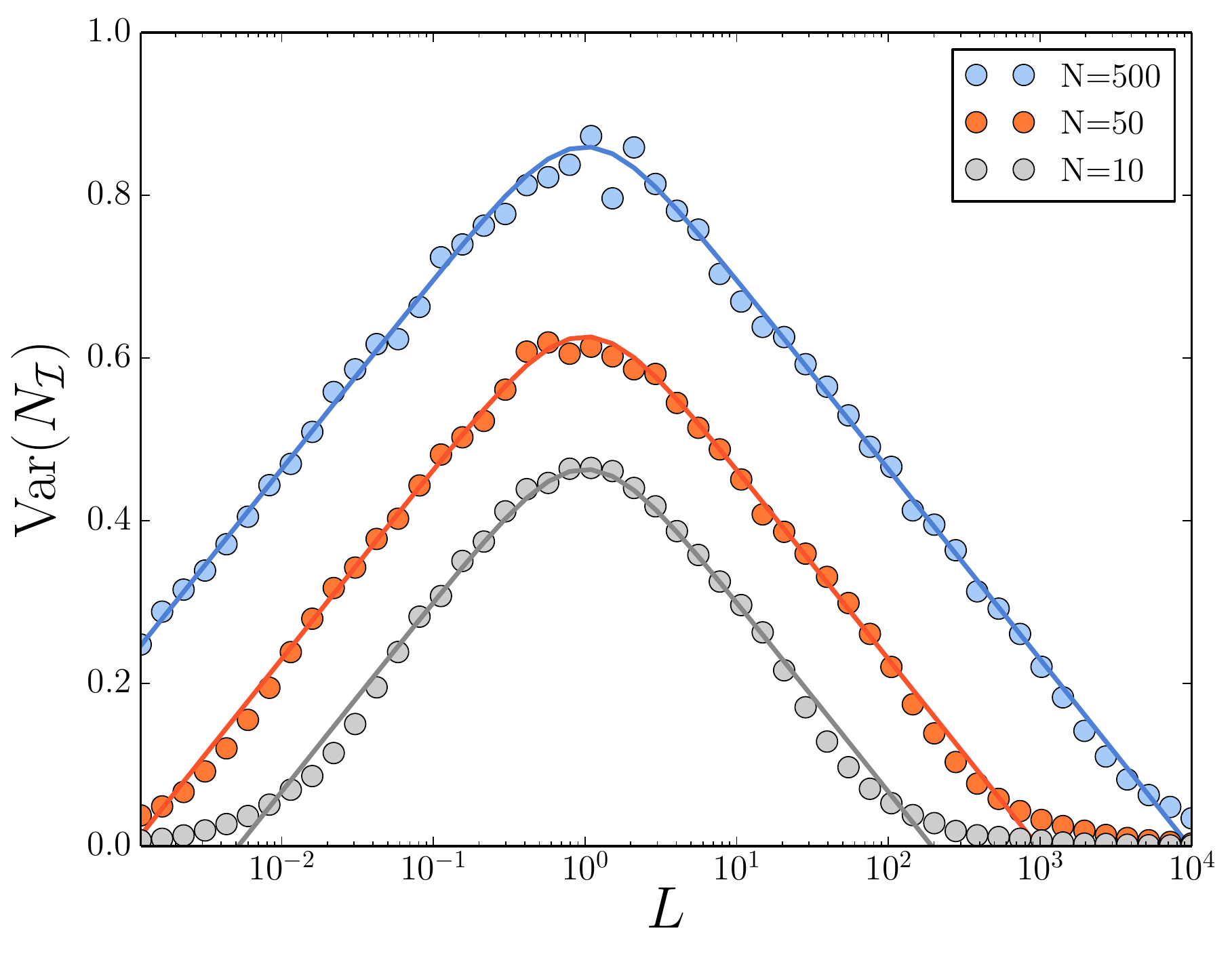}
    \caption{Results for the variance of $N_\I$ for the Cauchy ensemble for different values of $N$ when $\I=[-L,L]$. The theoretical result is in equation \eqref{var_cauchy_eq}.}
    \label{var_cauchy_fig}
\end{figure}
\subsection{Comparison with numerics}

We are able to simulate Cauchy eigenvalues by using the correspondence between the Cauchy and the circular ensemble \cite{MarMajSch14_2}. Comparing our prediction for the bulk regime in the Cauchy ensemble with numerical results, figure \ref{var_cauchy_fig}, we see perfect agreement within the theoretical result, specially the domain of validity of equation \eqref{var_cauchy_eq}. We notice how the theoretical prediction for the extended bulk in the three values of $N$ evaluated is no longer valid when $L>N$, which reinforces the idea of the presence of an effective edge at the average position of the largest eigenvalue.

\section{Conclusions}
\label{sec:conclusions}

In summary, we have considered $\beta$-ensembles of random matrices characterized by the confining potential $V(x)$. We have developed a general formalism (based on the combined application of the Coulomb gas and the resolvent technique in the presence of a wall) to determine the full distribution $\mathcal{P}_{\beta}^{(V)} (N_\I)$ (including large deviation tails) of the random variable $N_{\I}$, the number of eigenvalues contained in an interval $\I$ of the real line. We have shown that this probability in general scales for large $N$ as $\mathcal{P}_{\beta}^{(V)} (N_\I)\approx \exp\left(-\beta N^2 \psi^{(V)}(N_\I /N)\right)$, where $\beta$ is the Dyson index of the ensemble. The rate function $\psi^{(V)}(k_\I)$, independent of $\beta$, is computed in terms of single integrals for three classical random matrix ensembles: $\beta$-Gaussian (equation \eqref{action_gauss}), $\beta$-Wishart (equation \eqref{rate_wish}) and $\beta$-Cauchy (equation \eqref{rate_cauchy}). Expanding the rate function around its minimum, we have found that for small fluctuations around its average value, $N_\I$ has a Gaussian distribution modulated by a logarithmic behavior depending on both $N$ and $L$. As a result, for large $N$, the \emph{number variance} $\Var(N_\I)$ in the extended bulk regime behaves as 
\begin{equation}\label{scaling_form}
    \Var(N_\I)\approx \frac{2}{\beta\pi^2}\ln(N g^{(V)}(\I))+\mathcal{O}(1)\ ,
\end{equation}
where the function $g^{(V)}(\I)$ is non-universal. It depends on both the potential $V(x)$ and the interval $\I$, but is independent
of $\beta$. For example, for the $\beta$-Gaussian case, with $\I = [-L,L]$, one finds from Eq. (\ref{var_gauss_index})
\begin{equation}\label{g:Gauss}
g^{(G)}([-L,L]) \equiv g^{(G)}(L) = L\,(2-L^2)^{3/2} \;.
\end{equation}
Similarly, for the Wishart case, for an interval $\I = [1,1+l]$ one finds from Eq. (\ref{var_wishart_index}) 
\begin{equation}\label{g:Wishart}
g^{(W)}([1,1+l]) \equiv g^{(W)}(l) = l\,(3-l)^{3/4} \;.
\end{equation}
Finally, for the Cauchy case, with interval $\I = [-L,L]$, we get from Eq. (\ref{var_cauchy_eq})
\begin{equation}\label{g:Cauchy}
g^{(C)}([-L,L]) \equiv g^{(C)}(L) = \frac{L}{(1+L^2)} \;.
\end{equation}
%

\begin{figure}[!htbp]
    \includegraphics[width=0.60\linewidth]{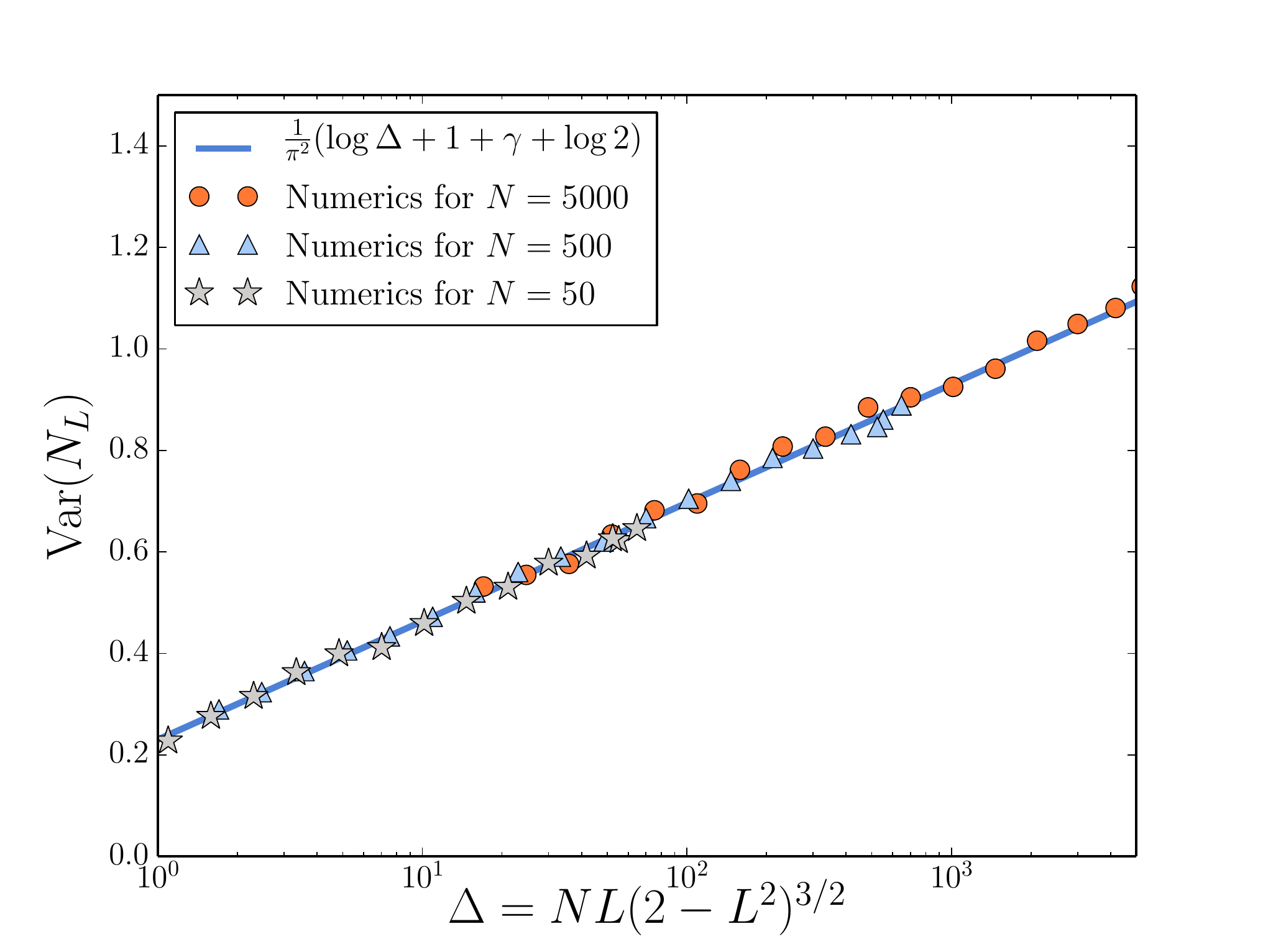}
    \caption{Numerical simulations of the number variance for GUE for different matrix sizes. The solid line represents $1/\pi^2 \ln(\Delta) + C_2$ where $C_2 = (1+\gamma+\ln 2)/\pi^2$ [see Eq. (\ref{C2})] is the Dyson-Mehta constant.}
    \label{fig:scaling comparison}
\end{figure}

This scaling form for the number variance as shown in Eq. (\ref{scaling_form}) seems to be quite generic for large $N$, and for all potentials. 
The main result of this paper is to obtain this function $g^{(V)}(\I)$ explicitly for the three classical RMT ensembles. Note that for intervals $[-L,L]$ (for the Gaussian and Cauchy ensembles) and $[1,1+l]$ (for the Wishart), our results reproduce the Dyson limit when $L \ll 1$ and $l \ll 1$. The full functional dependence of $g^{(G)}(L) = L (2-L^2)^{3/2}$ on $L$ (and similarly for the other two cases) however explains the non-monotonic growth of the variance as a function of $L$ for all $L$ up to the edge of the density (in the Gaussian and Wishart cases), thus going beyond the Dyson limit. This result is valid in the whole extended bulk region. For the Gaussian case, this scaling prediction that the number variance is the logarithm of a single scaling variable $\Delta$, $\Var(N_\I) \approx \frac{2}{\beta\pi^2}\log\Delta$, where $\Delta=Ng^{(G)}(L) = N\,L (2-L^2)^{3/2}$, has been verified by numerical simulations, for different values of $L$ and $N$ and collapsing all the curves on a single master curve with $\Delta$ as the $x$-axis (see Fig. \ref{fig:scaling comparison}).  

In the context of cold atoms our results for GUE make concrete predictions for the number variance and its dependence on the 
interval size $L$ at zero temperature for fermions trapped in a one-dimensional harmonic well. It would be nice if these results could be verified in cold atoms experiments. Recently, the random matrix techniques have been extended to study the properties of trapped fermions at finite temperature \cite{DeaLedMaj14}, as well as in higher dimensions \cite{DeaLedMaj15}. In both cases, a determinantal structure was found which may be exploited to obtain concrete predictions for the number variance of a domain at finite temperature and in higher dimensions.

\

\appendix

\section{Asymptotic analysis of Gaussian extended bulk regime}
\label{app}

We recall that the average density of eigenvalues for the Gaussian ensemble when the number of eigenvalues inside the interval $\I=[-L,L]$ is $k_\I$ is given by
\begin{equation}
    \rho^\star(x)=\frac{1}{\pi}\sqrt{\frac{\left(x^2-a^2\right)(b^2-x^2)}{x^2-L^2}},
\end{equation}
where $a$ and $b$ are the edges of the average density. When $k_\I=\overline{k_\I}$, this distribution is Wigner's semicircle, with $L=a$ and $b=\sqrt{2}$. We perturb the edges of the support by a small parameter with respect to Wigner's semicircle. The number of eigenvalues inside the interval will be $k_\I=\overline{k_\I}+\delta$ and the edges of the average density will shift by a small parameter
\begin{align}
    a=L+\epsilon, & & b=\sqrt{2}-\epsilon_2.
\end{align}
The normalization condition of the density imposes the following condition, to leading order in $\epsilon$ and $\epsilon_2$:
\begin{equation}
    \sqrt{2}\epsilon_2=L\epsilon.
\end{equation}

Our goal is to expand for the symmetric interval case $\I=[-L,L]$ the rate function action $\psi^{(G)}(\overline{k_\I}+\delta)$ into powers of $\delta$ and $\epsilon$, then calculate the relation between $\delta$ and $\epsilon$. We recall the expression of the Gaussian rate function for the symmetric interval 
\begin{align}
\nonumber   \psi^{(G)}(k_\I)=&\frac{1}{2}\int \frac{x^2}{2}\rho^\star(x)\dd x - \frac{\mu}{2}k_\I - \frac{\eta}{2}-S^{(G)}[\rho_{sc}]\\
\nonumber   =&\,\frac{1}{2}\intinf \rho^{\star}(x)\frac{x^2}{2}\dd x - \frac{1}{2}\left(\int_{L}^{a} G(x) \dd x-\frac{a^2}{2}+\frac{L^2}{2}\right)k_\I \\
    &- \frac{1}{2}\left[\ln b- \frac{b^2}{2} - \int_{b}^\infty \left(G(x)-\frac{1}{x}\right)\dd x\right]-\frac{3}{8}-\frac{\ln 2}{4}\ .\label{action_gauss_app}
\end{align}

We also recall  the resolvent for the symmetric case
\begin{align}
    G(z)=z\pm\sqrt{\frac{(z^2-b^2)(z^2-a^2)}{z^2-L^2}}.
\end{align}


For simplicity, we take the case $k_\I=\overline{k_\I}+\delta>\overline{k_\I}$, represented in figure \ref{fig:gaussian_density_larger}. We write the perturbed edges as $a=L+\epsilon$ and $b=\sqrt{2}-L\epsilon/\sqrt{2}$. We evaluate the integrals in \eqref{action_gauss_app} separately up to leading order of $\epsilon$ and $\delta$.

\begin{itemize}
    \item $I_1=\int x^2\rho^\star(x)\dd x$\\
    This integral is the second moment of the average density. We note that expanding $G(z)$ for large $z$ provides us with the following series
    \begin{equation} 
        G(z)=\int \frac{\rho^\star(x)}{z-x}\dd x = \sum_{n=0}^\infty \frac{1}{z^{n+1}}\int x^n\rho^\star(x)\dd x\ .
    \end{equation}

    Therefore $I_1$ is just given by the $z^{-3}$ coefficient in the expansion of $G(z)$ in powers of $1/z$. Applying the normalization condition \eqref{norm_cond_gauss_1} and \eqref{norm_cond_gauss_2} we find
    \begin{equation} 
        I_1=\intinf x^2\rho^\star(x)\dd x = \frac{1}{8}\left(a^4-2 a^2 b^2+2 a^2 L^2+b^4+2 b^2 L^2-3 L^4 \right)\ .
    \end{equation}
     
     Keeping only leading orders in $\epsilon$, we find
     \begin{equation} 
        I_1=\intinf x^2\rho^\star(x)\dd x = \frac{1}{2}+L(2-L^2)\epsilon +o(\epsilon)\ .
     \end{equation}

    \item $\mu=\int_{L}^{a} G(x) \dd x - \frac{a^2}{2}+\frac{L^2}{2}$
        
    Since the domain of integration is very small,  of size $\epsilon$, and we are keeping only the leading orders in $\epsilon$, we may treat separately the slow and the fast varying parts of the square-root. 
    \begin{align}
        \nonumber \mu=&\int_{L}^{a}G(x)\dd x -\frac{a^2}{2}+\frac{L^2}{2}=-\int_{L}^{a} \sqrt{\frac{(x^2-b^2)(x^2-a^2)}{x^2-L^2}}\dd x \\
        = & -\sqrt{\frac{(b^2-L^2)(L+a)}{2L}}\int_{L}^{a}\sqrt{\frac{a-x}{x-L}}\dd x + o(\epsilon) = -\frac{\pi}{2}\sqrt{2-L^2}\epsilon+o(\epsilon)\ .
    \end{align}

    \item $\eta=\ln b - \int_{b}^\infty \left(G(x)-\frac{1}{x}\right)\dd x-\frac{b^2}{2}$\\
    To compute the integral $\int \left(G(x)-1/x\right) \dd x$, we perform a suitable splitting of the domain of integration before expanding the integrand in powers of $\epsilon$ to keep only the leading orders in $\epsilon$.
    \begin{align}
        \nonumber\int_{b}^{\infty} \left(G(x)-\frac{1}{x}\right)\dd x =& \int_{b}^{\sqrt{2}} \left(G(x)-\frac{1}{x}\right)\dd x + \int_{\sqrt{2}}^\infty \left(G(x)-\frac{1}{x}\right)\dd x\\
        \nonumber =& \int_{b}^{\sqrt{2}} \left(x-\frac{1}{x}\right)\dd x -  \int_{b}^{\sqrt{2}} \sqrt{\frac{(x^2-b^2)(x^2-a^2)}{x^2-L^2}}+o(\epsilon)\\
        \nonumber &+\int_{\sqrt{2}}^\infty \left( x-\sqrt{x^2-2} - \frac{1}{x}\right) + \epsilon \int_{\sqrt{2}}^\infty \frac{L\left(2-L^2\right) }{\sqrt{x^2-2} \left(x^2-L^2\right)}\dd x\\
        =&\ln 2-\frac{1}{2}+\sqrt{2-L^2}\arcsin\left(\frac{L}{\sqrt{2}}\right)\epsilon+\frac{L\epsilon}{2}+o(\epsilon).
    \end{align}         
    The remaining terms in the definition of $\eta$ are calculated directly:
    \begin{equation} 
        \ln b - \frac{b^2}{2} = \frac{\ln 2}{2}-1+\frac{L\epsilon}{2}  + o(\epsilon).
    \end{equation}
    This yields
    \begin{equation} 
     \eta = -\frac{\ln 2}{2}-\frac{1}{2}-\sqrt{2-L^2}\arcsin\left(\frac{L}{\sqrt{2}}\right)\epsilon + o(\epsilon)\ .
    \end{equation}

    \item $k_{[-L,L]}=\int_{-L}^L \rho^\star(x) \dd x$\\
    The integral $\int_{-L}^L \rho^\star(x) \dd x$ may be evaluated asymptotically using
    \begin{equation} 
        \int_{-L}^L \rho^\star(x) \dd x = \int_{-L}^{-L+\epsilon}\rho^\star(x)\dd x + \int_{-L+\epsilon}^{L-\epsilon}\rho^\star(x)\dd x + \int_{L-\epsilon}^{L}\rho^\star(x)\dd x\ .\label{intintint}
    \end{equation}
    We then expand the second integral in powers of $\epsilon$ and integrate term by term to get
    \begin{equation} 
        \int_{\epsilon}^{L-\epsilon_2}\rho^\star(x)\dd x = \overline{k_{[-L,L]}} + \underbrace{\epsilon\frac{\sqrt{2-L^2}}{ \pi }\left(\ln \left( 2-L^2\right)+\ln L- \ln \epsilon \right)}_{=\delta} + \mathcal{O}(\epsilon)\ ,
        \label{middle_chunk}
    \end{equation}  
    the remaining integrals in \eqref{intintint} being $o(\epsilon)$ and hence sub-leading. The value of $\overline{k_{[-L,L]}}$ is simply given by
    \begin{equation}
        \overline{k_{[-L,L]}}=\frac{1}{\pi}\int_{-L}^L\sqrt{2-x^2}\dd x=\frac{L\sqrt{2-L^2} +2 \arcsin\left(\frac{L}{\sqrt{2}}\right)}{\pi }.
    \end{equation}
    From this expansion we obtain the relation between $\epsilon$ and $\delta$, to leading order
\begin{equation}
    \delta\approx\epsilon\frac{\sqrt{2-L^2}}{ \pi }\left(\ln \left( 2-L^2\right)+\ln L- \ln \epsilon \right)\label{delta_app},
\end{equation}
as in equation \eqref{delta}.
\end{itemize}

After a number of cancellations, we eventually obtain the leading term in $\epsilon$ and $\delta$ for the rate function for $\I=[-L,L]$
\begin{equation} 
    \psi^{(G)}(k_\I)=\frac{1}{4}\int x^2\rho^\star(x)\dd x - \frac{\eta}{2}-\frac{\mu}{2}k_I -\frac{3}{8}-\frac{\ln 2}{4} =  \frac{\pi}{4}\sqrt{2-L^2}\epsilon\delta + o(\epsilon^2\ln\epsilon)\ ,
\end{equation}
as in equation \eqref{rate_2}.

\bibliographystyle{apsrev4-1}

\end{document}